\documentclass[a4paper,11pt]{article}
\usepackage{graphicx,epsfig,pstricks,fancybox,fancyhdr,epic,rotating,color}
\usepackage{amsfonts}
\definecolor{blue-green}{rgb}{0.0, 0.87, 0.87}
\definecolor{bleudefrance}{rgb}{0.19, 0.55, 0.91}
\definecolor{darkpastelgreen}{rgb}{0.01, 0.75, 0.24}

\begin{document}

\begin{center}
{\color{black} \large \bf
Mass Generation in QCD - Oscillating Quarks and Gluons
}
\vspace*{0.1cm} \\
{\bf \color{black} based in part on 
'Tuning to harmonic numbers of oscimodes 
of baryons'}
\vspace*{0.3cm} \\
\vspace*{0.1cm}

\color{black}
{\bf Peter Minkowski}
\vspace*{0.2cm} \\
{\bf Albert Einstein Center for Fundamental Physics 
\vspace*{0.1cm} \\
ITP, University 
of Bern and CERN, PH-TH division}
\\
\vspace{0.2cm}

{\color{black} \bf
Abstract 
}
\vspace*{0.0cm} \\

{\color{black} 
\begin{tabular}{l}
The present lecture is devoted to embedding the approximate 
genuine 
\vspace*{-0.0cm} \\
harmonic oscillator structure 
of valence $\ q \ \overline{q} \ $ mesons and in more 
\vspace*{-0.0cm} \\
detail the $\ q \ q \ q \ $ configurations 
for u,d,s flavored baryons 
in QCD 
\vspace*{-0.0cm} \\
for three light flavors of quark. It includes notes, 
preparing the
\vspace*{-0.0cm} \\
counting of
'oscillatory modes of $\ N_{\ fl} \ = \ 3 \ $ light quarks, 
u , d , s 
\vspace*{-0.0cm} \\
in baryons', using the
$\mbox{SU}\left ( 2 N_{ fl} = 6 \right ) \times 
\mbox{SO3} 
 \left ( \vec{L} \right ) $ broken symmetry 
\vspace*{-0.0cm} \\
classification, extended to the 
harmonic oscillator symmetry of 3 
\vspace*{-0.0cm} \\
paired oscillator modes.
$ \vec{L} = \sum_{ n = 1}^{ N_{ fl}} \vec{L}_{ n}$ 
stands for the space rotation 
\vspace*{-0.0cm} \\
group generated by the sum 
of the 3 individual angular momenta
of 
\vspace*{-0.0cm} \\
quarks 
in their c.m. system. 
The oscillator extension to valence 
\vspace*{-0.0cm} \\
gauge boson states is not yet
developed 
to a comparable level .
\end{tabular}
\vspace*{0.0cm}

}

\end{center}

\newpage

{\color{black} 

\vspace*{-0.3cm}
\begin{center}
{\bf \color{black} List of contents}
\end{center}
\vspace*{-0.4cm}

\begin{tabular}{@{\hspace*{-2.7cm}}l@{\hspace*{-0.15cm}}l@{\hspace*{-0.1cm}}l
@{\hspace*{0.0cm}}}
1 & \begin{tabular}[t]{l}
Introduction
\end{tabular}
& ~\pageref{'1'}
\vspace*{-0.1cm} \\
2 & \begin{tabular}[t]{l}
Bose condensation of elementary
scalar fields;  the Brout-Englert-Higgs effect
\end{tabular}
& ~\pageref{'2'}
\vspace*{-0.1cm} \\
2-1 & \begin{tabular}[t]{l} 
Primary gauge breaking
\end{tabular}
& ~\pageref{'2-1'}
\vspace*{-0.1cm} \\
-- & \begin{tabular}[t]{l} 
the Majorana logic characterized by 
$\ {\cal{N}}_{\ e,\mu,\tau} \ $
\end{tabular}
& ~\pageref{'MajLogic'}
\vspace*{-0.1cm} \\
-- & \begin{tabular}[t]{l} 
specific notation for scalar field variables
\end{tabular}
& ~\pageref{'scalarvar'}
\vspace*{-0.1cm} \\
-- & \begin{tabular}[t]{l} 
to get an idea of the power of the set of scalars
\end{tabular}
& ~\pageref{'scpower'}
\vspace*{-0.1cm} \\
2-1-a & \begin{tabular}[t]{l} 
Primary gauge breaking - type a)
\end{tabular}
& ~\pageref{'2-1-a'}
\vspace*{-0.1cm} \\
2-1-b & \begin{tabular}[t]{l} 
Primary gauge breaking - type b)
\end{tabular}
& ~\pageref{'2-1-b'}
\vspace*{-0.1cm} \\
3 & \begin{tabular}[t]{l}
Mass Generation in QCD with 3 light flavors 
\vspace*{-0.1cm} \\
- oscillating Quarks and Gluons
\end{tabular}
& ~\pageref{'3'}
\vspace*{-0.1cm} \\
3-1 & \begin{tabular}[t]{l} 
embedding oscillator modes in u,d,s flavored
baryons in QCD
\end{tabular}
& ~\pageref{'3-1'}
\vspace*{-0.1cm} \\
3-1-1 & \begin{tabular}[t]{l}
Assembling elements of the QCD
Lagrangean density -- premises 
\end{tabular}
& ~\pageref{'3-1-1'}
\vspace*{-0.1cm} \\
3-1-2a & \begin{tabular}[t]{l} 
Gauge boson binary bilocal and adjoint ( here octet- ) string operators
\end{tabular}
& ~\pageref{'3-1-2a'}
\vspace*{-0.1cm} \\
2-3 & \begin{tabular}[t]{l} 
$\ \overline{q} \ q \ $ bilinears and triplet-string operators
\end{tabular}
& ~\pageref{'2-3'}
\vspace*{-0.1cm} \\
2-4 & \begin{tabular}[t]{l}
Connection and curvature - forms
\vspace*{-0.1cm} \\
preparing 
the ensuing analysis of regularity conditions
\end{tabular}
& ~\pageref{'2-4'}
\vspace*{-0.1cm} \\
2-5 & \begin{tabular}[t]{l}
The U1- or singlet axial current anomaly
\end{tabular}
& ~\pageref{'2-5'}
\vspace*{-0.1cm} \\
2-6 & \begin{tabular}[t]{l}
Quark masses and splittings : $\ m_{\ f} \ $ and $\ \Delta \ m_{\ f} 
\ = \ m_{\ f} - 
\ \left \langle \ m \ \right \rangle \ $
\end{tabular}
& ~\pageref{'2-6'}
\vspace*{-0.1cm} \\
2-7 & \begin{tabular}[t]{l} 
The scale- or trace- anomaly 
\end{tabular}
& ~\pageref{'2-7'}
\vspace*{-0.1cm} \\
2-8 & \begin{tabular}[t]{l} 
The two central anomalies alongside : scale- or trace- and 
U1-axial anomaly
\end{tabular}
& ~\pageref{'2-8'}
\vspace*{-0.1cm} \\
4 & \begin{tabular}[t]{l}
Ideas forging and foregoing - the dynamics of 
genuinely oscillatory
\vspace*{-0.1cm} \\
modes \cite{oscmod1980} , \cite{PMembedosc}
\end{tabular}
& ~\pageref{'4-1'}
\vspace*{-0.1cm} \\
-- & \begin{tabular}[t]{l}
Extension to include the $\ N \ q \ -$ bond  
\end{tabular}
& ~\pageref{'1-1N'}
\vspace*{-0.1cm} \\
5 & \begin{tabular}[t]{l} 
First results from counting 
oscillatory modes in u,d,s flavored baryons 
\end{tabular}
& ~\pageref{'5'}
\vspace*{-0.1cm} \\
1 & \begin{tabular}[t]{l}
Introduction'
\end{tabular}
& ~\pageref{''1''}
\vspace*{-0.1cm} \\
2 & \begin{tabular}[t]{l}
Factoring out the {\it approximate} symmetry group
\vspace*{-0.1cm} \\
in spin-flavor space as well as overall color
\end{tabular}
& ~\pageref{''2''}
\vspace*{-0.1cm} \\
6 & \begin{tabular}[t]{l}
Concluding remarks , outlook
\end{tabular}
& ~\pageref{'6'}
\vspace*{-0.1cm} \\
Ap1 & \begin{tabular}[t]{l}
Appendix 1: The spin (10) product representations
\end{tabular}
& ~\pageref{'E'}
\vspace*{-0.1cm} \\
-- & \begin{tabular}[t]{l}
Some conclusions from sections 1-1 and 1-2 
\end{tabular}
& ~\pageref{Sconc}
\vspace*{-0.1cm} \\
2-1+ & \begin{tabular}[t]{l} 
The Majorana logic 
\cite{FukuYana}
and mass from mixing -- setting \\
within the 'tilt to the left' or 'seesaw' of type I $\cdots$
characterized by $\ {\cal{N}}_{\ F} \ $
\end{tabular}
& ~\pageref{'2-1+'}
\vspace*{-0.1cm} \\
1-1a  & \begin{tabular}[t]{l} 
There does not exist a symmetry -- within
the standard model including
\vspace*{-0.1cm} \\
gravity
and containing only chiral spin $\frac{1}{2} \ $ 16 families of SO (10) 
\vspace*{-0.1cm} \\
which could enforce the vanishing of neutrino mass(es) .
\end{tabular}
& ~\pageref{'1-1a'}
\vspace*{-0.1cm} \\
1-2a  & \begin{tabular}[t]{l} 
There does not exist a symmetry -- within
the standard model including  
\vspace*{-0.1cm} \\
gravity and containing only chiral 16 families of SO (10),  
enforcing the 
\vspace*{-0.1cm} \\
vanishing of neutrino mass, 
yet chiral extensions can accomplish this 
\end{tabular}
& ~\pageref{'1-2a'}
\vspace*{-0.1cm} \\
R & \begin{tabular}[t]{l}
References
\end{tabular}
& ~\pageref{refmain}
\vspace*{-0.1cm} \\
R1 & \begin{tabular}[t]{l}
References for Appendix 1 and Supplementary material
\end{tabular}
& ~\pageref{rAp1}
\vspace*{0.4cm}
\end{tabular}

}

\newpage 

\begin{center}
\vspace*{-0.0cm}
{\bf \color{black} 1 - Introduction
 }
\label{'1'}
\end{center}
\vspace*{-0.3cm}

\noindent
The two main mass generation mechanisms within a general gauge field theory
-- in 3 + 1 uncurved space-time dimensions
-- henceforth called gravitationless gauge field theory --

\noindent
-- minimally the neutrino mass extended standard model based on the gauge group
$ SU3_{ c} \times SU2_{ L}  \times U1_{ {\cal{Y}}} $
and one scalar doublet with respect to $\ SU2_{ L} \ $ --

\noindent
form the basis of the present outline.
\vspace*{0.0cm}

\begin{description}

\item 1) the Bose condensation of some components of elementary scalar fields
\vspace*{-0.05cm} \\
scalar stands here for scalar and/or pseudoscalar Yukawa couplings

\item 2) the Bose condensation of the gauge field-strength bilinear
\vspace*{-0.05cm} \\
-- gauge- and renormilzation group invariant with respect to the 
{\it unbroken} gauge group $\ SU3_{\ c} \ $ --
\vspace*{-0.2cm}
\end{description}

\begin{center}
Table 1
\vspace*{-0.3cm}
\end{center}

\begin{displaymath}
\begin{array}{@{\hspace*{6.32cm}}l@{\hspace*{6.33cm}}}
\hline  \\
\end{array}
\vspace*{-0.6cm}
\end{displaymath}

\noindent
The key features of point 1) in table 1) are outlined in section 2.

\noindent
The main topic here, point 2) in table 1, is elaborated on in section 3,
and worked out in more detail in sections 4 - 6 and references quoted therein .
\vspace*{0.3cm}

\begin{center}
\vspace*{-0.3cm}
{\bf \color{black} 2 - Bose condensation of elementary
scalar fields ; the Brout-Englert-Higgs effect
 }
\label{'2'}
\end{center}
\vspace*{-0.3cm}

\noindent
Recent assessments can be found in the talks of Fran\c{c}ois Englert 
-- \cite{NobelFrancoisEnglert}
and, more historically oriented, Peter Higgs \cite{NobelPeterHiggs}
at the Nobel Prize 2013 awards ceremony.

\noindent
We base the general properties of spontaneous gauge breaking of an enveloping  
gravitationless gauge field theory , based on a gauge group 
$\ G_{\ env} \ \supset G_{\ min} \ $ in the sense of gauge group unification
beyond the minimal case 
$\ G_{\ min} \ = \ SU3_{c} \ \times SU2_{\ L} \ \times U1_{\ {\cal{Y}}} \ $, 
presented in the introduction.

\noindent
A minimal such enveloping gauge group is 

\vspace*{-0.3cm}
\begin{equation}
\label{eq:2-1}
\begin{array}{l}   
G_{\ env}^{\ min} \ = \ SO \ (10) \ \equiv \ spin \ (10)
\end{array}
\end{equation}

\noindent
as discussed in refs. \cite{SO10} and \cite{nuoriginPM} .

\noindent
Within larger {\it simple} enveloping groups the exceptional chain

\vspace*{-0.3cm}
\begin{equation}
\label{eq:2-2}
\begin{array}{l}   
G_{\ env} \ \rightarrow \ E6 \ \subset \ E8 \ \supset \ E6 \ \times 
\ SU \ (3) 
\end{array}
\end{equation}

\noindent
is singled out \cite{GurseyRamSik}, \cite{Gurseyoct} , most importantly because 
it offers the possibility of
canceling all gauge- and gravitational anomalies in the product
gauge group \cite{WittenAlv} , \cite{GreenSchwarzWitten2012}

\vspace*{-0.3cm}
\begin{equation}
\label{eq:2-3}
\begin{array}{l}
G_{\ env} \ = \ E8 \ \times \ E8
\end{array}
\end{equation}

\noindent
Following the hypothesis of an underlying unifying gauge group
the top down of gauge breaking is initiated by a primary breaking
followed last by the electroweak gauge breaking .



\vspace*{-0.3cm}
\begin{equation}
\label{eq:2-4}
\begin{array}{l}
M_{\ env} \ \gg \ v \ = 
\ \left ( \sqrt{\ \sqrt{2} \ G_{\ F} \ } \ \right )^{\ -1} 
\ = \ 246.220 \ \mbox{GeV}
\end{array}
\end{equation}

\noindent
In eq. \ref{eq:2-4} v denotes the v.e.v. of the unique doublet scalar field 
using the quaternion associated basis for the local scalar fields

\vspace*{-0.6cm}
\begin{equation}
\label{eq:2-5}
\begin{array}{l}
z \ ( \ x \ ) 
\ =
\ \left ( \ \begin{array}{lr}
\varphi^{\ 0}   & - \ \left ( \ \varphi^{\ -} \ \right )^{\ *}
\vspace*{0.1cm} \\
\varphi^{\ -}   & \left ( \ \varphi^{\ 0} \ \right )^{\ *}
\end{array}
\ \right ) \ ( \ x \ )
\vspace*{0.1cm} \\
\varphi^{\ 0} \ = \ 1 \ / \ \sqrt{2} 
\ \left ( \ Z_{\ 0} \ - \ i \ Z_{\ 3} \ \right )
\hspace*{0.2cm} ; \hspace*{0.2cm}
\varphi^{\ -} \ = \ 1 \ / \ \sqrt{2} 
\ \left ( \ Z_{\ 2} \ - \ i \ Z_{\ 1} \ \right )
\vspace*{0.1cm} \\
Z_{\ \mu} \ = \ \left ( \ Z_{\ \mu} \ \right )^{\ *} \ = 
\ \left ( \ Z_{\ 0} \ , \ Z_{\ 1} \ , \ Z_{\ 2} \ , \ Z_{\ 3} 
\ \right ) \ = \ \left ( \ Z_{\ 0} \ , \ \vec{Z} \ \right )
\vspace*{0.1cm} \\
\sigma_{\ 0} = 
\ \left ( \ \begin{array}{lr}
1 & 0
\vspace*{0.1cm} \\
0 & 1
\end{array} \ \right ) 
\ ,
 \sigma_{\ 1} =
 \left ( \ \begin{array}{lr}
0 & 1
\vspace*{0.1cm} \\
1 & 0
\end{array} \ \right )
\ ,
\ \sigma_{\ 2} =
\left ( \ \begin{array}{lr}
0 & - i
\vspace*{0.1cm} \\
i & 0
\end{array} \ \right )
\ ,
\vspace*{0.1cm} \\
\ \sigma_{\ 3} =
\left ( \ \begin{array}{lr}
1 & 0
\vspace*{0.1cm} \\
0 & - 1
\end{array} \ \right )
\vspace*{0.1cm} \\
\sigma_{\ \mu} \ = \ \left ( \ \sigma_{\ 0} \ , 
\ \sigma_{\ 1} \ , \ \sigma_{\ 2} \ , \ \sigma_{\ 3} 
\ \right ) \ = \ \left ( \ \sigma_{\ 0} \ , \ \vec{\sigma} \ \right )
\end{array}
\vspace*{-0.0cm}
\end{equation}

\noindent
In eq. \ref{eq:2-5} the symbol $\ ^{*}$ denotes hermitian conjugation of
individual complex and/or real field components.

\noindent
Thus the quantity z , defined in eq. \ref{eq:2-5} , shows its quaternionic
representation

\vspace*{-0.3cm}
\begin{equation}
\label{eq:2-6}
\begin{array}{l}
z \ ( \ x \ ) \ = \ \frac{1}{\sqrt{2}} \ \left (
\ Z_{\ 0} \ \sigma_{\ 0} 
\ + \ \sum_{\ k = 1}^{\ 3} \ Z_{\ k} \ \frac{1}{i} 
\ \sigma_{\ k} \ \right ) \ ( \ x \ )
\end{array}
\end{equation}

\noindent
The four $\ 2 \times 2 \ $ matrices , displayed in eq. \ref{eq:2-5} ,
form a 1 to 1 true representation of the base quaternions

\vspace*{-0.3cm}
\begin{equation}
\label{eq:2-7}
\begin{array}{l}
\sigma_{\ 0} \ \leftrightarrow \ q_{\ 0} \ = \ \P
\hspace*{0.2cm} ; \hspace*{0.2cm}
\frac{1}{i} \ \sigma_{\ m} \ \leftrightarrow \ q_{\ m} 
\hspace*{0.2cm} \mbox{for} \hspace*{0.2cm} m \hspace*{0.7cm} = \ 1,2,3
\vspace*{0.1cm} \\
q_{\ m} \ q_{\ n} \ = \ - \ \delta_{\ m n} \ q_{\ 0} \ + 
\ \varepsilon_{\ mnr} \ q_{\ r} 
\hspace*{0.2cm} \mbox{for} \hspace*{0.2cm} m,n,r \ = \ 1,2,3
\end{array}
\end{equation}

\noindent
As we will see , the final stage of the ( nu-mass extended - ) standard model
gauge breaking involving {\it just} 1 doublet of scalars with respect to the 
elecroweak part $\ SU (2)_{\ w} \ \times \ {\cal{Y}}_{\ w} \ $ represents a
case for perfectly semiclassical, driven Bose condensation , eventually
contrasting with intrinsic properties of primary breakdown.

\noindent
This is so, because there is precisely 1 invariant

\vspace*{-0.3cm}
\begin{equation}
\label{eq:2-8}
\begin{array}{l}
I \ ( \ z \ , \ z^{\ \dagger} \ ) \ = \ z \ z^{\ \dagger} \ =
\ \frac{1}{2} \ \sum_{\ \mu = 0}^{\ 3} \ \left ( \ Z_{\ \mu} \ \right )^{\ 2}
\end{array}
\end{equation}

\noindent
with respect to $\ SU (2)_{\ w} \ \times \ {\cal{Y}}_{\ w} \ $ of which
a general invariant is a function . 

\noindent
This would not remain true for more than one scalar doublet .

\noindent
As a consequence of eqs. \ref{eq:2-4} - \ref{eq:2-8} , the electroweak gauge 
breaking is {\it driven and semiclassical}

\vspace*{-0.3cm}
\begin{equation}
\label{eq:2-9}
\begin{array}{l}
\left \langle \ \Omega \ \right | \ z \ ( \ x \ ) 
\ \left | \ \Omega \ \right \rangle 
\ = 
\ \frac{1}{\sqrt{2}} \ \left ( \ \begin{array}{ll}
v & 0
\vspace*{0.1cm} \\
0 & v
\end{array} \ \right )
\hspace*{0.2cm} \mbox{with} \hspace*{0.2cm}
v \ = \ 246.220 \ \mbox{GeV} 
\vspace*{0.1cm} \\
\mbox{independent of} \ x
\end{array}
\end{equation}

\noindent
We shall discuss 2 types of primary gauge breaking denoted a) and b)
below. 

\noindent
The v.e.v. v in eq. \ref{eq:2-9} corresponds to the classical minimum of the
quartic potential , uniquely restricted to depend on two parameters

\vspace*{-0.3cm}
\begin{equation}
\label{eq:2-10}
\begin{array}{l}
V \ \left ( \ z \ , \ z^{\ \dagger} \ \right ) \ =
\ \left \lbrack \ - \ \mu^{2} \ z \ z^{\ \dagger} \ + \ \lambda 
\ \left ( \ z \ z^{\ \dagger} \ \right )^{\ 2} \ \right \rbrack_{\ 11}
\end{array}
\end{equation}

\noindent
The minimum conditions become

\vspace*{-0.3cm}
\begin{equation}
\label{eq:2-12}
\begin{array}{l}
\partial_{\ Z_{\ \nu}} \ V \ = 
\ \left ( \ - \ \mu^{\ 2} 
\ + \ \lambda 
\ \left | \ Z \ \right |^{\ 2} \ \right ) \ Z_{\ \nu} \ = \ 0
\hspace*{0.05cm} \longrightarrow \hspace*{0.05cm}
\left . z \ z^{\ \dagger} \ \right |_{ 11} = \frac{1}{2}
\ \mu^{\ 2} \ / \ \lambda 
\vspace*{0.1cm} \\
\left . V \ \right |_{\ min} \ =
\ - \ \frac{1}{4} \ \left ( \ \mu^{\ 4} \ / \ \lambda \ \right )
\end{array}
\end{equation}

\noindent
The second derivatives with respect to Z at the minimum of the potential become

\vspace*{-0.3cm}
\begin{equation}
\label{eq:2-13}
\begin{array}{l}
\frac{1}{2} \left . \partial_{\ Z_{\ \varrho}} 
\ \partial_{\ Z_{\ \sigma}} \ V \ =
\ \lambda \ Z_{\ \varrho} \ Z_{\ \sigma} \ \right |_{\ min}
\ = \ \lambda \ v^{2} \ \delta_{\ 0 \varrho} \ \delta_{\ 0 \sigma}
\vspace*{0.1cm} \\
\left . Z_{\ \nu} \ \right |_{\ min} \ = \ \left ( \ v \ , \ \vec{0} \ \right )
\end{array}
\end{equation}

\noindent
Expanding the deviation of the potential up to quadratic terms around the 
minimum thus yields

\vspace*{-0.3cm}
\begin{equation}
\label{eq:2-14}
\begin{array}{l}
\left ( \ \Delta \ V \ \right )^{\ (2)} \ = \ \lambda \ v^{\ 2}
\ \left ( \ \Delta \ Z_{\ 0} \ \right )^{\ 2}
\hspace*{0.2cm} ; \hspace*{0.2cm}
Z_{\ 0} \ ( \ x \ ) \ = \ v \ + \ \Delta \ Z_{\ 0} \ ( \ x \ )
\end{array}
\end{equation}

\noindent
It is customary to denote the shifted hermitian field 
$\ \Delta \ Z_{\ 0} \ ( \ x \ )$

\vspace*{-0.3cm}
\begin{equation}
\label{eq:2-15}
\begin{array}{l}
Z_{\ 0} \ = \ v \ + \ \Delta \ Z_{\ 0} 
\hspace*{0.2cm} ; \hspace*{0.2cm}
\Delta \ Z_{\ 0} \ ( \ x \ ) \ = \ H \ ( \ x \ )
\end{array}
\end{equation}

\noindent
From eq. \ref{eq:2-15} we read off the mass of the field $\ H \ ( \ x \ ) \ $
as well as the vanishing of the mass of the other three fields 

\vspace*{-0.3cm}
\begin{equation}
\label{eq:2-16}
\begin{array}{l}
m_{\ H}^{\ 2} \ = \ 2 \ \lambda \ v^{\ 2} \ = \ 2 \ \mu^{\ 2} 
\hspace*{0.2cm} ; \hspace*{0.2cm}
\left . m_{\ \vec{Z}} \ \right|_{\ from V} \ = \ 0
\vspace*{0.1cm} \\
\vec{Z} \ ( \ x \ ) \ = 
\ \left ( \ Z_{\ 1} \ , \ Z_{\ 2} \ , \ Z_{\ 3} \ \right ) \ ( \ x \ )
\end{array}
\end{equation}

\noindent
However in the case at hand the existence of would-be long range forces
represented by the $\ SU2_{\ L} \ \times U1_{\ {\cal{Y}}} \ $ gauge field
interactions does not permit the existence of goldstone-modes .

\noindent
The three would-be Goldstone fields, defined in eq. \ref{eq:2-16},
through their space-time gradient, mix with the gauge bosons to become massive,
vis. $\ W^{\ \pm} \ $ and $\ \widetilde{Z} \ $ 

\vspace*{-0.3cm}
\begin{equation}
\label{eq:2-17}
\begin{array}{l}
\partial_{\ x^{\ \tau}} \ Z_{\ m} \ ( \ x \ ) 
\ \leftrightarrow \ W_{\ \tau}^{\ \pm} \ ( \ x \ ) \ , 
\ \widetilde{Z}_{\ \tau} \ ( \ x \ )
\end{array}
\end{equation}

\noindent
in such a way as to obtain masses of the 3 massive gauge bosons
$\ m_{\ W} \ , m_{\ \widetilde{Z}} \ $ and physically form the longitudinal spin
components of the resulting massive states (resonances) .
In eq. \ref{eq:2-17} the neutral massive gauge boson is denoted 
$\ \widetilde{Z} \ $ not to confuse it with the sclar field components 
$\ Z_{\ 0} \ , \ \vec{Z} \ $.

\noindent
In tree approximation the mass-square of the H-field is twice the value
of the $\ Z_{\ 0} \ $ field in the unbroken case, i.e. for 
$\ \mu^{\ 2} \ \rightarrow \ - \ \mu^{\ 2}\ $

\vspace*{-0.3cm}
\begin{equation}
\label{eq:2-18}
\begin{array}{l}
m_{\ H}^{\ 2} \ = \ 2 \ \mu^{\ 2} \ \rightarrow
\ \mu \ = \ \frac{1}{\sqrt{2}} \ m_{\ H} \ = 88.388 \ \mbox{GeV}
\hspace*{0.2cm} \mbox{for} \hspace*{0.2cm} m_{\ H} \ = \ 125 \ \mbox{GeV}
\end{array}
\end{equation}

\noindent
The detailed description of the mixing as stated in eq. \ref{eq:2-17}
is not given here. A complete derivation can be found in the textbook
\cite{SchaeferPesk} .

\noindent
The simplicity of the presumably lowest in scale gauge breaking relative
to the SM gauge group $\ SU2_{\ L} \ \times U1_{\ {\cal{Y}}} \ $
prompted most discussions of primary gauge breaking to be of the
same type b) presented below, i.e. driven - semiclassical: 
for an example of primary gauge breakdown patterns see e.g. 
ref. \cite{MohaMarcia} .

\noindent
This brings us to the two types a) and b) --
of an enveloping gravitationless gauge field theory. 
For both  types the scalar field variables turn out to involve a {\it complex} 
ensemble of irreducible representations of the enveloping gauge group -- 
as e.g. $\ SO (10) \ $, in particular for the generation of
masses for neutrino flavors, both light and heavy, as discussed e.g. in refs. 
4 -- \cite{nuoriginPM} and 10 -- \cite{MohaMarcia} through Yukawa couplings
to basic fermion bilinears .

\noindent
To this end we introduce notation for the ensemble of scalar fields
adapted to primary gauge breaking, generalizing $\ SU2_{\ L} \ $ doublets
as defined in eqs. \ref{eq:2-5} - \ref{eq:2-7} . 

\noindent
Lets fix for definiteness  

\vspace*{-0.3cm}
\begin{equation}
\label{eq:2-19}
\begin{array}{l}   
G_{\ env}^{\ min} \ = \ SO \ (10) \ \equiv \ spin \ (10)
\end{array}
\end{equation}

\noindent
in the following. A general irreducible representation of $\ SO \ (10) \ $
shall be denoted $\ \left \lbrack {\cal{D}} \right \rbrack \ $, where 
$\ {\cal{D}} \ $ is equivalenced to its dimension.
As entry point we take the $\ \left \lbrack 16 \right \rbrack \ $
representation for one fermion family 
( the lightest in mass ) in the left chiral basis

\vspace*{-0.1cm}
\begin{equation}
\label{eq:2-20}
\begin{array}{c}
\left \lbrack 16 \right \rbrack \ :
\ \left ( 
\begin{array}{llll@{\hspace*{-0.0cm}} l llll}
\color{red} u^{\ 1} & \color{green} u^{\ 2} & \color{cyan} u^{\ 3}
& \color{blue} \nu_{\ e} & \color{blue} | & \color{blue} 
{\cal{N}}_{\ e}
& \color{cyan} \widehat{u}^{\ 3} & \color{green} \widehat{u}^{\ 2}
& \color{red} \widehat{u}^{\ 1}
\vspace*{0.2cm} \\
\color{red} d^{\ 1} & \color{green} d^{\ 2} & \color{cyan} d^{\ 3}
& \color{blue} e^{\ -} \hspace*{0.2cm} & \color{blue} | & \color{blue} e^{\ +}
& \color{cyan} \widehat{d}^{\ 3} & \color{green} \widehat{d}^{\ 2}
& \color{red} \widehat{d}^{\ 1} \
\end{array} \color{blue} \right ) \vspace*{-0.5cm} \hspace*{-0.3cm}
\begin{array}{l}
\vspace*{-1.5cm} \\
^{\ \color{blue} \dot{\gamma} \ \rightarrow \ L}
\end{array}
\vspace*{0.5cm} \\
\color{blue}
= \ \left ( \ f \ \right )^{\ \dot{\gamma}}
\end{array}
\end{equation}

\begin{center}
\vspace*{-0.3cm}
{\bf \color{black} 2-1 - Primary gauge breaking
 }
\label{'2-1'}
\end{center}
\vspace*{0.0cm}

\noindent
Following the hypothesis of an underlying unifying gauge group
the top down of gauge breaking is initiated by a primary breaking
followed last by the electroweak gauge breaking .

\noindent
Primary gauge beaking is linked to the unifying gauge group scale 
$\ M_{\ env} \ $ assumed and also restricted by limits on direct observation of
baryon decays and lepton flavor violation to be much larger than the
electroweak scale, defined in eq. \ref{eq:2-4}, repeated below 

\vspace*{-0.3cm}
\begin{equation}
\label{eq:2-4+}
\begin{array}{l}
M_{\ env} \ \gg \ v \ = 
\ \left ( \sqrt{\ \sqrt{2} \ G_{\ F} \ } \ \right )^{\ -1} 
\ = \ 246.220 \ \mbox{GeV}
\end{array}
\end{equation}

\noindent
In eq. \ref{eq:2-4+} v denotes the v.e.v. of the unique doublet scalar field 
using the quaternion associated basis for the local scalar fields,
defined in eq. \ref{eq:2-5}, where
the symbol $\ ^{*}$ denotes hermitian conjugation of
individual complex and/or real field components.

\noindent
Thus the quantity z , defined in eq. \ref{eq:2-5} , shows its quaternionic
representation, as defined in eq. \ref{eq:2-6} repeated below

\vspace*{-0.3cm}
\begin{equation}
\label{eq:2-6+}
\begin{array}{l}
z \ ( \ x \ ) \ = \ \frac{1}{\sqrt{2}} \ \left (
\ Z_{\ 0} \ \sigma_{\ 0} 
\ + \ \sum_{\ k = 1}^{\ 3} \ Z_{\ k} \ \frac{1}{i} 
\ \sigma_{\ k} \ \right ) \ ( \ x \ )
\end{array}
\end{equation}

\noindent
The four $\ 2 \times 2 \ $ matrices , displayed in eq. \ref{eq:2-5} ,
form a 1 to 1 true representation of the base quaternionsi as given in eq.
\ref{eq:2-7} , repeated below

\vspace*{-0.3cm}
\begin{equation}
\label{eq:2-7+}
\begin{array}{l}
\sigma_{\ 0} \ \leftrightarrow \ q_{\ 0} \ = \ \P
\hspace*{0.2cm} ; \hspace*{0.2cm}
\frac{1}{i} \ \sigma_{\ m} \ \leftrightarrow \ q_{\ m} 
\hspace*{0.2cm} \mbox{for} \hspace*{0.2cm} m \hspace*{0.7cm} = \ 1,2,3
\vspace*{0.1cm} \\
q_{\ m} \ q_{\ n} \ = \ - \ \delta_{\ m n} \ q_{\ 0} \ + 
\ \varepsilon_{\ mnr} \ q_{\ r} 
\hspace*{0.2cm} \mbox{for} \hspace*{0.2cm} m,n,r \ = \ 1,2,3
\end{array}
\end{equation}

\noindent
As we will see , the final stage of the ( nu-mass extended - ) standard model
gauge breaking involving {\it just} 1 doublet of scalars with respect to the 
elecroweak part $\ SU (2)_{\ w} \ \times \ {\cal{Y}}_{\ w} \ $ represents a
case for perfectly semiclassical, driven Bose condensation , eventually
contrasting with intrinsic properties of primary breakdown.
 
\noindent
This is so, because there is precisely 1 invariant

\vspace*{-0.3cm}
\begin{equation}
\label{eq:2-8+}
\begin{array}{l}
I \ ( \ z \ , \ z^{\ \dagger} \ ) \ = \ z \ z^{\ \dagger} \ =
\ \frac{1}{2} \ \sum_{\ \mu = 0}^{\ 3} \ \left ( \ Z_{\ \mu} \ \right )^{\ 2}
\end{array}
\end{equation}

\noindent
with respect to $\ SU (2)_{\ w} \ \times \ {\cal{Y}}_{\ w} \ $ of which
a general invariant is a function . 

\noindent
This would not remain true for more than one scalar doublet .

\noindent
As a consequence of eqs. \ref{eq:2-4} - \ref{eq:2-8} , the electroweak gauge 
breaking is {\it driven and semiclassical}

\vspace*{-0.3cm}
\begin{equation}
\label{eq:2-9+}
\begin{array}{l}
\left \langle \ \Omega \ \right | \ z \ ( \ x \ ) 
\ \left | \ \Omega \ \right \rangle 
\ = 
\ \frac{1}{\sqrt{2}} \ \left ( \ \begin{array}{ll}
v & 0
\vspace*{0.1cm} \\
0 & v
\end{array} \ \right )
\hspace*{0.2cm} \mbox{with} \hspace*{0.2cm}
v \ = \ 246.220 \ \mbox{GeV} 
\vspace*{0.1cm} \\
\mbox{independent of} \ x
\end{array}
\end{equation}

\noindent
We shall discuss 2 types of primary gauge breaking denoted a) and b)
below . 

\noindent
The simplicity of the presumably lowest in scale gauge breaking relative
to the SM gauge group $\ SU2_{\ L} \ \times U1_{\ {\cal{Y}}} \ $
prompted most discussions of primary gauge breaking to be of the
same type b) presented below, i.e. driven - semiclassical: 
for an example of primary gauge breakdown patterns see e.g. 
ref. \cite{MohaMarcia} .

\noindent
This brings us to the two types a) and b) --
of an enveloping gravitationless gauge field theory. 
For both  types the scalar field variables turn out to involve a {\it complex} 
ensemble of irreducible representations of the enveloping gauge group -- 
as e.g. $\ SO (10) \ $, in particular for the generation of
masses for neutrino flavors, both light and heavy, as discussed e.g. in refs. 
4 -- \cite{nuoriginPM} and 10 -- \cite{MohaMarcia} through Yukawa couplings
to basic fermion bilinears .

\noindent
To this end we introduce notation for the ensemble of scalar fields
adapted to primary gauge breaking, generalizing $\ SU2_{\ L} \ $ doublets
as defined in eq. \ref{eq:2-20} , repeated below, 

\noindent
fixing for definiteness  

\vspace*{-0.3cm}
\begin{equation}
\label{eq:2-19+}
\begin{array}{l}   
G_{\ env}^{\ min} \ = \ SO \ (10) \ \equiv \ spin \ (10)
\end{array}
\end{equation}

\noindent
in the following. A general irreducible representation of $\ SO \ (10) \ $
shall be denoted $\ \left \lbrack {\cal{D}} \right \rbrack \ $, where 
$\ {\cal{D}} \ $ is equivalenced to its dimension.
As entry point we take the $\ \left \lbrack 16 \right \rbrack \ $
representation for one fermion family 
( the lightest in mass ) in the left chiral basis as defined in eq. 19, repeated below

\vspace*{-0.1cm}
\begin{equation}
\label{eq:2-20a}
\begin{array}{c}
\left \lbrack 16 \right \rbrack \ :
\ \left ( 
\begin{array}{llll@{\hspace*{-0.0cm}} l llll}
\color{red} u^{\ 1} & \color{green} u^{\ 2} & \color{cyan} u^{\ 3}
& \color{blue} \nu_{\ e} & \color{blue} | & \color{blue} 
{\cal{N}}_{\ e}
& \color{cyan} \widehat{u}^{\ 3} & \color{green} \widehat{u}^{\ 2}
& \color{red} \widehat{u}^{\ 1}
\vspace*{0.2cm} \\
\color{red} d^{\ 1} & \color{green} d^{\ 2} & \color{cyan} d^{\ 3}
& \color{blue} e^{\ -} \hspace*{0.2cm} & \color{blue} | & \color{blue} e^{\ +}
& \color{cyan} \widehat{d}^{\ 3} & \color{green} \widehat{d}^{\ 2}
& \color{red} \widehat{d}^{\ 1} \
\end{array} \color{blue} \right ) \vspace*{-0.5cm} \hspace*{-0.3cm}
\begin{array}{l}
\vspace*{-1.5cm} \\
^{\ \color{blue} \dot{\gamma} \ \rightarrow \ L}
\end{array}
\vspace*{0.5cm} \\
\color{blue}
= \ \left ( \ f \ \right )^{\ \dot{\gamma}}
\end{array}
\end{equation}

\begin{center}
\vspace*{-0.05cm}
{\bf \color{black} the Majorana logic characterized by 
$\ {\cal{N}}_{\ e,\mu,\tau} \ $
}
\label{'MajLogic'}
\end{center}
\vspace*{-0.0cm}

\noindent
We illustrate the use of the basis defined in 
eq. \ref{eq:2-20a} considering the group decomposition

\vspace*{-0.3cm}
\begin{equation}
\label{eq:2-21+}
\begin{array}{c}
\mbox{spin} \ (10) \ \rightarrow \ \mbox{SU5} \ \times \ \mbox{U1}_{\ J_{\ 5}}
\end{array}
\vspace*{-0.2cm}
\end{equation}

\noindent
Among the 3 generators of spin (10) commuting with $\mbox{SU3}_{\ c}$ ,
$ I_{\ 3 \ L} \ , \ I_{\ 3 \ R} $ and 
Cartan subalgebra of spin (10)
there is one combination, denoted $\ J_{\ 5} \ $ in eq. \ref{eq:2-21+},
commuting with its largest unitary subgroup SU5 .
The charges $\ Q \ ( \  J_{\ 5} \ ) $ form the pattern as in eq. \ref{eq:2-22+}

\vspace*{-0.3cm}
\begin{equation}
\label{eq:2-22+}
\begin{array}{c}
\left \lbrack 16 \right \rbrack \ : 
\hspace*{0.2cm} Q \ ( \  J_{\ 5} \ ) \ =
\ \left ( 
\begin{array}{llll@{\hspace*{0.4cm}} l rrrr}
\color{red} 1 & \color{green} 1 & \color{cyan} 1
& \color{blue} -3 & \color{blue} | & \color{blue} 
5
& \color{cyan} 1 & \color{green} 1
& \color{red} 1
\vspace*{0.2cm} \\
\color{red} 1 & \color{green} 1 & \color{cyan} 1
& \color{blue} -3  & \color{blue} | & \color{blue} 1
& \color{cyan} -3 & \color{green} -3
& \color{red} -3
\end{array} \color{blue} \right )
\end{array}
\end{equation}

\noindent
$\ Q \ ( \ J_{\ \mu} \ ) \ $ with charges as given in eq. \ref{eq:2-22+}
represents a {\it hermitian} generator of the Cartan subalgebra of
$\ spin \ (10) \ $, unique up to a (real) multiplicative factor,
which commutes with the the SU5 subgroup of
$\ spin \ (10) \ $. The flavors within one family sharing the same Q-charges
in fact form irreducible representations of SU5 , which shall be labeled 

\vspace*{-0.3cm}
\begin{equation}
\label{eq:2-23+}
\begin{array}{c}
{\cal{D}} \ ( \ SU5 \ ) \ \rightarrow \left \lbrace {\cal{D}} \ ( \ SU5 \ )
\right \rbrace_{\ Q}
\end{array}
\vspace*{-0.1cm}
\end{equation}

\noindent
with $\ {\cal{D}} \ ( \ SU5 \ ) \ $ equivalenced with the dimension of the
representation. The suffix Q is added in eq. \ref{eq:2-23+}, since in the 
present context SU5 multiplets necessarily occur {\it embedded} in 
$\ SO \ (10) \ $ representations.

\noindent
Thus the Q values of the $\ \left \lbrack 16 \right \rbrack \ $ representation
on the right han side of eq. \ref{eq:2-22+} translate to

\vspace*{-0.3cm}
\begin{equation}
\label{eq:2-24+}
\begin{array}{l}
\left \lbrack 16 \right \rbrack \ =
\ \left \lbrace 1 \right \rbrace_{\ 5}
\ + \ \left \lbrace {10} \right \rbrace_{\ 1}
\ + \ \left \lbrace \overline{5} \right \rbrace_{\ -3}
\end{array}
\vspace*{-0.1cm}
\end{equation}

\noindent
The sequence of Q values : $\ Q \ = \ 5 \ , \ 1 \ , -3 \ $ as arranged in the
sequence on the right hand side of eq. \ref{eq:2-24+} -- decreasing in 
steps of 4 -- is related to the properties of binomial coefficients

\vspace*{-0.3cm}
\begin{equation}
\label{eq:2-25+}
\begin{array}{l}
\left ( \begin{array}{c}
10 
\vspace*{0.1cm} \\
n
\end{array} \right )
\hspace*{0.2cm} \mbox{for} \hspace*{0.2cm}
n \ = \ 0 \ , \ 4 \ , 8 
\hspace*{0.2cm} \mbox{and} \hspace*{0.2cm}
n \ = \ 10 \ , \ 6 \ , 2
\end{array}
\vspace*{-0.0cm}
\end{equation}

\noindent
This is derived in ref. 4 -- \cite{nuoriginPM} -- and reproduced in part
in Fig. 3 at the end of Appendix 1.

\noindent
It is instructive to decompose the Q-value pattern in eq. \ref{eq:2-22+}
into an $SU2_{L + R} $ invariant part and the remainder proportional
to $\ I_{\ 3 \ L + R} \ $

\vspace*{-0.0cm}
\begin{equation}
\label{eq:2-26+}
\begin{array}{l}
\ \left ( 
\begin{array}{llll@{\hspace*{0.4cm}} l rrrr}
\color{red} 1 & \color{green} 1 & \color{cyan} 1
& \color{blue} -3 & \color{blue} | & \color{blue} 5
& \color{cyan} 1 & \color{green} 1
& \color{red} 1
\vspace*{0.2cm} \\
\color{red} 1 & \color{green} 1 & \color{cyan} 1
& \color{blue} -3  & \color{blue} | & \color{blue} 1
& \color{cyan} -3 & \color{green} -3
& \color{red} -3
\end{array} \color{blue} \right )
\ =
\vspace*{0.1cm} \\
\ \left ( 
\begin{array}{llll@{\hspace*{0.4cm}} l rrrr}
\color{red} 1 & \color{green} 1 & \color{cyan} 1
& \color{blue} -3 & \color{blue} | & \color{blue} 3
& \color{cyan} - 1 & \color{green} - 1
& \color{red} - 1
\vspace*{0.2cm} \\
\color{red} 1 & \color{green} 1 & \color{cyan} 1
& \color{blue} -3  & \color{blue} | & \color{blue} 3
& \color{cyan} -1 & \color{green} -1
& \color{red} -1
\end{array} \color{blue} \right )
\ +
\vspace*{0.1cm} \\
\ \left ( 
\begin{array}{lllr@{\hspace*{0.4cm}} l rrrr}
\color{red} 0 & \color{green} 0 & \color{cyan} 0
& \color{blue} \hspace*{0.2cm} 0 & \color{blue} | & \hspace*{-0.2cm} 
\color{blue} 2
& \color{cyan} 2 & \color{green} 2
& \color{red} 2
\vspace*{0.2cm} \\
\color{red} 0 & \color{green} 0 & \color{cyan} 0
& \color{blue} \hspace*{0.2cm} 0  & \color{blue} | & \hspace*{-0.2cm} 
\color{blue} -2
& \color{cyan} -2 & \color{green} -2
& \color{red} -2
\end{array} \color{blue} \right )
\end{array}
\end{equation}

\noindent
From eq. \ref{eq:2-26+} we obtain the identification of the $\ spin \ (10) \ $
Cartan subalgebra hermitian components \\
( charges ) 

\vspace*{-0.3cm}
\begin{equation}
\label{eq:2-27+}
\begin{array}{l}
Q \ = \ 3 \ ( \ B \ - L \ ) \ - \ 4 \ I_{\ 3 \ R}
\end{array}
\end{equation}

\noindent
In eq. \ref{eq:2-27+} \ B and L denote baryon and lepton number respectively .

\noindent
Two remarks shall follow , concerning the recognizable key features inherent to
spontaneous gauge breaking of an enveloping gravitationless gauge
field theory , based on a gauge group $\ G_{\ env} \ = \ spin \ (10) \ $

\begin{description}
\item 1) primary gauge breakdown 

must be much different than on the
lowest -- i.e. electroweak -- scale level pertaining to 
$\ G_{\ SM} \ = \ SU3_{\ c} \ \times \ SU2_{\ L} \ \times \ U1_{\ {\cal{Y}}} 
\ $.  

This is so because empirically well established candidate symmetries , like
baryon and lepton number conservation are broken on the primary level and imply
very large scale of unification 
$M_{env} = O \ \left ( 10^{16} \mbox{GeV} \right ) $.

As examples let me quote the upper limits of the
$\ \mu^{\ +} \ \rightarrow \ e^{\ +} \ \gamma \ $
and $\ \mu^{\ +} \  \rightarrow \  e^{\ +} \ + \ e^{\ +} \ e^{\ -} \ $
branching fractions

\vspace*{-0.3cm}
\begin{equation}
\label{eq:2-28+}
\begin{array}{l}
Br \ ( \mu^{+} \rightarrow e^{+} + \gamma )
\ < \ 2.4 \ . \  10^{\ - 12} 
\vspace*{0.1cm} \\
Br \ ( \mu^{+} \rightarrow e^{+} + e^{+} e^{-} ) 
\ < \ 1.0 \ . \  10^{\ - 12}
\hspace*{0.1cm} \mbox{in ref. 11 -- \cite{conslaws}}
\end{array}
\end{equation}

\item 2) the power of the set of scalar fields

involved in primary gauge breaking does not follow any principle of
minimal selection of $\ spin \ (10) \ $ representations pertaining
to scalar fields . 

\vspace*{-0.4cm}
\end{description}

\begin{center}
\vspace*{-0.0cm}
{\bf \color{black} specific notation for scalar field variables
 }
\label{'scalarvar'}
\end{center}
\vspace*{0.0cm}

\noindent
We proceed defining notation for scalar field variables suitable for primary
gauge breaking

\vspace*{-0.3cm}
\begin{equation}
\label{eq:2-29+}
\begin{array}{l}
{\displaystyle \lbrace} \underline{z} {\displaystyle \rbrace}
= \underline{z}
\left \lbrack  
\begin{array}{l}
\vspace*{-0.35cm} \\
\begin{array}{l} \left ( 
\begin{array}{lll} {\cal{D}}^{(1)} & , & \otimes n_{1}
\end{array} 
\right )
\vspace*{0.1cm} \\
\left ( 
\begin{array}{lll} {\cal{D}}^{(2)} & , & \otimes n_{2}
\end{array} 
\right )
\vspace*{0.1cm} \\
\left ( 
\begin{array}{lll} \cdots \hspace*{0.0cm} & , & \otimes \cdots 
\end{array} 
\right )
\vspace*{0.2cm} \\
\end{array} 
\end{array} \ \right \rbrack
\hspace*{0.1cm} | \mbox{broken down to real coordinates}
\end{array}
\end{equation}

\noindent
In eq. \ref{eq:2-29+} $\ {\cal{D}}^{\ (\nu)} \ ; \ \nu \ = \ 1,2,\cdots \ $
denote a complete set of unitary irreducible representations of
$\ spin \ (10) \ $, finite dimensional; constructively defined
through the method of Peter and Weyl \cite{PetWeyl} .\\
$\ n_{\ \nu} \ $ stands for the multiplicity of a given representation
$\ {\cal{D}}^{\ (\nu)} \ $.

\noindent
For $\ {\cal{D}}^{\ (\nu)} \ , \ \overline{{\cal{D}}}^{\ (\nu)} \ $
beeing a pair of inequivalent , relative complex conjugate representations
the coordinates $\ \underline{z} \ ( \ {\cal{D}}^{\ (\nu)} \ )
\ , \ \underline{\overline{z}} \hspace*{0.1cm} ( \ \overline{{\cal{D}}}^{\ (\nu)} \ ) \ $
broken down to real and imaginary parts count as 
$\ 2 \ (complex) \ dim \ ( \ {\cal{D}} \ ) \ $ components over real numbers.
This is the meaning of the attribute 'broken down to real coordinates' on the
right hand side of eq. \ref{eq:2-29+} .

\noindent
Thus choosing real values for the components of 
$\ \underline{z} \ \in \ R^{\ M} \ $
as defined in eq. \ref{eq:2-29+} it follows

\vspace*{-0.3cm}
\begin{equation}
\label{eq:2-30+}
\begin{array}{l}
\underline{z} \ = \ \left ( \ z_{\ 1} \ , \ \cdots \ , \ \ z_{\ r} 
\ , \ \cdots \ , z_{\ M}
\ \right ) 
\hspace*{0.2cm} ; \hspace*{0.2cm}
z_{\ r} \ : \ \mbox{hermitian fields}
\end{array}
\vspace*{-0.4cm}
\end{equation}

\noindent
we find

\vspace*{-0.3cm}
\begin{equation}
\label{eq:2-31+}
\begin{array}{l}
M \ = \ \sum_{\ \nu} \ n_{\ \nu} \ \left \lbrace
\begin{array}{l}
\dim \ \left ( \ {\cal{D}}^{\ (\nu)} \right ) \ \mbox{for}
\ {\cal{D}}^{\ (\nu)} \ \mbox{real}
\vspace*{0.1cm} \\
2 \ \mbox{complex} \ dim \ \left ( \ {\cal{D}}^{\ (\nu)} \ \right )
\ \mbox{for}
\ {\cal{D}}^{\ (\nu)} \ \mbox{complex}
\end{array} \right \rbrace
\ < \ \infty
\end{array}
\end{equation}

\begin{center}
\vspace*{-0.0cm}
{\bf \color{black} to get an idea of the power of the set of scalars
 }
\label{'scpower'}
\end{center}
\vspace*{0.0cm}

\noindent
We illustrate the order of M , in eqs. \ref{eq:2-30+} , \ref{eq:2-31+} by the 
representations
of the fermion bilinears from left and right chiral bases, adapting the scalar
variables to their definition in eq. \ref{eq:2-30+} , allowing for complex
linear combinations applied to complex representations (from Appendix 1 and
ref. 4, \cite{nuoriginPM} \hspace*{0.0cm}).

\vspace*{-0.5cm}
\begin{equation}
\label{eq:2-32+}
\begin{array}{l}
\hspace*{2.5cm} {\cal{H}}_{\ fermion \ mass} \ \longleftarrow 
\vspace*{0.2cm} \\
\left ( \overline{z}^{\overline{126} \ F \ G} \right )_{ \overline{\xi}}
\left ( f_{a \ 16 \ F} \right )_{\dot{\gamma}} 
\left ( f_{b \ 16 \ G} \right )^{\dot{\gamma}}
C \ \left ( \begin{array}{c|cc}
\begin{array}{c} 126
\\
\xi
\end{array} 
& \begin{array}{c} 16
\\
a
\end{array}
& \begin{array}{c} 16
\\
b
\end{array}
\end{array} \right )
+ \ h.c.
\vspace*{0.2cm} \\
\left ( \overline{z}^{\overline{126} \ F \ G} \right )_{ \overline{\xi}}
: \mbox{(pseudo-) scalar fields in the } \overline{126}
\mbox{ representation of SO (10)}
\vspace*{0.1cm} \\
F , G = I,II,III : \mbox{fermion family indices}
\end{array}
\end{equation}

\noindent
In eq. \ref{eq:2-32+} $\ C \ \left ( \ \begin{array}{c|cc}
\begin{array}{c} 126
\\
\xi
\end{array}
& \begin{array}{c} 16
\\
a
\end{array}
& \begin{array}{c} 16
\\
b
\end{array}
\end{array} \right ) \ $
\vspace*{0.2cm}

\noindent
denotes the $\ spin \ (10) \ $ Clebsch-Gordan coefficients projecting
the product of two ( fermionic ) 16 representations on
irreducible spin (10) representations.

\noindent
We reproduce the 4 product representations of 
$\ 16 \ \mbox{and} \ \overline{16} \ $ representations
from ref. 4 ( Appendix E ) --  \cite{nuoriginPM} \hspace*{0.01cm} ) 

\vspace*{-0.3cm}
\begin{equation}
\label{eq:2-33+}
\begin{array}{l}
\begin{array}{@{\hspace*{0.0cm}}c@{\hspace*{0.1cm}}|@{\hspace*{0.1cm}}c
@{\hspace*{0.1cm}}|@{\hspace*{0.1cm}}c@{\hspace*{0.0cm}}}
& \left \lbrack 16 \right \rbrack & \left \lbrack \overline{16} \right \rbrack
\vspace*{0.0cm} \\
\hline \vspace*{-0.4cm} \\
\left \lbrack 16 \right \rbrack  
& s \ : \ \begin{array}{l}
\left \lbrack 10 \right \rbrack \ +
\vspace*{0.1cm} \\
\left \lbrack 126 \right \rbrack
\end{array}
\hspace*{0.1cm} , \hspace*{0.1cm}
a \ : \ \left \lbrack 120 \right \rbrack
& \begin{array}{l} \left \lbrack 1 \right \rbrack \ + 
\ \left \lbrack 45 \right \rbrack \ +
\vspace*{0.1cm} \\
\left \lbrack 210 \ \right \rbrack
\end{array}
\\ \hline \vspace*{-0.4cm} \\
\left \lbrack \overline{16} \right \rbrack
& \begin{array}{l} \left \lbrack 1 \right \rbrack \ +
\ \left \lbrack 45 \right \rbrack \ +
\vspace*{0.1cm} \\
\left \lbrack 210 \ \right \rbrack
\end{array}
& s \ : \ \begin{array}{l}
\left \lbrack 10 \right \rbrack \ +
\vspace*{0.1cm} \\
\left \lbrack \overline{126} \right \rbrack
\end{array}
\hspace*{0.1cm} , \hspace*{0.1cm}
a \ : \ \left \lbrack 120 \right \rbrack
\end{array}
\end{array}
\end{equation}

\noindent
The real and complex representations in the multiplication table in eq.
\ref{eq:2-33+} are denoted $\ {\cal{D}}_{\ R} \ , \ {\cal{D}}_{\ C}  $
respectively

\vspace*{-0.2cm}
\begin{equation}
\label{eq:2-34+}
\begin{array}{l}
{\cal{D}}_{\ R} \ : \ \left \lbrack 10 \right \rbrack_{\ R} \ ,
\ \left \lbrack 120 \right \rbrack_{\ R} \ ,
\ \left \lbrack 1 \right \rbrack_{\ R} \ ,
\ \left \lbrack 45 \right \rbrack_{\ R} \ ,
\ \left \lbrack 210 \ \right \rbrack_{\ R}
\vspace*{0.1cm} \\
{\cal{D}}_{\ C} \ : \ \left \lbrack 126 \right \rbrack_{\ C} \ ,
\ \left \lbrack \overline{126} \right \rbrack_{\ C}
\end{array}
\vspace*{-0.0cm}
\end{equation}

\noindent
Choosing minimally multiplicities 2 for real and 1 for complex
representations in eqs.
\ref{eq:2-33+} and \ref{eq:2-34+} yields

\vspace*{-0.3cm}
\begin{equation}
\label{eq:2-35+}
\begin{array}{l}
M \ = \ 772 \ + \ 252 \ = \ 1028 
\end{array}
\end{equation}

We conclude from the example multiplicites leading to eq. \ref{eq:2-35+}

\vspace*{-0.3cm}
\begin{equation}
\label{eq:2-36+}
\begin{array}{l}
M \ \geq \ O \ ( \ 1000 \ )
\end{array}
\end{equation}

\begin{center}
\vspace*{-0.0cm}
{\bf \color{black} 2-1-a - Primary gauge breaking - type a)
 }
\label{'2-1-a'}
\end{center}
\vspace*{0.0cm}

\noindent
The gauge breaking in type a) {\it gravitationless gauge field theory} is

\begin{description}

\item 1) driven 

by pre-established quadratic, cubic and quartic scaler field self interactions

\item 2) {\it not} reducible to semiclassical approximation 

for vacuum expected values for scalar variables and their composite operators,
necessarily include gauge variant ones in order to qualify for
{\it gauge breaking}

\vspace*{-0.1cm}
\end{description}

\noindent
For clarity let me remark that condition 2) above is necessary, since
in the case of {\it exclusively} gauge invariant vev's for composite scalar 
operators they
are part of an alternative case -- in conjunction with other gauge invariant
composite field variables --
of spontaneous mass generation without gauge breaking. 
This will be discussed within QCD with 3 light flavors
-- without scalars -- in section 3.

\begin{center}
\vspace*{-0.0cm}
{\bf \color{black} 2-1-b - Primary gauge breaking - type b)
 }
\label{'2-1-b'}
\end{center}
\vspace*{0.0cm}

\noindent
The gauge breaking in type b) {\it gravitationless gauge field theory} is

\begin{description}

\item 1) driven

by pre-established quadratic, cubic and quartic scaler field self interactions
like in type a)

\item 2) reducible to semiclassical approximation 

for vacuum expected values for scalar variables and their composite operators.
This is the usual case discussed in the literature, as e.g. in refs. 10
-- \cite{MohaMarcia} -- and 13 -- \cite{johnellis} , while in the latter
reference the main topic is electroweak gauge breaking . 

\end{description}

\noindent
The semiclassical approximation -- with respect to vev's of scalar fields and
their composite local operators as well as composite operators involving other
fields -- means

\vspace*{-0.3cm}
\begin{equation}
\label{eq:2-37+}
\begin{array}{l}
\left \langle \ \Omega \ \right |
\ f \ \left ( \ \underline{z} \ \right ) 
\ \left | \ \Omega \ \right \rangle \ = 
\ f 
\ \left (
\begin{array}{l} 
\vspace*{-0.55cm} \\
\left \langle \ \Omega \ \right | 
\ \underline{z} 
\ \left | \ \Omega  \ \right \rangle
\vspace*{0.0cm} \\
\end{array}
\ \right )
\end{array}
\end{equation}

\begin{center}
\vspace*{-0.1cm}
{\bf \color{black} 3 - Mass Generation in QCD with 3 light flavors 
- oscillating Quarks and Gluons
 }
 \label{'3'}
\end{center}
\vspace*{0.0cm}

\noindent
In this section we turn to the main topic of this lecture : Mass generation 
in QCD for three light flavors of quark u,d,s
-- spontaneous and through the trace anomaly , persisting in the chiral limit
$\ M_{\ u,d,s} \ \rightarrow \ 0 \ $.  

\noindent
Here the oscillatory modes of valence quark-antiquark states as well
as three valence quark baryon \\
( antibaryon ) states will be discussed
and new results on counting these modes presented. 

\noindent
The next subsections are devoted to embed oscillatory modes of quarks in
QCD following lectures given by the author in Erice 2013, ref. 14,
\cite{PMembedosc}.

\begin{center}
\vspace*{-0.1cm}
{\bf \color{black} 3-1 - embedding oscillator modes in u,d,s flavored
baryons in QCD
 }
 \label{'3-1'}
\end{center}
\vspace*{0.0cm}

\noindent
In order to put the main topic of this lecture into perspective let me begin
citing ref. \cite{oscmod1980} , my construction of Poincar\'{e} 
invariant oscillatory modes of three valence quarks ( u , d ) restricted to the
two nonstrange flavors in nonstrange baryons . A small collection of references 
to previous work in this direction is given there ( \cite{oscmod1980} ) .
The disappearing of perfectly gauge invariant explicit dependence on color
of quark- and gauge boson-fields is by the confined nature of the oscillator
wave functions -- restricted to the center of mass system relative space
coordinates -- relegated to an outer factor 

\vspace*{-0.3cm}
\begin{equation}
\label{eq:3-1}
\begin{array}{l}
\varepsilon_{\ c_{\alpha} c_{\beta} c_{\gamma}} 
\hspace*{0.2cm} ; \hspace*{0.2cm} 
c_{\ 1,2,3} \ = \ \mbox{red , green , blue}  
\vspace*{0.0cm} \\
\alpha, \beta, \gamma \ = \ 1 , 2 , 3 
\hspace*{0.2cm} : \hspace*{0.2cm} 
\mbox{numbering individual quark {\it positions}}
\end{array}
\vspace*{-0.15cm}
\end{equation}

\noindent
The color factor in eq. \ref{eq:3-1} : 
$\ \varepsilon_{\ c_{\alpha} c_{\beta} c_{\gamma}} \ $ , antisymmetric in
its three color indices ,
must be gauge invariant with respect
to the local $\ SU3_{\ c} \ $ gauge group and thus
reduced from the 3 positions $\ \vec{x}_{\ 1,2,3} \ $ 
to a common space point
through parallel transport $\ q \ q \ q $  
( triple ) QCD string factors .The detailed form of the QCD string factors is
discussed in section 2 -- premises , for which I cite my previous two Erice
lectures in refs. \cite{PMErice2011} , \cite{PMErice2012} in order to maintain
consistency of notation .

\noindent
A sketch of the $\ q \ \overline{q} \ $ ( bilocal- or double- ) 
and $\ q \ q \ q $   ( triple- ) QCD string factors , also called 'bond
structures' , is given in figure 1 - I \hspace*{0.1cm} below

\vspace*{-0.1cm}
\begin{center}
\hspace*{0.0cm}
\begin{figure}[htb]
 \label{fig-1}
\vskip -0.3cm
\hskip -0.0cm
\includegraphics[angle=0,width=7.5cm]{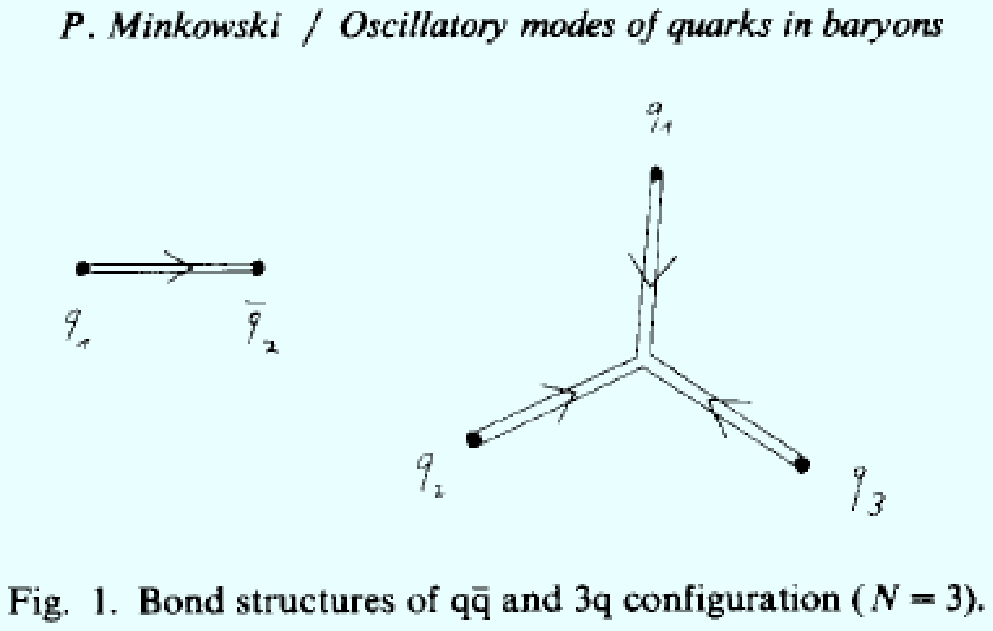}
\vskip -0.0cm
\vspace*{+0.20cm}
{\color{black} \hspace*{0.0cm}
\begin{tabular}{c} Fig. 1 - I : 
Bond structures of $\ q \overline{q} \ $ and $\ 3 q \ $ configurations
\vspace*{0.0cm} \\
$\ \left ( \ N = 3 \ \right ) \ $ from ref. \cite{oscmod1980}
{\color{black} $ \longleftrightarrow$}
\end{tabular}
}
\end{figure}
\end{center}
\vspace*{-0.2cm}

\noindent
The next step revived questions related to oscillatory modes of a pair of
independent oscillators and their eventual connection to Bogoliubov 
transformations in November 2010 in ref. \cite{pairmodes2010}, an olympic
cycle.

\noindent
Finally on the road leading to the present outline a discussion deserves
mention, with the group 
of Willibald Plessas from the University of Graz during the 
Oberw\"{o}lz Symposium  2012 :
’Quantum Chromodynamics: History and Prospects’
516. WE-Heraeus-Seminar , Oberw\"{o}lz, Styria, Austria. 3. - 8. September 
2012 .

\noindent
The discussion arose as to whether the baryon modes of light flavors u, d, s 
of quarks with low total spin were presently more or less
completely accounted for 
-- according to the new PDG-review \cite{AmsGrandKru} -- contrasting with the
situation in 1980 with respect to only u and d flavors .

\noindent
Looking at the exhaustive tables of ref. \cite{AmsGrandKru} {\it included
today} the answer is obviously to the negative , yet these tables were not
available in the web-version of the PDG tables \cite{PDG2012} upon my
last search before the Oberw\"{o}lz Symposium 2012 .
Out of this situation the challenge took shape to count the oscillatory modes 
in baryons in their own right in analytic ways . This work is 'in progres'
in collaboration with Sonia Kabana . First results
have been reported in ref. 28 -- \cite{Countoscimodes-fl6-A4a-SKPM} .
\vspace*{0.2cm}

\noindent
{\color{black} With the exception of subsection 3-1-1, quark masses are
denoted by capital letters $\ M_{\ \alpha} \ $ as at the beginning of section
3, whereas the induced , position dependent mass functions pertaining to
the oscillatory modes of three valence quarks in baryons are denoted
by small letters $\ m_{\ \alpha} \ $.}
\vspace*{0.2cm}

\noindent
In subsection 3-1-1 and section 4 below the (sub-) section numbering is 
taken over from ref. 14 -- \cite{PMembedosc} .

\begin{center}
\vspace*{-0.0cm}
{\bf \color{black} 3-1-1 -- Assembling elements of the QCD
Lagrangean density -- premises
}
\label{'3-1-1'}
\end{center}
\vspace*{0.0cm}

\noindent
We face the theoretical abstraction of QCD with $N_{\ fl} \ = \ 6$ ,
representing strong interactions -- adaptable to two or three light flavors 
u , d , s of quarks
and antiquarks. \hspace*{0.4cm} $\color{black} \leftrightarrow$
\vspace*{0.1cm}

\begin{tabular}{l@{\hspace*{0.05cm}}l@{\hspace*{0.1cm}}l}
\color{black} quarks  :  \color{black} color is counted in $\pi^{\ 0} 
\ \rightarrow \ \gamma \ \gamma$ & &
\vspace*{0.1cm} \\
{\color{black} $\left ( 
\mbox{\begin{tabular}{@{\hspace*{0.1cm}}l@{\hspace*{0.1cm}}}
assuming global color- and 
\vspace*{-0.05cm} \\
flavor-projections to commute
\end{tabular}} \right ) $ yet see ref. \cite{BaerWie}} & &
\vspace*{0.1cm} \\
\color{black} spin and flavor are clearly seen
in $q \overline{q}$ and 
$3 q \ , \ 3 \overline{q}$ spectroscopy \hspace*{0.0cm} &
\vspace*{0.0cm} \\ 
{\color{black} $\left (
\mbox{\begin{tabular}{@{\hspace*{0.1cm}}l@{\hspace*{0.1cm}}}
a pre-condition 
\vspace*{-0.20cm} \\
to count color
\end{tabular}} \right ) $ } .  
\end{tabular}

\vspace{-0.0cm}

\begin{equation}
\label{eq:3-1-1}
\begin{array}{c}
{\cal{L}} \ = 
\begin{array}{c}
 \left \lbrack  \begin{array}{c}
\overline{q}^{\ \dot{c}'}_{\ \dot{{\cal{S}}'} \ \dot{f}}
\ \left \lbrace \ \begin{array}{c}
\frac{i}{2} \ \stackrel{\leftharpoondown \hspace*{-0.3cm} \rightharpoonup}{\partial}_{\ \mu} 
\ \delta_{\ c' \dot{c}}
\vspace*{0.1cm} \\
+ \ W_{\ \mu}^{\ r} \ \left ( \ \frac{1}{2}  \lambda_{\ r} 
\ \right )_{\ c' \dot{c}}
\end{array}                     
\right \rbrace
\ \gamma^{\ \mu}_{\ \dot{{\cal{S}}'} {\cal{S}}} \ q^{\ c}_{\ {\cal{S}} \ f} 
\vspace*{0.1cm} \\
\hspace*{0.5cm}
- \ m_{\ f} \ \overline{q}^{\ \dot{c}}_{\ \dot{{\cal{S}}} \ \dot{f}} 
\  q^{\ c}_{\ {\cal{S}} \ f}
\end{array}
 \right \rbrack
\vspace*{0.1cm} \\
- \ \frac{1}{4 \ \color{black} g^{\ 2}} \ \color{black} B^{\ \mu \nu \ r} 
\ B_{\ \mu \nu}^{\ r}
\hspace*{0.1cm} 
+ \ \Delta \ {\cal{L}}
\end{array}
\vspace*{0.1cm} \\
W_{\ \mu}^{\ r} \ \equiv \ - \ v_{\ \mu}^{\ r} 
\hspace*{0.1cm} : \hspace*{0.1cm} 
\mbox{for identification of convention for potentials}
\vspace*{0.1cm} \\
\color{black} \mbox{quarks} \ : \color{black}  c' \ , \ c  =  1,2,3 
\ \mbox{color}
\hspace*{0.2cm} , \hspace*{0.2cm}
f \ = \ 1,\cdots,6 \ \mbox{flavor}
\vspace*{0.0cm} \\
{\cal{S}'} , {\cal{ S}} \ = \ 1,\cdots,4 \ \mbox{spin} \ , \  m_{\ f} \ \mbox{mass}
\end{array}
\vspace*{-0.0cm}
\end{equation}

\vspace*{-0.1cm}
\noindent
In eq. \ref{eq:3-1-1} the ${\cal{D}}$  
related gauge connection fields, where 
${\cal{D}} \ = \ {\cal{D}} \ \left ( \ {\cal{G}} \ \right ) $
denotes a general, irreducible representation of the local gauge group
${\cal{G}} = SU3_{c} $, appear in the form appropriate
for quarks : $\ {\cal{D}} \ = \ \left \lbrace \ 3 \ \right \rbrace \ $,
and antiquarks : 
$\ {\cal{D}} \ = \ \left \lbrace \ \overline{3} \ \right \rbrace \ $
respectively

\vspace*{-0.2cm}
\begin{equation}
\label{eq:3-1-2}
\begin{array}{|l|}
\hline \vspace*{-0.4cm} \\
\left ( \ {\cal{W}}_{\ \mu} \ ( \ {\cal{D}} \ ) \ \right )_{\ \alpha \beta}
\ ( \ x \ )
\ = \ W^{\ r}_{\ \mu} \ ( \ x \ ) \ \left ( \ d_{\ r} 
\ \right )_{\ \alpha \beta}
\vspace*{0.1cm} \\
\hspace*{0.0cm} \leftrightarrow \hspace*{0.2cm}
{\cal{W}}_{\ \mu} \ ( \ {\cal{D}} \ ) \ = \ - \ {\cal{W}}_{\ \mu} \ ( \
{\cal{D}} \ )^{\ \dagger}
\\ \vspace*{-0.3cm} \\
d_{\ r} \ = \ - \ d_{\ r}^{\ \dagger} \ = \ \frac{1}{i} \ J_{\ r}
\ \in \ Lie \ ( \ {\cal{D}} \ )
\hspace*{0.2cm} ; \hspace*{0.2cm}
\left \lbrack \ d_{\ p} \ , \ d_{\ q} \ \right \rbrack \ = \ f_{\ p q r}
\ d_{\ r}
\\ \vspace*{-0.4cm} \\
r \ , \ p \ , \ q \ = \ 1, \ \cdots , \ dim \ {\cal{G}} 
\hspace*{0.2cm} ; \hspace*{0.2cm}
\alpha \ , \ \beta \ = \ 1, \ \cdots , \ dim \ {\cal{D}}
\vspace*{-0.4cm} \\
\vspace*{-0.0cm} \\ \hline
\end{array}
\end{equation}

\noindent
For $\ {\cal{D}} \ ( \ SU3_{\ c} \ ) = \ \left \lbrace \ 3 \ (\overline{3}) 
\ \right \rbrace \ $
the representation matrices become ( the Gell-Mann matrices \cite{eightfway} ) 

\vspace*{-0.3cm}
\begin{equation}
\label{eq:3-1-3}
\begin{array}{l}
\left ( \ d_{\ r} \ (3) \ = \ \frac{1}{i} \ \frac{1}{2} \ \lambda_{\ r} 
\ \right )_{\ \alpha \beta}
\hspace*{0.1cm} ; \hspace*{0.1cm}
r \ = \ 1, \ \cdots , \ 8
\vspace*{0.1cm} \\
\left ( \ \alpha \ , \ \beta \ \right ) \ \leftrightarrow
\left ( \ c^{'} \ , \ \dot{c} \ \right )
= \ 1, \ \cdots , \ 3
\hspace*{0.1cm} ; \hspace*{0.1cm}
d_{\ r} \ (\overline{3}) \ = \ \overline{d}_{\ r} \ (3)
\vspace*{0.1cm} \\
\mbox{with the {\it conventional} normalization conditions : }
- \ tr \ d^{\ r} \ d^{\ s} \ = \ \frac{1}{2} \ \delta^{\ r s}  
\end{array}
\vspace*{-0.1cm}
\end{equation}

\noindent
The quantity proportional to the gauge potentials $\ W_{\ \mu}^{\ r} \ $
for the $\ \overline{q} \ q \ $ in eq. \ref{eq:3-1-1}
is thus identified as

\vspace*{-0.2cm}
\begin{equation}
\label{eq:3-1-4}
\begin{array}{l}
\left \lbrack \ W_{\ \mu}^{\ r} \ \left ( \ \frac{1}{2}  \lambda_{\ r} 
\ \right )_{\ c' \dot{c}} \ = \ i 
\ \left ( \ {\cal{W}}_{\ \mu} \ ( \ {\cal{D}} \ = 
\ \left \lbrace \ 3 \ \right \rbrace \ )
\ \right )_{\ c' \dot{c}} \ \right \rbrack \ ( \ x \ )
\end{array}
\end{equation}

\noindent
Here we postpone the discussion of complete connections 
and extend the QCD Lagrangean density to include the term quadratic in the 
field strengths $\ B_{\ \mu \nu}^{\ r} \ $ and 
$\ \Delta \ {\cal{L}} \ $ in eq. \ref{eq:3-1-1}, in Fermi gauges.
\vspace*{-0.0cm}

\vspace{-0.5cm}
\begin{equation}
\label{eq:3-1-5}
\begin{array}{c}
\begin{array}{l}
\hspace*{-2.8cm}
\color{black} \mbox{gauge bosons} :
\hspace*{1.0cm} \color{black}
{\cal{L}}_{\ B} \ = 
\ - \ \frac{1}{4 \ \color{black} g^{\ 2}} \ \color{black} B^{\ \mu \nu \ r} 
\ B_{\ \mu \nu}^{\ r}
\end{array}
\vspace*{0.1cm} \\
B_{\ \mu \nu}^{\ r} \ =
\ \partial_{\ \mu} \ W_{\ \nu}^{\ r} \ -
\ \partial_{\ \nu} \ W_{\ \mu}^{\ r} \ + \ f_{\ r s t} \ W_{\ \mu}^{\ s}
\ W_{\ \nu}^{\ t}
\hspace*{0.2cm}
{\color{black} \longleftarrow}
\hspace*{0.1cm}
\left ( \ W_{\ \mu}^{\ r} \ \equiv \ - \ v_{\ \mu}^{\ r} \ \right )
\vspace*{0.1cm} \\
r, s, t \ = \ 1,\cdots,dim \ ( \ {\cal{G}} \ = \ SU3_{\ c} \ ) \ = \ 8
\vspace*{0.1cm} \\ 
\mbox{Lie algebra labels}  ,
 \left \lbrack  \frac{1}{2} \ \lambda^{\ r} \ , \ \frac{1}{2} \ \lambda^{\ s}
 \right \rbrack \ = \ i \ f_{\ r s t} \ \frac{1}{2} \ \lambda^{\ t}
\vspace*{0.1cm} \\
\color{black} \mbox{perturbative rescaling :}
\vspace*{0.0cm} \\
W_{\ \mu}^{\ r} \ = \ \color{black} g \ \color{black} W_{\ \mu \ pert}^{\ r} 
\ , \ B_{\ \mu \nu}^{\ r} \ = \ \color{black} g 
\ \color{black} B_{\ \mu \nu \ pert}^{\ r}
\end{array}
\end{equation}

\noindent
Degrees of freedom are seen in jets , in (e.g.) the energy momentum sum rule in 
deep inelastic scattering 
but not clearly in spectroscopy.

\noindent
Completing $\Delta \ {\cal{L}}$ in Fermi gauges

\vspace{-0.2cm}
\begin{equation}
\label{eq:3-1-6}
\begin{array}{l}
\Delta \ {\cal{L}} \ =
\ \left \lbrace \begin{array}{c}
- \ \frac{1}{2 \ \eta \ \color{black} g^{\ 2}}
\ \color{black} \left ( \ \partial_{\ \mu} \ W^{\ \mu \ r} \ \right )^{\ 2}
\vspace*{0.1cm} \\
+ \ \partial^{\ \mu} \ \overline{c}^{\ r} \ ( \ D_{\ \mu} \ c \ )^{\ r}
\end{array}
\right \rbrace
\hspace*{0.2cm} ; \hspace*{0.2cm}
\eta \ : \ \mbox{gauge parameter}
\vspace*{0.1cm} \\
\color{black} \mbox{ghost fermion fields :} \ \color{black} c \ , 
\ \overline{c}
\hspace*{0.2cm} ; \hspace*{0.2cm}
( \ D_{\ \mu} \ c \ )^{\ r} \ = \ \partial_{\ \mu} \ c^{\ r} \ + \ f_{\ r s t} 
\ W_{\ \mu}^{\ s}
\ c^{\ t}
\vspace*{0.1cm} \\
\color{black} \mbox{gauge fixing constraint :} \ \color{black}
C^{\ r} \ = \ \partial_{\ \mu}  \ W^{\ \mu \ r} 
\end{array}
\vspace*{-0.3cm}
\end{equation}

\begin{center}
\vspace*{-0.3cm}
{\bf \color{black}
3-1-2a -- Gauge boson binary bilocal and adjoint ( here octet- ) string operators
}
\label{'3-1-2a'}
\end{center}
\vspace*{0.1cm}

\noindent
One goal is, to identify -- not just some candidate resonance -- gluonic mesons,
binary and higher modes, and to relate them to the base quantities within QCD .
Here we follow ref. \cite{PMErice2011} .

\vspace*{-0.3cm}
\begin{equation}
\label{eq:4}
\begin{array}{l}
B_{\ \left \lbrack \ \mu_{1} \ \nu_{1} \  \right \rbrack
\ , \ \left \lbrack  \ \mu_{2} \ \nu_{2}  \ \right \rbrack}
\ ( \ x_{\ 1} \ , \ x_{\ 2} \ )
\ =
\vspace*{0.1cm} \\
\hspace*{0.2cm}
= \ B_{\ \left \lbrack \ \mu_{1} \ \nu_{1} \ \right \rbrack}^{\hspace*{0.2cm} r}
\ ( \ x_{\ 1} \ )
\ U \ ( \ x_{\ 1} \ , \ r \ ; \ x_{\ 2} \ , \ s \ ) 
\ B_{\ \left \lbrack \ \mu_{2} \ \nu_{2} \ \right \rbrack}^{\hspace*{0.2cm} s}
\ ( \ x_{\ 2} \ )
\vspace*{0.1cm} \\
r \ , \ s \ , \cdots \ = \ 1, \cdots , 8
\hspace*{0.2cm} ; \hspace*{0.2cm}
\mbox{no flavor but spin}
\end{array}
\vspace*{-0.2cm}
\end{equation}

\noindent
$\ B_{\ \left \lbrack \ \mu \ \nu \ \right \rbrack}^{\hspace*{0.2cm} r} 
\ ( \ x \ ) \ $
denote the local color octet of field strengths.

\noindent
The quantity $ U \ ( \ x \ , \ r \ ; \ y \ , \ s \ )$
in eq. (\ref{eq:4}) denotes the octet string operator,
i. e. the path ordered exponential
over a straight line path ${\cal{C}}$ from y to x

\vspace*{-0.3cm}
\begin{equation}
\label{eq:5}
\begin{array}{l}
\begin{array}{@{\hspace*{0.0cm}}l@{\hspace*{0.0cm}}l@{\hspace*{0.0cm}}l}
U \ ( x , r ; y , s )
& = &
\left . P \ \exp 
\ \left ( 
\ \left . 
{\displaystyle{\int}}_{\ y}^{\ x} 
\ \right |_{\ {\cal{C}}}
\ d \ z^{\ \mu}
\ \frac{1}{i} \ v_{\ \mu}^{\ t} \ ( \ z \ ) \ {\cal{F}}_{\ t}
\ \right ) \right |_{\ r \ s}
\vspace*{0.1cm} \\
& = &
\left . P \ \exp 
\ \left ( 
\ \left . 
- \ {\displaystyle{\int}}_{\ y}^{\ x} 
\ \right |_{\ {\cal{C}}}
\ d \ z^{\ \mu}
\ W_{\ \mu}^{\ t} \ ( \ z \ ) \ \left ( \ \frac{1}{i} 
\ {\cal{F}}_{\ t} \ \right )
\ \right ) \right |_{\ r \ s}
\end{array}
\vspace*{0.1cm} \\
\left ( \ {\cal{F}}_{\ t} \ \right )_{\ r s} \ = \ i \ f_{\ r \ t \ s}
\hspace*{0.2cm} ; \hspace*{0.2cm} 
\left ( \ ad_{\ t} \ \right )_{\ r s} = \ \frac{1}{i} 
\ \left ( \ {\cal{F}}_{\ t} \ \right )_{\ r s} \ = \ f_{\ r \ t \ s}
\end{array}
\end{equation}

\noindent
The path ordered exponential as a matrix function of the argument is 
to be performed before the matrix elements, denoted $\ |_{\ ..} \ $ 
in eq. \ref{eq:5} , are taken. The  local limit becomes

\vspace*{-0.5cm}
\begin{equation}
\label{eq:6}
\begin{array}{l}
B_{\ \left \lbrack \ \mu_{1} \ \nu_{1} \  \right \rbrack
\ , \ \left \lbrack  \ \mu_{2} \ \nu_{2}  \ \right \rbrack}
\ ( \ x_{\ 1} \ = \ x_{\ 2} \ = \ x \ ) 
\ = 
\vspace*{0.1cm} \\
\hspace*{1.5cm} = \  (:) \ B_{\ \left \lbrack \ \mu_{1} \ \nu_{1} \ \right \rbrack}^{\ r} 
\ ( \ x \ ) \ B_{\ \left \lbrack \ \mu_{2} \ \nu_{2} \ \right \rbrack}^{\ r} 
\ ( \ x \ ) (:)
\hspace*{0.1cm} ; \hspace*{0.1cm}
{\color{black} \mbox{no flavor but spin}}
\end{array}
\end{equation}

\begin{center}
\vspace*{-0.1cm}
{\bf \color{black}
2-3 -- $\ \overline{q} \ q \ $ bilinears and triplet-string operators
}
\label{'2-3'}
\end{center}
\vspace*{0.0cm}

\vspace*{-0.5cm}
\begin{equation}
\label{eq:7}
\begin{array}{l}
B^{\ q}_{\ \left \lbrack \ \dot{{\cal{A}}} \ \dot{f}_{\ 1} \ , 
\ {\cal{B}} \ f_{\ 2} \  \right \rbrack} \ ( \ x_{\ 1} \ , \ x_{\ 2} \ )
\ = 
\vspace*{0.1cm} \\
\hspace*{1.0cm} = 
\ \overline{q}^{\ \dot{c}_{\ 1}}_{\ {\cal{B}} \ \dot{f}_{\ 2}} 
\ ( \ x_{\ 1} \ )
\ U \ ( \ x_{\ 1} \ , \ c_{\ 1} \ ; \ x_{\ 2} \ , \ \dot{c}_{\ 2} \ )
\ q^{\ c}_{\ {\cal{A}} \ f_{\ 1}} \ ( \ x_{\ 2} \ )
\vspace*{0.1cm} \\
{\color{black} \mbox{flavor {\it and} spin}}
\vspace*{0.1cm} \\
\begin{array}{@{\hspace*{0.0cm}}l@{\hspace*{0.0cm}}l@{\hspace*{0.0cm}}l}
U ( x , c_{1} ; y , \dot{c}_{2} )
& = & \left . P \ \exp
\ \left (
\ \left .
{\displaystyle{\int}}_{\ y}^{\ x}
\ \right |_{\ {\cal{C}}}
 d \ z^{\ \mu}
\ \frac{1}{i} \ v_{\ \mu}^{\ t} \ ( \ z \ ) \ \frac{1}{2} \ \lambda_{\ t}
\ \right ) \right |_{\ c_{ 1} \ \dot{c}_{ 2}}
\vspace*{0.15cm} \\
& = &
\left . P \ \exp
\ \left (
\ \left .
- \ {\displaystyle{\int}}_{\ y}^{\ x}
\ \right |_{\ {\cal{C}}}
 d \ z^{\ \mu}
\ W_{\ \mu}^{\ t} \ ( \ z \ ) \ \left ( \ \frac{1}{i} \ \frac{1}{2} \ \lambda_{\ t} 
\ \right )
\ \right ) \right |_{\ c_{ 1} \ \dot{c}_{ 2}}
\end{array}
\end{array}
\vspace*{-0.2cm}
\end{equation}

\noindent
with the local limit

\vspace*{-0.2cm}
\begin{equation}
\label{eq:8}
\begin{array}{l}
B^{q}_{\left \lbrack \dot{{\cal{B}}} \dot{f}_{2} ,
{\cal{A}} \ f_{1} \right \rbrack} 
\ ( x_{1} = x_{2} = x )
= (:) \overline{q}^{\dot{c}}_{{\cal{B}} \dot{f}_{2}}
\ ( x ) 
\ q^{c}_{{\cal{A}} f_{1}} \ ( x ) (:) 
\end{array}
\end{equation}

\noindent
The symbols $\ (:) \ $ in eqs. \ref{eq:6} and \ref{eq:8} should indicate that
normal ordering of regulating the local limits is required and further that
such normal ordering is {\it not} unique , related to Bogoliubov
transformations , and dependent on quark masses
in the case of the $\ \overline{q} \ q \ $ bilinears.

\begin{center}
\vspace*{-0.1cm}
{\bf \color{black}
2-4 -- Connection and curvature - forms
\vspace*{-0.1cm} \\
preparing 
the ensuing analysis of regularity conditions
}
\label{'2-4'}
\end{center}
\vspace*{0.0cm}

\noindent
Lets begin this (sub-)section rewriting the bilocal (formally) unitary
operators forming the gauge connection dependent octet- ( eq. \ref{eq:5} )
and triplet ( eq. \ref{eq:7} ) QCD strings , substituting an equivalent , matrix
oriented notation

\vspace*{-0.3cm}
\begin{equation}
\label{eq:9}
\begin{array}{l}
\hspace*{-0.3cm} \begin{array}{@{\hspace*{0.0cm}}l@{\hspace*{-0.0cm}}l
@{\hspace*{0.0cm}}}
\mbox{{\color{black} \begin{tabular}{c}
octet
\vspace*{-0.2cm} \\
string
\end{tabular} \hspace*{-0.3cm} :
}}
& 
U ( x , r \hspace*{0.0cm} ; \hspace*{0.0cm} y \ , \ s \ )
\hspace*{0.0cm} =
\left . P \exp 
\left ( 
\left . 
- {\displaystyle{\int}}_{ y}^{ x} 
\ \right |_{{\cal{C}}}
d \ z^{\ \mu}
\ W_{\ \mu}^{\ t} \ ( z ) \left ( \frac{1}{i} 
{\cal{F}}_{\ t} \right )
\ \right ) \right |_{\ r s}
\vspace*{0.2cm} \\
\mbox{{\color{black} \begin{tabular}{c}
triplet
\vspace*{-0.2cm} \\
string
\end{tabular} \hspace*{-0.3cm} :
}}
&
U ( x , c_{ 1}  ; y ,  \dot{c}_{ 2} )
=  \left . P  \exp
\left (
\left .
- {\displaystyle{\int}}_{\ y}^{\ x}
 \right |_{{\cal{C}}}
d \ z^{\ \mu}
\ W_{\ \mu}^{\ t} \ ( z ) \ \left ( \ \frac{1}{i} \frac{1}{2} 
\lambda_{\ t} \right )
\right ) \right |_{\ c_{1} \dot{c}_{2}}
\end{array}
\vspace*{0.0cm} \\
\mbox{with the substitutions} \hspace*{0.2cm} \longrightarrow
\vspace*{0.0cm} \\
\hspace*{-0.4cm} \left . \begin{array}{l@{\hspace*{+0.1cm}}l}
\mbox{{\color{black} \begin{tabular}{c}
octet
\vspace*{-0.2cm} \\
string
\end{tabular} \hspace*{-0.3cm} :
}}
& 
U ( x , r \hspace*{0.0cm} ; \hspace*{0.0cm} y , s )
\ \rightarrow \ \left ( 
\ U \ \left ( x \stackrel{C}{\leftarrow} y \right )
\ \right )_{ r s}
\vspace*{0.1cm} \\
\mbox{{\color{black} \begin{tabular}{c}
triplet
\vspace*{-0.2cm} \\
string
\end{tabular} \hspace*{-0.3cm} :
}}
& 
U \ ( x , c_{ 1} ; y , \dot{c}_{ 2} )
\ \rightarrow \ \left (
U \left ( x \stackrel{C}{\leftarrow} y \right )
\right )_{ c_{1} \dot{c}_{2}}
\end{array}
\hspace*{-0.1cm} \right \rbrace \rightarrow
\hspace*{-0.4cm} 
\begin{array}[t]{r} \left . U ( x , C , y ; {\cal{D}} ) 
\right ._{\alpha \beta}
\vspace*{0.1cm} \\
\in   
\hspace*{-0.0cm} 
{\cal{D}} \ ( {\cal{G}} )
\end{array}
\vspace*{0.1cm} \\
\mbox{with} {\cal{G}} = \mbox{simple compact gauge group}
\hspace*{0.1cm} ; \hspace*{0.1cm}
\vspace*{0.1cm} \\
{\cal{D}} : \mbox{general irreducible representation of} \ {\cal{G}}
\end{array}
\end{equation}

\noindent
Here $\ {\cal{G}} \ = \ SU3_{\ c} \ $ and $\ {\cal{D}} \ $ is the octet-
, triplet representation for the respective QCD $\ {\cal{D}} \ $- strings.

\noindent
Further let us consider matrix valued connection 1-forms , which define
the bilocal matrix valued operators $\  \left ( \ U \ ( \ x \ , C \ , y 
\ ; \ {\cal{D}} \ ) \ \right )_{ \alpha \beta}
\ \in \ {\cal{D}} \ ( \ {\cal{G}} \ ) $ as given in eq. \ref{eq:9} .
To this end the form of octet and triplet strings in eq. \ref{eq:9} is repeated below

\vspace*{-0.5cm}
\begin{equation}
\label{eq:10}
\begin{array}{l}
\hspace*{-0.4cm} \begin{array}{l@{\hspace*{-0.0cm}}l}
\mbox{{\color{black} \begin{tabular}{c}
octet
\vspace*{-0.2cm} \\
string
\end{tabular} \hspace*{-0.3cm} :
}}
& 
U ( x , r \hspace*{0.0cm} ; \hspace*{0.0cm} y , s )
\hspace*{0.0cm} =
\left . P \exp 
\left ( 
\left . 
- {\displaystyle{\int}}_{\ y}^{\ x} 
\right |_{{\cal{C}}}
d z^{\mu}
W_{\mu}^{t} \ ( z ) \left ( \frac{1}{i} 
{\cal{F}}_{t} \right )
\ \right ) \right |_{\ r \ s}
\vspace*{0.2cm} \\
\mbox{{\color{black} \begin{tabular}{c}
triplet
\vspace*{-0.2cm} \\
string
\end{tabular} \hspace*{-0.3cm} :
}}
&
U ( x , c_{1} ; y , \dot{c}_{2} )
= \left . P \exp
\left (
\left .
- {\displaystyle{\int}}_{\ y}^{\ x}
\right |_{{\cal{C}}}
d \ z^{\ \mu}
W_{\mu}^{t} \ ( z ) \left ( \frac{1}{i} \frac{1}{2} 
\lambda_{t} \right )
\right ) \right |_{c_{1} \dot{c}_{2}}
\end{array}
\end{array}
\end{equation}
 
\noindent
The two matrices in brackets to the right of the integrand expressions in 
eq. \ref{eq:10} form an antihermitian basis of the Lie algebra representation
$\ Lie \ ( \ {\cal{D}} \ ) \ $ for $\ {\cal{D}} \ = \ \mbox{adjoint} \ $
and $\ {\cal{D}} \ = \ \mbox{triplet} \ $ representations of 
$\ {\cal{G}} \ = \ SU3_{\ c} \ $ respectively

\vspace*{-0.3cm}
\begin{equation}
\label{eq:11}
\begin{array}{l}
d_{t} \equiv d_{t} \ ( {\cal{D}} ) 
\leftrightarrow \left ( d_{t} \right )_{\alpha \beta}
= \left \lbrace \begin{array}{cll}
\left ( \frac{1}{i}
{\cal{F}}_{t} \right )_{r s} & \mbox{for} & Lie \ ( {\cal{D}} )
= \mbox{adjoint} 
\vspace*{0.1cm} \\
\left ( \frac{1}{i} \frac{1}{2}
\lambda_{t} \right )_{c_{1} \dot{c}_{2}}
& \mbox{for} & Lie \ ( {\cal{D}} )
= \mbox{triplet}
\end{array} \right .
\vspace*{0.1cm} \\
d_{t} = - d_{t}^{\dagger}
\hspace*{0.1cm} ; \hspace*{0.1cm}
t = 1,\cdots,dim \ {\cal{G}} 
\hspace*{0.1cm} ; \hspace*{0.1cm}
\alpha , \beta = 1,\cdots,dim \ {\cal{D}
\hspace*{0.2cm} \mbox{for general}} \ {\cal{D}}
\end{array}
\end{equation}

\noindent
From eqs. \ref{eq:10} and \ref{eq:11} we construct a -- hopefully -- consistent
notation as appropriate for matrix valued $\ {\cal{D}} \ $ connections, 1-forms
and strings, as well as derived 2- and higher forms. First eq. \ref{eq:11}
is subject to the ( matrix- ) commutation relations

\vspace*{-0.3cm}
\begin{equation}
\label{eq:12}
\begin{array}{l}
\left \lbrack d_{r} , d_{s} \right \rbrack 
= f_{r s t} d_{t} 
\hspace*{0.1cm} ; \hspace*{0.1cm} 
\forall \ {\cal{D}} \ ( {\cal{G}} )
\hspace*{0.1cm} ; \hspace*{0.1cm} 
r,s,t = 1,\cdots,dim \ {\cal{G}}
\hspace*{0.1cm}
\longrightarrow
\vspace*{0.1cm} \\
\left ( d_{t} \ ( {\cal{D}} 
= \mbox{adjoint representation} )
\right )_{s r} = \left ( ad_{t} \right )_{s r}
= f_{s t r}
\hspace*{0.1cm} : \hspace*{0.1cm}
\mbox{independent of } {\cal{D}}
\vspace*{0.1cm} \\
f_{s t r} : \mbox{totally antisymmetric, real structure constants of} 
\ Lie \ ( {\cal{G}} )
\end{array}
\end{equation}

\noindent
In physics the antihermitian matrix code with respect to the
representations $\ Lie \ ( \ {\cal{D}} \ ) \ $ is ( mostly ) replaced by the 
hermitian one \footnote{{\color{black} \begin{tabular}[t]{l}
A ( partial ) collection of  
historical and textbook references to the topics
\vspace*{-0.1cm} \\
pertining to
'Continous
transformation groups and
differential geometry'  
\vspace*{-0.1cm} \\
is presented under
R-H  references
and labelled by the symbols 1H, 2H $\cdots$ 
\vspace*{-0.1cm} \\
{\color{black} 
in refs. 14 -- \cite{PMembedosc} and 16 -- \cite{PMErice2012}}.
\end{tabular}
}
}

\vspace*{-0.3cm}
\begin{equation}
\label{eq:13}
\begin{array}{l}
\left .
\left ( d_{t} \equiv \frac{1}{i} h_{t} \right ) 
\left ( Lie ( {\cal{D}} ) \right ) \right |_{\alpha \beta}
\hspace*{0.1cm} ; \hspace*{0.1cm}
h_{t} = h_{t}^{\dagger}
\hspace*{0.1cm} ; \hspace*{0.1cm}
\left \lbrack h_{r} , h_{s} \right \rbrack = 
i f_{r s t} h_{t}
\vspace*{0.1cm} \\
\alpha,\beta = 1,\cdots,dim \ {\cal{D}}
\end{array}
\end{equation}

\noindent
Eq. \ref{eq:12} serves to define 
matrix valued connections built from a basis of $\ Lie \ ( \ {\cal{D}} \ ) \ $
representation matrices as defined in eq. \ref{eq:12} for general
irredcible representations $\ {\cal{D}} \ ( \ {\cal{G}} \ ) \ $
denoted $\ {\cal{W}}_{\ \mu} \ ( \ z \ , \ {\cal{D}} \ ) \ $

\vspace*{-0.3cm}
\begin{equation}
\label{eq:14}
\begin{array}{l}
\left .
{\cal{W}}_{\mu} ( z , {\cal{D}} ) 
\right |_{\alpha \beta}
= W_{\mu}^{r} ( z ) \left ( d_{r} 
\right )_{\alpha \beta \ Lie \ ( {\cal{D}} ) }
\hspace*{01cm} ; \hspace*{0.1cm}
\vspace*{0.2cm} \left \lbrack \begin{array}{l}
r = 1,\cdots,dim \ ( {\cal{G}} )
\vspace*{-0.1cm} \\
\alpha , \beta = 1,\cdots,dim \ ( {\cal{D}} )
\end{array} \right \rbrack
\vspace*{0.1cm} \\
\left .
{\cal{W}}_{\mu} ( z , {\cal{D}} ) \right |_{\alpha \beta}
\hspace*{0.1cm} \longrightarrow \hspace*{0.1cm} 
{\cal{W}}_{\mu} \hspace*{0.1cm} \mbox{for compact matrix notation}
\end{array}
\vspace*{-0.1cm}
\end{equation}

\noindent
In the following it is to be understood, that $\ {\cal{W}}_{\ \mu} \ $
is extended to a general collection of representations
$\ \bigcup \ {\cal{D}} \ $ -- thought to be carried by 
real and spurious spin $\ \frac{1}{2} \ $
fields -- care beeing taken that asymptotic freedom
in the ultraviolet is not upset. 

\noindent
From eq. \ref{eq:14} we define the associated matrix valued connection 1-form
displayed alongside the base definition repeated from eq. \ref{eq:14} in
eq. \ref{eq:15} below

\vspace*{-0.1cm}
\begin{equation}
\label{eq:15}
\begin{array}{l}
\begin{array}{@{\hspace*{0.0cm}}l@{\hspace*{0.0cm}}l@{\hspace*{0.0cm}}l 
@{\hspace*{0.0cm}}l@{\hspace*{0.0cm}}l@{\hspace*{0.0cm}}}
\left . {\cal{W}}^{(1)} ( z , {\cal{D}} )
\right |_{\alpha \beta}
& = & d \ z^{\mu}  
\left .
{\cal{W}}_{\mu} ( z , {\cal{D}} ) \right |_{\alpha \beta}
& \longrightarrow &
{\cal{W}}^{(1)} 
\vspace*{0.1cm} \\
\left .
{\cal{W}}_{\mu} ( z , {\cal{D}} ) \right |_{\alpha \beta}
& = & W_{\mu}^{r} ( z ) \left ( d_{r}
\right )_{\alpha \beta \ Lie \ ( {\cal{D}} ) }
& \longrightarrow &
{\cal{W}}_{\mu}
\end{array}
\end{array}
\end{equation}

\noindent
and the matrix valued field-strength tensor 

\vspace*{-0.6cm}
\begin{equation}
\label{eq:16}
\begin{array}{l}
\left . {\cal{W}}_{\mu \nu} ( z , {\cal{D}} ) 
\right |_{\alpha \beta}
= \left \lbrace \begin{array}{c}
\partial_{\mu} {\cal{W}}_{\nu} ( z , {\cal{D}} )
- \partial_{\nu} {\cal{W}}_{\mu} ( z , {\cal{D}} )
+
\vspace*{0.1cm} \\
+ \left \lbrack {\cal{W}}_{\mu} ( z , {\cal{D}} ) ,
{\cal{W}}_{\nu} ( z , {\cal{D}} ) \right \rbrack
\end{array} \right \rbrace_{\alpha \beta}
\hspace*{0.1cm} \longrightarrow \hspace*{0.1cm}
{\cal{W}}_{\mu \nu}
\end{array} 
\vspace*{-0.0cm}
\end{equation}

\noindent
together with their associated curvature 2-form

\vspace*{-0.5cm}
\begin{equation}
\label{eq:17}
\begin{array}{l}
\begin{array}{lll @{\hspace*{0.0cm}}l@{\hspace*{0.1cm}}l}
\left . {\cal{W}}^{(2)} ( z , {\cal{D}} )
\right |_{\alpha \beta}
& = & \frac{1}{2} d \ z^{\mu} \wedge d \ z^{\mu}
\left .
{\cal{W}}_{\mu \nu} ( z , {\cal{D}} ) \right |_{\alpha \beta}
& \longrightarrow &
{\cal{W}}^{(2)}
\vspace*{0.2cm} \\
\left . {\cal{W}}_{\mu \nu} ( z , {\cal{D}} ) 
\right |_{\alpha \beta}
& = & \left \lbrace \begin{array}{c}
\partial_{\mu} {\cal{W}}_{\nu} ( z , {\cal{D}} )
- \partial_{\nu} {\cal{W}}_{\mu} ( z , {\cal{D}} )
+
\vspace*{0.1cm} \\
+ \left \lbrack {\cal{W}}_{\mu} ( z , {\cal{D}} ) ,
{\cal{W}}_{\nu} ( z , {\cal{D}} ) \right \rbrack
\end{array} \right \rbrace_{\alpha \beta}
& \longrightarrow &
{\cal{W}}_{\mu \nu}
\vspace*{0.1cm} \\
& = &
W_{\mu \nu}^{r} ( z ) \left ( \ d_{r}
\right )_{\alpha \beta \ Lie \ ( {\cal{D}} ) }
& &
\end{array} 
\vspace*{0.2cm} \\ \hline \vspace*{-0.2cm} \\
W_{\mu \nu}^{r} = 
\partial_{\mu} W_{\nu}^{r} - \partial_{\nu} W_{\mu}^{r}
+ f_{r s t} W_{\mu}^{s} W_{\nu}^{t}
\hspace*{0.1cm} ; \hspace*{0.1cm}
\mbox{independent of } {\cal{D}}
\vspace*{0.2cm}
\end{array} 
\vspace*{0.0cm}
\end{equation}

\noindent
Two remarks are in place here

\vspace*{-0.2cm}
\begin{description}
\item 1) In order to distinguish field strengths from potentials ( connections )
the following equivalent but different notations for the field strength shall
be used

\vspace*{-0.4cm}
\begin{equation}
\label{eq:18}
\begin{array}{l}
{\cal{W}}_{\ \mu \nu} \ \equiv \ {\cal{B}}_{\ \mu \nu}
\hspace*{0.2cm} ; \hspace*{0.2cm}
{\cal{W}}^{\ (2)} \ \equiv \ {\cal{B}}^{\ (2)}
\hspace*{0.2cm} ; \hspace*{0.2cm}
W_{\ \mu \nu}^{\ r} \ \equiv \ B_{\ \mu \nu}^{\ r}
\end{array}
\end{equation}

\item 2) From the last relation in eq. \ref{eq:17} it may appear redundant 
to extend  connections and curvatures to matrix valued form with respect to
a wide collection of irreducible representations 
$\ {\cal{D}} \ ( \ {\cal{G}} \ ) \ $.
This however is tantamount to neglecting nontrivial 
global regularity conditions in the infrared . 
\vspace*{-0.3cm}
\end{description}

\noindent
We end this subsection ( 2-4 ) displaying the bilocal ( parallel transport- ) operators
defined in eq. \ref{eq:9} using the shorthand notation in eq. \ref{eq:17}

\vspace*{-0.3cm}
\begin{equation}
\label{eq:19}
\begin{array}{l}
\begin{array}{@{\hspace*{0.0cm}}c@{\hspace*{0.0cm}}l@{\hspace*{0.0cm}}c
@{\hspace*{0.0cm}}}
\left ( U ( x , C , y ; {\cal{D}} )
\right )_{\alpha \beta} 
& = &
\left . P \exp
\left (
\left .
- {\displaystyle{\int}}_{y}^{x}
\right |_{{\cal{C}}}
{\cal{W}}^{(1)} ( z , {\cal{D}} )
\right ) \right |_{\alpha \beta}
\vspace*{0.1cm} \\
\downarrow & & \downarrow
\vspace*{0.1cm} \\
U ( x , C , y ) 
& = &
\left . P \exp
\left (
- {\displaystyle{\int}}_{y}^{x}
\right |_{\ {\cal{C}}}
{\cal{W}}^{(1)} \right )
\hspace*{0.1cm} \left \lbrack \mbox{for} \hspace*{0.1cm}
\left ( \bigcup {\cal{D}} \right ) \right \rbrack
\end{array}
\end{array}
\end{equation}

\begin{center}
\vspace*{-0.3cm}
{\bf \color{black}
2-5 -- The U1- or singlet axial current anomaly
}
\label{'2-5'}
\end{center}
\vspace*{0.1cm}

\noindent
The U1-axial central anomaly involves the local chiral current projections
from $\ B^{\ q}_{\ \left \lbrack \ \dot{{\cal{B}}} \ \dot{f}_{\ 2} \ ,
\ {\cal{A}} \ f_{\ 1} \  \right \rbrack}
\ (  x \ ) \ $ in eq. \ref{eq:8}

\vspace*{-0.3cm}
\begin{equation}
\label{eq:9b}
\begin{array}{l}
\begin{array}{lll}
\left ( j_{\mu}^{\pm} \right )_{\dot{f}_{2} f_{1}} \ ( x ) & =
& B^{q}_{\left \lbrack \dot{{\cal{B}}} \dot{f}_{2} ,
{\cal{A}} f_{1} \right \rbrack}
\ (  x ) \ \left ( \ \gamma_{\mu} \frac{1}{2} \left ( \P \pm \gamma_{5} \right )
\right )_{{\cal{B}} \dot{{\cal{A}}}} 
\vspace*{0.1cm} \\
& = & (:) \overline{q}^{\ \dot{c}}_{\ \dot{f}_{\ 2}} \ \gamma_{\ \mu}^{\ \pm}
\ q^{\ c}_{\ f_{\ 1}} \ ( x ) (:)
\end{array}
\vspace*{0.1cm} \\
\gamma_{\ 5} \ =  \ \gamma_{\ 5 \ R} \ = \ \frac{1}{i} \ \gamma_{\ 0} \ \gamma_{\ 1}
\ \gamma_{\ 2} \ \gamma_{\ 3}
\hspace*{0.2cm} ; \hspace*{0.2cm}
\gamma_{\ \mu}^{\ \pm} \ = 
\ \gamma_{\ \mu} \ \frac{1}{2} \ \left ( \ \P \ \pm \ \gamma_{\ 5} \
\right )
\end{array}
\end{equation}

\noindent
The equations of motion for the fermion fields are {\it and superficially imply}
(upon $f_{\ 1} \ \leftrightarrow \ f_{\ 2} $)

\vspace*{-0.5cm}
\begin{equation}
\label{eq:10b}
\begin{array}{l}
\begin{array}{@{\hspace*{0.0cm}}l@{\hspace*{0.0cm}}ll@{\hspace*{0.0cm}}
ll@{\hspace*{0.0cm}}l}
\slash \hspace*{-0.15cm} \partial \ q^{\ c}_{f_{2}} & = & \frac{1}{i} 
\left ( \ \slash 
\hspace*{-0.15cm} v^{\ c \ \dot{c}'}
+ \delta^{c \dot{c}'} m_{f_{2}} \right ) q^{c'}_{f_{2}}
\vspace*{0.1cm} \\
\overline{q}^{\dot{c}}_{\dot{f}_{1}} 
\stackrel{\leftarrow}{\slash \hspace*{-0.15cm} \partial}
& = &  \overline{q}^{\dot{c}'}_{\dot{f}_{1}} \frac{1}{i} 
\left ( - \ \slash
\hspace*{-0.15cm} v^{c' \dot{c}} - \delta^{c' \dot{c}} m_{f_{1}} \right )
\end{array}
\hspace*{0.1cm} ; \hspace*{0.1cm}
\mbox{no sums over} \dot{f}_{1} , f_{2} \hspace*{0.2cm} {\color{black} 
\rightarrow}
\vspace*{0.2cm} \\
\partial^{\mu} \left ( j_{\mu}^{\pm} \right )_{\dot{f}_{1} f_{2}} =
\frac{1}{2 i} \left ( \left ( m_{f_{2}} - m_{f_{1}} \right ) 
\ S_{\dot{f}_{1} f_{2}} \mp \left ( m_{f_{2}} + m_{f_{1}} \right )
\ P_{\dot{f}_{1} f_{2}} \right )
\vspace*{0.1cm} \\
S_{\dot{f}_{1} f_{2}} = (:) \overline{q}^{\dot{c}}_{\dot{f}_{1}} q^{c}_{f_{2}}
(:)
\hspace*{0.1cm} , \hspace*{0.1cm}
P_{\dot{f}_{1} f_{2}} = (:) \overline{q}^{\dot{c}}_{\dot{f}_{1}} \gamma_{5} 
\hspace*{0.1cm} q^{c}_{f_{2}} (:)
\end{array}
\end{equation}

\noindent
In eq. \ref{eq:10b} $ \ m_{\ f} \ $ denotes the {\it real , nonnegative} quark 
mass for flavor f.

\noindent
From eq. \ref{eq:10b} the relations for vector and axial vector currents {\it superficially} follow

\vspace*{-0.3cm}
\begin{equation}
\label{eq:11b}
\begin{array}{l}
\left ( \ j_{\ \mu} \ \right )_{\ \dot{f}_{\ 1} \ f_{\ 2}}
\ = \ \left ( \ j_{\ \mu}^{\ +} \ \right )_{\ \dot{f}_{\ 1} \ f_{\ 2}}
\ + \  \left ( \ j_{\ \mu}^{\ -} \ \right )_{\ \dot{f}_{\ 1} \ f_{\ 2}}
\vspace*{0.1cm} \\
\left ( \ j_{\ \mu}^{\ 5} \ \right )_{\ \dot{f}_{\ 1} \ f_{\ 2}}
\ = \ \left ( \ j_{\ \mu}^{\ +} \ \right )_{\ \dot{f}_{\ 1} \ f_{\ 2}}
\ - \  \left ( \ j_{\ \mu}^{\ -} \ \right )_{\ \dot{f}_{\ 1} \ f_{\ 2}}
\vspace*{0.2cm} \\
\partial^{\ \mu} \ \left ( \ j_{\ \mu} \ \right )_{\ \dot{f}_{\ 1} \ f_{\ 2}}
\ = \ \frac{1}{i} \ \left ( \ m_{\ f_{\ 2}} \ - \ m_{\ f_{\ 1}} \ \right ) \ S_{\ \dot{f}_{\ 1} \ f_{\ 2}}
\vspace*{0.1cm} \\
\partial^{\ \mu} \ \left ( \ j_{\ \mu}^{\ 5} \ \right )_{\ \dot{f}_{\ 1} \ f_{\ 2}}
\ = \hspace*{0.4cm} \left ( \ m_{\ f_{\ 2}} \ + \ m_{\ f_{\ 1}} \ \right ) \ i \ P_{\ \dot{f}_{\ 1} \ f_{\ 2}}
\end{array}
\end{equation}

\noindent
As it follows from the original derivation by Adler and Bell and Jackiw 
\cite{AdlBeJa1} in QED,
the vecor current Ward identities in eq. \ref{eq:11b} can be implemented also in QCD , leaving the axial current ones reduced to the flavor non-singlet case, 
leaving the U1 axial current divergent anomalous

\vspace*{-0.3cm}
\begin{equation}
\label{eq:12b}
\begin{array}{l}
\partial^{\ \mu} \ \left ( \ j_{\ \mu} \ \right )_{\ \dot{f}_{\ 1} \ f_{\ 2}}
\ = \ \frac{1}{i} \ \left ( \ m_{\ f_{\ 2}} \ - \ m_{\ f_{\ 1}} \ \right ) \ S_{\ \dot{f}_{\ 1} \ f_{\ 2}}
\hspace*{0.2cm} {\color{red} \surd}
\vspace*{0.4cm} \\
\left \lbrace \begin{array}{l}
j_{\ \mu}^{\ 5}
\vspace*{0.1cm} \\
P
\end{array} \right \rbrace_{\ \dot{f}_{\ 1} \ f_{\ 2}}^{\ NS} \ =
\ \left \lbrace \begin{array}{l}
j_{\ \mu}^{\ 5}
\vspace*{0.1cm} \\
P
\end{array} \right \rbrace_{\ \dot{f}_{\ 1} \ f_{\ 2}} \ - 
\ \frac{1}{N_{\ fl}} \ \delta_{\ \dot{f}_{\ 1} \ f_{\ 2}} 
\ \sum_{\ f} \ \left \lbrace \begin{array}{l}
j_{\ \mu}^{\ 5}
\vspace*{0.1cm} \\
P
\end{array} \right \rbrace_{\ \dot{f} \ f} 
\end{array}
\end{equation}

\noindent
and similarly

\vspace*{-0.3cm}
\begin{equation}
\label{eq:13b}
\begin{array}{l}
\left \lbrace \begin{array}{l}
j_{\ \mu}^{\ 5}
\vspace*{0.1cm} \\
P
\end{array} \right \rbrace_{\ \dot{f}_{\ 1} \ f_{\ 2}}^{\ S} \ =
\ \sum_{\ f} \ \left \lbrace \begin{array}{l}
j_{\ \mu}^{\ 5}
\vspace*{0.1cm} \\
P
\end{array} \right \rbrace_{\ \dot{f} \ f}
\end{array}
\end{equation}

\begin{center}
\vspace*{-0.0cm}
{\bf \color{black}
2-6 -- Quark masses and splittings : $\ m_{\ f} \ $ and $\ \Delta \ m_{\ f} 
\ = \ m_{\ f} - 
\ \left \langle \ m \ \right \rangle \ $
}
\label{'2-6'}
\end{center}
\vspace*{0.0cm}

\noindent
In the subtitle above $ \ \left \langle \ m \ \right \rangle \ $ stands for the mean quark mass

\vspace*{-0.3cm}
\begin{equation}
\label{eq:14b}
\begin{array}{l}
\left \langle \ m \ \right \rangle \ = \ \frac{1}{N_{\ fl}} \ \sum_{\ f} \ m_{\ f}
\end{array}
\end{equation}

\noindent
The identities for vector currents in eqs. \ref{eq:11b} and \ref{eq:12b} can be extended separating
the conributions proportional to $\ \Delta \ m_{\ f} \ $ and $ \ \left \langle 
\ m \ \right \rangle \ $

\vspace*{-0.5cm}
\begin{equation}
\label{eq:15b}
\begin{array}{l}
\partial^{\ \mu} \ \left ( \ j_{\ \mu} \ \right )_{\ \dot{f}_{\ 1} \ f_{\ 2}}
\ = \ \frac{1}{i} \ \left ( \ \Delta \ m_{\ f_{\ 2}} \ - \ \Delta \ m_{\ f_{\ 1}} \ \right ) 
\ S_{\ \dot{f}_{\ 1} \ f_{\ 2}}
\hspace*{0.2cm} {\color{black} \surd}
\vspace*{0.1cm} \\
\partial^{\ \mu} \ \left ( \ j_{\ \mu}^{\ 5} \ \right )_{\ \dot{f}_{\ 1} \ f_{\ 2}}^{\ NS}
\ = \hspace*{0.4cm} \left ( \ \Delta \ m_{\ f_{\ 2}} \ + \ \Delta \ m_{\ f_{\ 1}} \ \right ) 
\ i \ P_{\ \dot{f}_{\ 1} \ f_{\ 2}}^{\ NS}
\hspace*{0.2cm} {\color{black} \surd}
\vspace*{0.1cm} \\
\partial^{\ \mu} \ \left ( \ j_{\ \mu}^{\ 5} \ \right )_{\ \dot{f}_{\ 1} \ f_{\ 2}}^{\ S}
\ = \ 2 \ \left \langle \ m \ \right \rangle \ i \ P^{\ S}
\hspace*{0.2cm} {\color{black} \slash \hspace*{-0.2cm} \surd} \ \left \lbrack 
\ \longrightarrow \ + \ \delta_{\ 5} \ \right \rbrack
\vspace*{0.1cm} \\
\delta_{\ 5} = \left ( \ 2 \ N_{\ fl} \right ) \ \frac{1}{32 \pi^{2}} 
\ \left . B_{\ \mu \ \nu}^{\ r} \ \widetilde{B}^{\ \mu \ \nu \ r}
\right |_{\rightarrow ren.gr.inv}
\hspace*{0.1cm} ; \hspace*{0.1cm}
\widetilde{B}_{\mu \ \nu}^{r} = \frac{1}{2} \ \varepsilon_{\mu \nu \sigma \tau}
\ B^{\sigma \ \tau \ r}
\end{array}
\vspace*{-0.4cm}
\end{equation}
\footnote{{\color{black} \hspace*{0.1cm} $\delta_{\ 5} \ $ was, as far as 
I know, introduced
by Murray Gell-Mann in lectures $\sim $ 1970 in
\hspace*{0.7cm}
Hawaii.}}

\noindent
We shall return to the question of how the local operator
$\ \ ch_{2} \ ( \ B \ ) \ \equiv \ \frac{1}{32 \pi^{\ 2}} 
\ (:) \ B_{\ \mu \ \nu}^{\ r} \ \widetilde{B}^{\ \mu \ \nu \ r} \ (:) \ $
is to be normalized and rendered renormalization group invariant
\cite{PMDob} . Here we just assume this to have been achieved and denote the
U1-axial anomaly, the first of the central two,  in its general form \\
( eq. \ref{eq:15b} )

\vspace*{-0.3cm}
\begin{equation}
\label{eq:16b}
\begin{array}{l}
\left \lbrace 
\ \partial^{\ \mu} \ \left ( \ j_{\ \mu}^{\ 5} \ \right )^{\ S}
\ = \ 2 \ \left \langle \ m \ \right \rangle \ i \ P^{\ S}
\ + \ \delta_{\ 5} 
\ \right \rbrace \ ( \ x \ )
\vspace*{0.1cm} \\
\delta_{\ 5} \ = \ \left ( \ 2 \ N_{\ fl} \ \right ) \ \frac{1}{32 \pi^{\ 2}}
\ \left . (:) \ B_{\ \mu \ \nu}^{\ r} \ \widetilde{B}^{\ \mu \ \nu \ r}
\ (:)
\ \right |_{\ \rightarrow ren.gr.inv}
\end{array}
\end{equation}

\noindent
From here it is conceptually clear how the scale- (or trace-) anomaly arises
but strictly within QCD . The renormalizability of a field theory in the
limit of uncurved space-time gives rise to a local , symmetric and {\it
conserved} energy momentum tensor , implying exact Poincar\'{e} invariance

\vspace*{-0.3cm}
\begin{equation}
\label{eq:17b}
\begin{array}{l}
\left \lbrace \ \vartheta_{\ \mu \ \nu} \ = \ \vartheta_{\ \nu \ \mu}
\ \right \rbrace \ ( \ x \ )
\vspace*{0.1cm} \\
\partial^{\ \nu} \ \vartheta_{\ \mu \ \nu} \ = \ 0 
\end{array}
\end{equation}

\noindent
In connection with the normal ordering questions it is important to admit
in the precise form of the energy momentum tensor a nontrivial vacuum
expected value , which
in view of exact Poincar\'{e} invariance must be of the form

\vspace*{-0.3cm}
\begin{equation}
\label{eq:18b}
\begin{array}{l}
\left \langle \Omega \right | \vartheta_{\mu \ \nu} \ ( x )
\left | \Omega \right \rangle =
\frac{1}{4} \ \eta_{\ \mu \ \nu} \ \tau
\vspace*{0.1cm} \\
\left \lbrace \begin{array}{c}
\eta_{\ \mu \ \nu} = diag \ \left ( 1 , -1 , -1 , -1 
\right )
\vspace*{0.1cm} \\
\tau
\end{array} \right \rbrace
\hspace*{0.1cm} \mbox{independent of} \ x 
\hspace*{0.1cm} {\color{black} \longrightarrow}
\vspace*{0.2cm} \\
\Delta \ \vartheta_{\mu \ \nu} \ ( x ) = 
\vartheta_{\mu \ \nu} \ ( x ) - 
\left \langle \Omega \right | \ \vartheta_{\mu \ \nu} \ ( x )
\left | \ \Omega \right \rangle
\times \left \lbrace \begin{array}{l} 
\widehat{\P} 
\vspace*{0.1cm} \\
\mbox{or} \hspace*{0.1cm} \left | \Omega \right \rangle
\left \langle \Omega \right |
\end{array} \right .
\vspace*{0.1cm} \\
\mbox{with} \hspace*{0.1cm} \partial^{\nu} \Delta \ \vartheta_{\mu \ \nu}
\ ( x ) = 0
\hspace*{0.1cm} ; \hspace*{0.1cm}
\left \langle \Omega \right | \Delta \ \vartheta_{\mu \ \nu} 
\ ( x ) \left | \Omega \right \rangle
= 0
\end{array}
\end{equation}

\noindent
In eq. \ref{eq:18b} $\ \widehat{\P} \ $ denotes the unit operator in the entire
Hilbert space of states , while 
$\ P_{\ \Omega} \ = \ \left | \ \Omega \ \right \rangle
\ \left \langle \ \Omega \ \right | \ $ stands for the projector on the ground
state .

\noindent
Furthermore from the two local, {\it conserved} tensors in eq. \ref{eq:18b}
only $\Delta \ \vartheta_{\mu \ \nu} \ ( x ) $ with vanishing vacuum
expected value is acceptable as  representing the conserved 4 momentum 
{\it operators} in the integral form

\vspace*{-0.2cm}
\begin{equation}
\label{eq:19b}
\begin{array}{l}
\widehat{P}_{\ \mu} \ = \ {\displaystyle{\int}}_{\ t} \ d^{\ 3} \ x
\ \Delta \ \vartheta_{\ \mu \ \ 0} \ ( \ t \ , \ \vec{x} \ )
\end{array}
\end{equation}

\noindent
All these arguments {\it notwithstanding} to subtract any eventual
vacuum expected values of local operators , often put forward 
as mathematical prerequisites , it is wise {\it not to do so}
in the presence of spontaneous
parameters , the dynamical origin of spontaneous symmetry breaking, e.g.
chiral symmetries in the limit or neighbourhood 
of some $\ m_{\ f} \ \rightarrow \ 0 \ $ .

\noindent
Using the (classical) equations of motion pertaining to the Lagrangean in
eqs. \ref{eq:3-1} - \ref{eq:3-1-1} 

\vspace*{-0.2cm}
\begin{equation}
\label{eq:20}
\begin{array}{l}
{\color{black}
\left ( \ D_{\ \nu} \ B^{\ \mu \nu} \ \right )^{\ r} \ = 
\ j^{\ \mu \ \ r} \ ( \ \overline{q} \ , \ q \ )
\hspace*{0.2cm} ; \hspace*{0.2cm} B \ \rightarrow \ B_{\ pert}
}  
\vspace*{0.1cm} \\
( \ D_{\ \varrho} \ B^{\ \mu \ \nu} \ )^{\ r} \ = 
\ \partial_{\ \varrho} \ B^{\ \mu \ \nu \ r} \ + \ f_{\ r s t}
\ W_{\ \varrho}^{\ s}
\ B^{\ \mu \ \nu \ t}
\vspace*{0.1cm} \\
j_{\ \mu}^{\ r} \ ( \ \overline{q} \ , \ q \ )
\ = \ g \ \overline{q}^{\ \dot{c}}_{\ \dot{{\cal{A}}} \ \dot{f}} 
\ \left ( \ \gamma_{\ \mu} \ \right )_{\ \dot{{\cal{A}}} \ {\cal{B}}}
\ \frac{1}{2} \ \left ( \ \lambda^{\ r} \ \right )_{\ c \dot{c}'}
q^{\ c'}_{\ {\cal{A}} \ f}
\vspace*{0.2cm} \\
{\color{black}
i \ \left ( \ \gamma^{\ \mu} \ D_{\ \mu} \ q 
\ \right )^{\ c}_{\ {\cal{A}} \ f}
\ = \ m_{\ f} \ q^{\ c}_{\ {\cal{A}} \ f}
\hspace*{0.2cm} \mbox{and} \hspace*{0.2cm}
q \ \rightarrow \ \overline{q}
}
\vspace*{0.1cm} \\
\left ( \ D_{\ \mu} \ q \ \right )^{\ c}_{\ {\cal{A}} \ f}
\ = \ \left \lbrack \ \partial_{\ \mu} \ \delta_{\ c \dot{c}'} \ + 
\ \frac{1}{i} 
\ W_{\ \mu}^{\ t} \ \frac{1}{2} 
\ \left ( \ \lambda^{\ t} \ \right )_{\ c \dot{c}'} \ \right \rbrack
\ q^{\ c'}_{\ {\cal{A}} \ f}
\vspace*{0.2cm} \\ \hline \vspace*{-0.2cm} \\
W_{\ \mu}^{\ r} \ \equiv \ - \ v_{\ \mu}^{\ r} \ = \ g \ \left ( 
\ W_{\ \mu}^{\ r} \ \right )_{\ pert} \ \equiv
\ - \ g \ \left ( \ v_{\ \mu}^{\ r} \ \right )_{\ pert}
\end{array}
\end{equation}

\noindent
the associated form of the energy momentum becomes

\vspace*{-0.2cm}
\begin{equation}
\label{eq:21}
\begin{array}{l}
\vartheta_{\ \mu \ \nu}^{\ (cl)} \ =
\ \left \lbrack \begin{array}{c}
\ \frac{1}{4 \ \color{black} g^{\ 2}}
\ \left \lbrack \begin{array}{c}
B_{\ \mu \ \varrho}^{\ t} \ B^{\ \varrho \ t}_{\hspace*{0.3cm} \nu} 
\ - \ \frac{1}{4}
\ \eta_{\ \mu \ \nu} \ B_{\ \sigma \ \varrho}^{\ t} \ B^{\ \varrho \ \sigma \ t}
\end{array} \ \right \rbrack
\ + 
\vspace*{0.1cm} \\
+ \ \frac{1}{2} \ \left \lbrack
\ \overline{q}_{\ \dot{f}} \ \gamma_{\ \mu} \ \frac{i}{2} \
\stackrel{\leftharpoondown \hspace*{-0.3cm} \rightharpoonup}{D_{\ \nu}}
\ q_{\ f} \ + \ \mu \ \leftrightarrow \ \nu \ \right \rbrack
\end{array} \right \rbrack
\end{array}
\end{equation}

\noindent
and using once more the fermion part of the equations of motion
the trace of the classical energy momentum tensor becomes

\vspace*{-0.2cm}
\begin{equation}
\label{eq:22}
\begin{array}{l}
\vartheta^{\ \mu \ (cl)}_{\hspace*{0.3cm} \mu}
\ = \ \sum_{\ f} m_{\ f} \ S_{\ \dot{f} \ f}
\vspace*{0.1cm} \\
S_{\ \dot{f}_{\ 1} \ f_{\ 2}} \ = \ (:) \ \overline{q}^{\ \dot{c}}_{\
\dot{f}_{\ 1}} \ q^{\ c}_{\ f_{\ 2}}
\ (:)
\end{array}
\end{equation}

\begin{center}
\vspace*{-0.0cm}
{\bf \color{black}
2-7 -- The scale- or trace- anomaly 
}
\label{'2-7'}
\end{center}
\vspace*{0.0cm}

\noindent
From the classical soft fermionic contribution to the trace of the energy
momentum tensor there is a clear conjecture , also by Murray
Gell-Mann , of the anomalous contribution , which subsequently became 
the scale- or trace- anomaly within QCD

\vspace*{-0.2cm}
\begin{equation}
\label{eq:23}
\begin{array}{l}
\vartheta^{\ \mu}_{\hspace*{0.3cm} \mu}
\ = \ \sum_{\ f} m_{\ f} \ S_{\ \dot{f} \ f} \ + \ \delta_{\ 0} 
\vspace*{0.1cm} \\
\delta_{\ 0} \ = \ - 
\ \left ( \ - 2 \ \beta \ ( \ g \ ) \ / \ g^{\ 3} \ \right )
\ \left \lbrack \ \frac{1}{4} \ (:) \ B_{\ \mu \ \nu}^{\ t} 
\ B^{\ \mu \ \nu \ t} \ (:)
\ \right \rbrack_{\ \rightarrow \ ren.gr.inv}
\end{array}
\end{equation}

\begin{center}
\vspace*{-0.0cm}
{\bf \color{black}
2-8 -- The two central anomalies alongside : scale- or trace- and 
U1-axial anomaly
}
\label{'2-8'}
\end{center}
\vspace*{-0.1cm}

\noindent
We collect the two anomalous identities in eqs. \ref{eq:23} and \ref{eq:16}

\vspace*{-0.3cm}
\begin{equation}
\label{eq:24}
\begin{array}{l}
\left \lbrace
\ \vartheta^{\ \mu}_{\hspace*{0.3cm} \mu}
\hspace*{0.95cm}  = \ \sum_{\ f} m_{\ f} \ S_{\ \dot{f} \ f} \ + \ \delta_{\ 0}
\ \right \rbrace \ ( \ x \ )
\vspace*{0.1cm} \\
\left \lbrace
\ \partial^{\ \mu} \ \left ( \ j_{\ \mu}^{\ 5} \ \right )^{\ S}
\ = \ 2 \ \left \langle \ m \ \right \rangle \ i \ P^{\ S}
\hspace*{0.20cm} + \ \delta_{\ 5}
\ \right \rbrace \ ( \ x \ )
\vspace*{0.2cm} \\
\delta_{\ 0} \ = \ - \ \left ( \ - 2 \ \beta \ ( \ g \ ) \ / \ g^{\ 3} \ \right )
\ \left \lbrack \ \frac{1}{4} \ (:) \ B_{\ \mu \ \nu}^{\ t}
\ B^{\ \mu \ \nu \ t} \ (:)
\ \right \rbrack_{\ \rightarrow \ ren.gr.inv}
\vspace*{0.1cm} \\
\delta_{\ 5} \ = \hspace*{0.8cm}  \left ( \ 2 \ N_{\ fl} \ \right ) 
\ \frac{1}{8 \pi^{\ 2}}
\ \left \lbrack \ \frac{1}{4} \ (:) \ B_{\ \mu \ \nu}^{\ t} 
\ \widetilde{B}^{\ \mu \ \nu \ t}
\ (:)
\ \right \rbrack_{\ \rightarrow ren.gr.inv}
\vspace*{0.2cm} \\ \hline \vspace*{-0.3cm} \\
- \ \beta \ / g^{\ 3} \ = \ \frac{1}{16 \pi^{\ 2}} \ b_{\ 0} \ + \ O \ ( \ X \ )
\hspace*{0.2cm} ; \hspace*{0.2cm} X \ = \ g^{\ 2} \ / \ ( \ 16 \ \pi^{\ 2} 
\vspace*{0.1cm} \\
\beta \ ( \ g \ ) \hspace*{0.2cm} : \hspace*{0.2cm} 
\mbox{Callan-Symanzik rescaling function in QCD}
\end{array}
\vspace*{-0.15cm}
\end{equation}

\noindent
The qualification 'central' for the anomalies in eq. \ref{eq:24} stands for the
property that in rendering the square coupling constant and the associated $\ \vartheta \ - \ $
parameter in the gauge boson {\it renormalized} Lagrangean density x dependent

\vspace*{-0.5cm}
\begin{equation}
\label{eq:25}
\begin{array}{l}
{\cal{L}}_{\ g.b.} \ = \ - \ \frac{1}{g^{\ 2}} \ \frac{1}{4} 
\ (:) \ B_{\ \mu \ \nu}^{\ t}
\ B^{\ \mu \ \nu \ t} \ (:) \ + \ \vartheta \ \frac{1}{8 \pi^{\ 2}}
\ \frac{1}{4} \ (:) \ B_{\ \mu \ \nu}^{\ t} \ \widetilde{B}^{\ \mu \ \nu \ t}
\hspace*{0.2cm} {\color{red} \longrightarrow}
\vspace*{0.1cm} \\
g^{\ 2} \ \rightarrow \ g^{\ 2} \ ( \ x \ )
\hspace*{0.2cm} ; \hspace*{0.2cm}
\vartheta \ \rightarrow \ \vartheta \ ( \ x \ )
\end{array}
\vspace*{-0.2cm}
\end{equation}

\noindent
maintains perturbative renormalizability and acts together with suitable 
boundary- -- more generally -- regularity conditions 
as external sources for the scalar and pseudoscalar local field strength 
bilinears

\vspace*{-0.3cm}
\begin{equation}
\label{eq:26}
\begin{array}{l}
\frac{1}{4} \ (:) \ B_{\ \mu \ \nu}^{\ t}
\ B^{\ \mu \ \nu \ t} \ (:)
\hspace*{0.2cm} , \hspace*{0.2cm}
\frac{1}{4} \ (:) \ B_{\ \mu \ \nu}^{\ t} \ \widetilde{B}^{\ \mu \ \nu \ t}
\end{array}
\vspace*{-0.15cm}
\end{equation}

\noindent
We will use the following definitions relative to the rescaling function 
$\ \beta \ $

\vspace*{-0.3cm}
\begin{equation}
\label{eq:27}
\begin{array}{l}
- \ \beta \ / \ g \ = \ X \ B \ ( \ X \ )
\hspace*{0.2cm} ; \hspace*{0.2cm} B \ ( \ X \ ) \ = \ b_{\ 0} \ A \ ( \ X \ )
\vspace*{0.1cm} \\
B \ ( \ X \ ) \ \sim \ \sum_{\ n=0}^{\ \infty} \ b_{\ n} \ X^{\ n}
\hspace*{0.2cm} , \hspace*{0.2cm}
A \ ( \ X \ ) \ \sim \ \sum_{\ n=0}^{\ \infty} \ a_{\ n} \ X^{\ n}
\vspace*{0.1cm} \\
\kappa \ = \ g^{\ 2} \ / \ ( \ 16 \ \pi^{\ 2} \ )
\hspace*{0.2cm} \mbox{generic} \hspace*{0.2cm} \longrightarrow
\hspace*{0.2cm} X \ , \ Y
\vspace*{0.2cm} \\
b_{\ 0} \ = \ \frac{1}{3} \ \left ( \ 33 \ - \ 2 \ N_{\ fl} \ \right )
\hspace*{0.2cm} , \hspace*{0.2cm} a_{\ 0} \ = \ 1 
\hspace*{0.2cm} , \hspace*{0.2cm} a_{\ n} \ = \ b_{\ n} \ / \ b_{\ 0}
\vspace*{0.1cm} \\
b_{\ 1} \ = \ \frac{2}{3} \ \left ( \ 153 \ - \ 19 \ N_{\ fl} \ \right )
\vspace*{0.1cm} \\
b_{\ 2} \ = \ \frac{1}{54} \ \left ( \ 77139 \ - \ 15099 \ N_{\ fl} \ + \ 325 \ N_{\ fl}^{\ 2}
\ \right )
\vspace*{0.1cm} \\
b_{\ 3} \ \sim \ 29243 \ - \ 6946.3 \ N_{\ fl} \ + \ 405.089 \ N_{\ fl}^{\ 2} \ +
\ 1.49931 \ N_{\ fl}^{\ 3}
\end{array}
\vspace*{-0.15cm}
\end{equation}

\noindent
References for this section ( 2 - premises ) are presented
in five (partial) collections in ref. 16 -- \cite{PMErice2012} :

\vspace*{-0.2cm}
\begin{description} 

\item 1 : (R) directly related to the two central anomalies

\vspace*{-0.15cm}
\item 2 : (rBsquare) establishing the one renormalization group invariant
quantity of dimension $\ \left \lbrack \ M^{\ 4} \ \right \rbrack \ $

\vspace*{-0.15cm}
\item 3 : (r-sp-1) a recent paper by Guido Altarelli and references cited
therein

\vspace*{-0.15cm}
\item 4 : (r-A2x) a selection of papers and textbooks for the entire realm 
of QCD

\vspace*{-0.15cm}
\item 5 : (r-condx) : Condensation phenomena and field theory realizations
\vspace*{-0.5cm}
\end{description}

\vspace*{-0.2cm}
\begin{center}
\hspace*{0.0cm}
\begin{figure}[htb]
\vskip -0.2cm
\hskip -0.0cm
\includegraphics[angle=0,width=7.0cm]{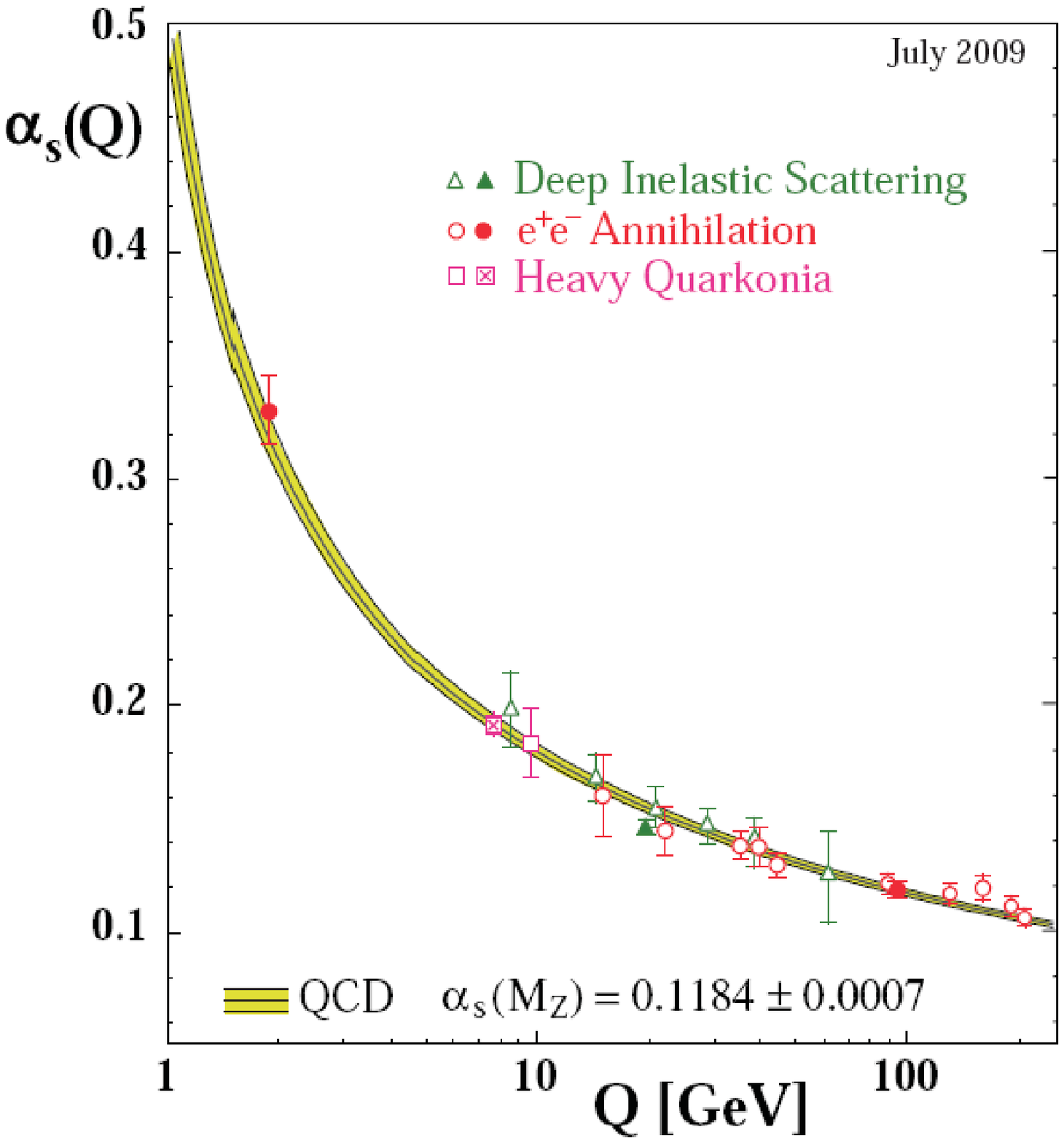}
\vskip -0.2cm
\label{figA21a}
\vspace*{0.10cm}
{\bf \color{black} \hspace*{0.0cm}
\begin{tabular}{c} Fig. A21 : $\ \alpha_{\ s} \ ( \ Q \ ) \ =
\ 4 \pi \ \kappa_{\ \overline{\mu} \ = \ Q} \ $
from ref. \cite{Bethke4} .
\end{tabular}
}
\end{figure}
\vspace*{-0.3cm}
\end{center}

\begin{center}
{\color{black} This ends section 2 -- premises} \vspace*{-0.35cm} \\
{\color{black}
\begin{tabular}{c@{\hspace*{9cm}}c}
 & 
\vspace*{0.0cm} \\ \hline 
\end{tabular}
}
\vspace*{-0.1cm}
\end{center}

\begin{center}
\vspace*{-0.0cm}
{\bf \color{black} 4 -- Ideas forging and foregoing - the dynamics of 
genuinely oscillatory modes \cite{oscmod1980}
} 
\end{center}
\label{'4-1'}
\vspace*{0.1cm}

\noindent
It is worth noting , that Erwin Schr\"{o}dinger turned to the discussion
of oscillatory modes of single- and by reduction of c.m. coordinates -- of 
a pair mode of oscillatory motion , in the last ( 4th ) paper in \\
ref. \cite{ESchr} . 

\noindent
However in the above paper he ( E.S. ) makes the assumption , that associated
forces arise from the exchange of a photon , i.e. involve a local
electromagnetic exchange interaction as giving rise to an equally local second
order wave equation , responsible for oscillatory (pair-) modes . This is
incorrect , contrary to the structure embedded in QCD , up to the present
incomplete level of completion , which remains a future task .

\noindent
Thus we concentrate on the present topic and lay out the ideas in ref. 
\cite{oscmod1980} .

\noindent
We relate the bond structure of quark-antiquark ( meson ) and $\ N- \ $quark
( baryon ) systems, subject to an SU($N$) unbroken color gauge group ,
to the long-range dynamics involving the oscillatory modes in the phase space 
of the center of mass position and momentum variables 
$ \left \lbrack \ N \ \equiv \ N_{\ c} \ \rightarrow \ 3 \ \right \rbrack $
to be clear. 
\vspace*{0.2cm}

\noindent
These canonical barycentric 3-vector variables are shown in eq. \ref{eq:1-1}

\vspace*{-0.5cm}
\begin{equation}
\label{eq:1-1}
\hspace*{-0.2cm} \begin{array}{l}
\begin{array}{@{\hspace*{0.0cm}}c@{\hspace*{0.0cm}}l@{\hspace*{0.0cm}}c
@{\hspace*{0.0cm}}}
\vec{\pi}_{1} = \frac{1}{\sqrt{2}} ( \ \vec{p}_{1} - 
\ \vec{p}_{2} \ )
& , &
\vec{z}_{1} = \frac{1}{\sqrt{2}} ( \ \vec{x}_{1} -
\ \vec{x}_{2} \ )
\vspace*{0.1cm} \\
\vec{\pi}_{2} = \frac{1}{\sqrt{6}} \ ( \ \vec{p}_{1} \ +
\ \vec{p}_{2} - 2 \ \vec{p}_{3} \ )
& , &
\vec{z}_{2} = \frac{1}{\sqrt{6}} ( \ \vec{x}_{1} +
\vec{x}_{\ 2} - 2 \ \vec{x}_{3} \ )
\vspace*{0.1cm} \\
. &   & .
\vspace*{0.0cm} \\
. &   & .
\vspace*{0.1cm} \\
\vec{\pi}_{\nu}  = (\nu ( \nu + 1 ))^{-1/2}
\left ( \begin{array}{l}
\sum_{\alpha = 1}^{\nu} \ \vec{p}_{\alpha} )
\vspace*{0.1cm} \\
 - \nu \ \vec{p}_{\nu + 1} 
\end{array}
\right )
& , & 
\vec{z}_{\ \nu} = (\nu ( \nu + 1 ))^{-1/2}
\left ( \begin{array}{l}
\sum_{\alpha = 1}^{\nu} \ \vec{x}_{\alpha}
\vspace*{0.1cm} \\
- \nu \ \vec{x}_{\nu + 1} 
\end{array}
\right )
\vspace*{0.0cm} \\
. &   & .
\vspace*{0.0cm} \\
. &   & .
\vspace*{0.1cm} \\
\vec{\pi}_{N - 1} = \cdots
& , &
\vec{z}_{N - 1} = \cdots
\vspace*{0.2cm} \\ \hline \vspace*{-0.3cm} \\
\vec{\pi}_{N} = N^{-1/2} 
\sum_{\alpha = 1}^{N} \ \vec{p}_{\alpha}
\ \rightarrow \ 0
& , &
\vec{z}_{N} = N^{-1/2}
\sum_{\alpha = 1}^{N} \ \vec{x}_{\alpha}
\ \rightarrow \ 0
\end{array}
\end{array}
\end{equation}
\vspace*{0.1cm}

\noindent
The last line in eq. \ref{eq:1-1} refers to c.m. momentum and position .

\vspace*{-0.0cm}
\noindent
The bond structures of $\ q \overline{q} \ $ and $\ 3 q \ $ configurations
are shown in figure 1 - I , repeated below

\vspace*{-0.1cm}
\begin{center}
\hspace*{0.0cm}
\begin{figure}[htb]
\vskip -0.0cm
\hskip -0.0cm
\includegraphics[angle=0,width=7.5cm]{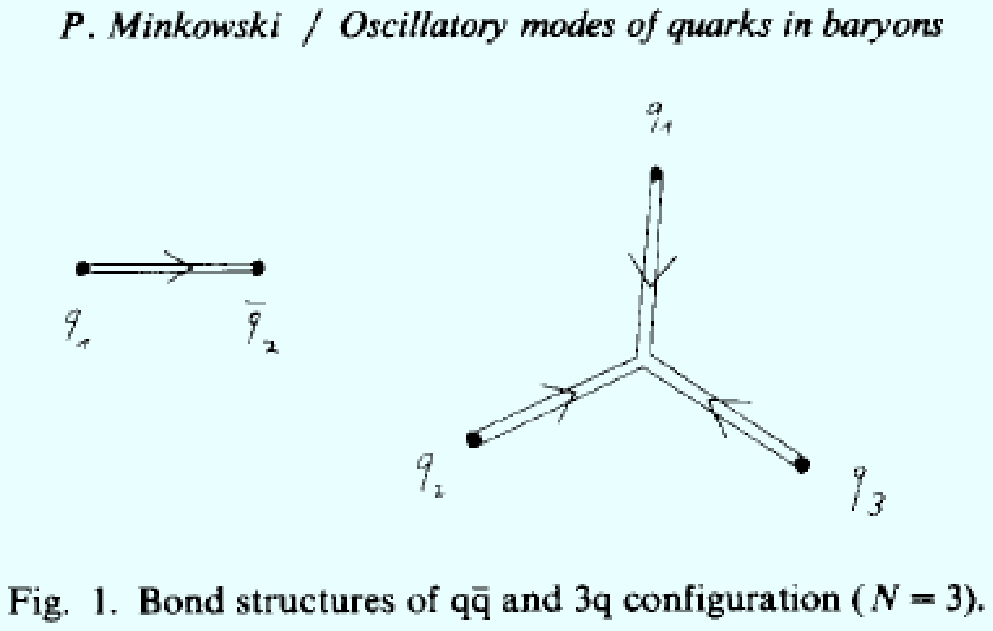}
\vskip -0.0cm
\vspace*{+0.20cm}
{\bf \color{black} \hspace*{0.0cm}
\begin{tabular}{c} Fig. 1 - I : 
Bond structures of $\ q \overline{q} \ $ and $\ 3 q \ $ configurations
\vspace*{0.0cm} \\
\hspace*{0.9cm}
$\left ( \ N = 3 \ \right ) $
{\color{black} $ \longleftrightarrow$}
\end{tabular}
}
\end{figure}
\end{center}
\vspace*{-0.0cm}

\noindent
Of course we are interested in $\ N = 3 \ $ but one goal of this investigation
is to shed light on the N dependence of the ratio of
baryonic to mesonic inverse Regge slopes 

\vspace*{-0.3cm}
\begin{equation}
\label{eq:1-2}
\begin{array}{l}
\Lambda_{N} / \ / \Lambda \ = \ 1 
\hspace*{0.2cm} \mbox{for vanishing quark masses} \hspace*{0.2cm}
M_{\ \alpha} \ \rightarrow \ 0
\hspace*{0.2cm} ; \hspace*{0.2cm}
\alpha \ = \ 1,\cdots,N
\end{array}
\vspace*{-0.4cm}
\end{equation}

\vspace*{-0.3cm}
\begin{equation}
\label{eq:1-3}
\begin{array}{l}
\begin{array}{lll}
\Delta \ {\cal{M}}_{\ baryon}^{\ 2} 
& = & 
2 \ \Lambda_{\ N} \ \sum_{\ \alpha = 1}^{\ 3 N - 3} \ \Delta \ \nu_{\ \alpha}
\vspace*{0.0cm} \\
& & \hspace*{3.5cm} \Delta \ \nu_{\ \alpha} \ = 0 \ , \ \pm 1 \ , \ \pm 2 \ , \cdots
\vspace*{0.0cm} \\
\Delta \ {\cal{M}}_{\ meson}^{\ 2}
& = &
2 \ \Lambda \hspace*{0.5cm} \sum_{\ \alpha = 1}^{\ 3 } 
\ \Delta \ \nu_{\ \alpha}
\end{array}
\end{array}
\end{equation}

\noindent
In eq. \ref{eq:1-2} quark masses are denoted 
$\ M_{\ \alpha} \ ; \ \alpha \ \mbox{: quark flavor} \ $ , as throughout
section 4 , in order to distinguish them from
-- the oscillator state 
configuration space variable dependent masses --  
denoted $ \ m \ , \ \overline{m} \ \cdots \ $.

\noindent
The connection of this ratio to the intermediary range quark-antiquark
potential -- mainly studied for the $\ c \overline{c} \ $ charmonium 
( binding ) spectrum -- determined by the conditions

\vspace*{-0.3cm}
\begin{equation}
\label{eq:1-4}
\begin{array}{l}
Min_{\ \alpha = 1,\cdots,N} \ M_{\ \alpha} 
\ \ll \ \left | \ V_{\ NR} \ ( \ z_{\ \beta} \ ) \ \right |
\ \ll \ Min_{\ \gamma = 1,\cdots,N-1} \ \left | \ z_{\ \gamma} 
\ \right |^{\ -1}
\end{array}
\end{equation}

\noindent
We reformulate as starting point the snowball effect for a 
$\ q \overline{q} \ $ equal mass pair , described ( in the c.m. system )
by the Lagrangean

\vspace*{-0.3cm}
\begin{equation}
\label{eq:1-5}
\begin{array}{l}
\begin{array}{lll}
{\cal{L}}_{\ (2)} 
& = & 
- \ m_{\ 1}  \ \left ( \ 1 \ - \ v_{\ 1}^{\ 2} \ \right )^{\ -1/2}
\ - \ m_{\ 2}  \ \left ( \ 1 \ - \ v_{\ 2}^{\ 2} \ \right )^{\ -1/2}
\vspace*{0.1cm} \\
& = & 
- \ 2 \ m \ \left ( \ 1 \ - \ v^{\ 2} \ \right )^{\ -1/2}
\end{array}
\vspace*{0.1cm} \\
\vec{v}_{\ j} = \dot{\vec{x}}_{\ j} \ ; j \ = \ 1,2
\hspace*{0.1cm} ; \hspace*{0.1cm}
m = \ m \ ( z )
\hspace*{0.1cm} \mbox{for (just here)}  \hspace*{0.1cm}
M_{q_{1}} = M_{q_{2}} = 0
\end{array}
\end{equation}

\noindent
Eq. \ref{eq:1-5} is only valid in the c.m. frame , where the analog
of energy conservation takes the form

\vspace*{-0.3cm}
\begin{equation}
\label{eq:1-6}
\begin{array}{l}
{\cal{H}}_{\ (2)} \ = 
\ \vec{v} \ {\cal{L}}_{\ (2) \ , \ \vec{v}} \ - \ {\cal{L}}_{\ (2)}
\ = \ \begin{array}{c}
2 \ m 
\vspace*{0.1cm} \\ \hline \vspace*{-0.4cm} \\
\sqrt{ \ 1 \ - \ v^{\ 2} \ }
\end{array}
\vspace*{0.0cm} \\
( \vec{p} )_{\ 1} \ - \ ( \vec{p} )_{\ 2} \ = \ 2 \ \vec{p}_{\ c.m.}
\ = \ {\cal{L}}_{\ (2) \ , \ \vec{v}} \ = \ {\cal{H}}_{\ (2)} \ \vec{v} 
\vspace*{0.2cm} \\
\left ( \ {\cal{H}}_{\ (2)} \ \right )^{\ 2} \ v^{\ 2} 
\ = \ \left ( \ {\cal{H}}_{\ (2)} \ \right )^{\ 2} \ - \ 4 \ m^{\ 2}
\ = \ 4 \ p_{\ c.m.}^{\ 2}
\end{array}
\end{equation}

\noindent
Eqs. \ref{eq:1-5} and \ref{eq:1-6} are understood as approximations for large
distances . They can be interpreted classically {\it or}
quantum mechanically .

\noindent
We note that in the discussion ongoing of $\ q \overline{q} \ $ oscillator
modes, we do not use the orthogonal normalization as displayed in eq. 
\ref{eq:1-1} . This is so because other conventions had been used before,
as for myself to the year 1976, while working at Caltech.
The definitions used are shown in the next equation 

\vspace*{-0.3cm}
\begin{equation}
\label{eq:1-7}
\begin{array}{l}
\overline{m} \ = \ 2 \ m 
\hspace*{0.2cm} , \hspace*{0.2cm}
- \ \Delta_{\ z} \ + \ \overline{m}^{\ 2} \ ( \ z \ ) \ = \ H^{\ 2}_{\ (2)}
\hspace*{0.2cm} , \hspace*{0.2cm}
z \ \rightarrow \ \vec{y}
\vspace*{0.1cm} \\
\left ( \ {\cal{H}}_{\ (2)} \ \vec{v} \ \right )^{\ .}
\ = \ {\cal{H}}_{\ (2)} \ \frac{1}{2} \ \stackrel{..}{\vec{y}} 
\ = \ - \ \begin{array}{c} 
1
\vspace*{0.1cm} \\ \hline \vspace*{-0.4cm} \\
{\cal{H}}_{\ (2)}
\end{array}
\hspace*{0.2cm} grad_{\ y} \ \overline{m}^{\ 2}
\hspace*{0.2cm} ; \hspace*{0.2cm}
\vec{y} \ = \ \vec{x}_{\ 1} \ - \ \vec{x}_{\ 2}
\vspace*{0.1cm} \\
\longrightarrow \ {\cal{H}}_{\ (2)}^{\ 2} \ \frac{1}{4} 
\ ( \ \dot{\vec{y}} \ )^{\ 2} \ = \ 4 \ p_{\ c.m.}^{\ 2} \ +
\ \overline{m}^{\ 2} \ = \ {\cal{M}}^{\ 2} \ = \ \mbox{constant}
\end{array}
\end{equation}

\noindent
adopting the 
long range approximate nature of the harmonic oscillator relations
-- for the $\ q \overline{q} \ -$ bond

\vspace*{-0.3cm}
\begin{equation}
\label{eq:1-8}
\begin{array}{l}
\overline{m}^{\ 2} \ \sim_{\  | y | \rightarrow \infty}
\hspace*{0.2cm} \left ( \frac{1}{2} \ \Lambda \ \right )^{\ 2} \ y^{\ 2}
\ \left \lbrack \ 1 \ + \ O \ \left ( \ M_{\ q} \ / \ | \ y \ | \ \right ) 
\ + \ \cdots \ \right \rbrack
\end{array}
\end{equation}

\noindent
In eq. \ref{eq:1-7} we have substituted the variable $\ \vec{y} \ $ for
$\ z \ \rightarrow \ \vec{z} \ $.

\noindent
A few remarks shall follow

\begin{description}

\item 1) The lessons from the $\ q \overline{q} \ -$ bond are limited

The exclusive relative {\it distance}-dependence contained in the 
asymptotic term in eq. \ref{eq:1-8}

\vspace*{-0.3cm}
\begin{equation}
\label{eq:1-9}
\begin{array}{l}
\overline{m}^{\ 2} \ \sim
\ \frac{1}{2} \ \Lambda^{\ 2} \ y^{\ 2}
\end{array}
\end{equation}

generates genuine oscillatory modes for the $\ q \overline{q} \ -$ bond ,
yet no multi-position dependent generalization can accomplish the same
for the $\ 3 \ q \ -$ $\ \left ( \ N \ q \ - \ \right ) \ $ bonds .

\item 2) but not empty

For the $\ q \overline{q} \ -$ bond it follows

\vspace*{-0.3cm}
\begin{equation}
\label{eq:1-10}
\begin{array}{l}
\begin{array}{lll}
\stackrel{..}{\vec{y}} \hspace*{0.2cm} =
\ - \ \left ( \begin{array}{c}
\Lambda
\vspace*{0.1cm} \\ \hline \vspace*{-0.4cm} \\
{\cal{H}}_{\ (2)}
\end{array} \right )
\hspace*{0.2cm} y
& , &
\Lambda \ \left \lbrack \ - \ \Delta_{\ \xi} \ + \ \xi^{\ 2} \ \right \rbrack
\ = \ H_{\ (2)}^{\ 2}
\vspace*{0.1cm} \\
&  &
\xi \ = \ \frac{1}{2} \ \left ( \ \Lambda \ \right )^{\ 1/2} \ y
\vspace*{0.1cm} \\
\longrightarrow 
\ \omega_{\ cl} \ = 
\ \begin{array}{c}
\Lambda
\vspace*{0.1cm} \\ \hline \vspace*{-0.4cm} \\
{\cal{H}}_{\ (2)}
\end{array}
& , &
H_{(2)}^{2} \left ( \left \lbrace \nu \right \rbrace \right )
\ = \ 2 \ \Lambda \ \sum_{\alpha = 1}^{3} \nu_{\alpha} 
\ + \ 3 \ \Lambda
\end{array}
\vspace*{0.2cm} \\
\nu_{\ \beta} \ = \ 0 \ , \ 1 \ , \ \cdots
\hspace*{0.2cm} ; \hspace*{0.2cm}
\beta \ = \ 1,2,3
\hspace*{0.2cm} ; \hspace*{0.2cm}
\mbox{oscillator occupation numbers}
\end{array}
\end{equation}



\item 3) The color quantum number has vanished from the description

Vacuum - vacuum amplitudes of two colored local operators are not gauge 
invariant , provided local gauge invarince is not conserved 'completely' ,
to be defined including appropriate generalized boundary conditions , in QCD .

The wave functions on the other hand are not local .

\item 4) Which are the dependences on quantum numbers like (light-) flavors 
and spin ?

The quantity $\ \Lambda \ $ in eq. \ref{eq:1-9} , of dimension 
$\ \mbox{mass}^{\ 2} \ $, is considered universal ,
i.e. does not depend on any other quantum numbers neither on quark masses,
except on the occupation numbers of the oscillatory modes at hand .

\end{description} 

\begin{center}
\vspace*{-0.0cm}
{\bf \color{black} Extension to include the $\ N \ q \ -$ bond  
 }
\label{'1-1N'}
\end{center}
\vspace*{0.0cm}

\noindent
The key idea arose upon a discussion initiated by H. R. Dicke , concerning the
feasibility and appropriateness to envisage a revision of the errors , as
established by Lor\'{a}nd  ( Roland  v. ) E\"{o}tv\"{o}s in 1918 , in his 
famous experiments 
in ref. \cite{Eotvos}
measuring the equality of inertial and gravitational mass with the help
of rotating springs , to which test bodies are attached ,
while the springs are fixed to one point , say atop a rotating rod .
It should be added that the springs must be elongating under the
centrifugal force only in one longitudinal direction . 

\noindent
Dicke reports in ref. \cite{Eotvos} that E\"{o}tv\"{o}s , 
in his description of the
experiments mentions an important obstacle to overcome , 
consisting in a precise
separation of the mass of the test bodies from a combination of a part of the
spring mass with them .

\noindent
Hence the (my) conclusion from the above situation to the question envisaged 
was , that in presence of position dependent mass this mass and inertial
mass were {\it not the same} .

\noindent
This led to the Ansatz , as I followd in ref. \cite{oscmod1980} 

\vspace*{-0.3cm}
\begin{equation}
\label{eq:1-11}
\begin{array}{l}
{\cal{L}}_{\ N} \ = \ - \ \sum_{\ \alpha = 1}^{\ N} 
\ \left \lbrack \ m_{\ \alpha}^{\ 2} \ - \ Q^{\ \alpha}_{\ \beta \gamma}
\ \vec{v}_{\ \beta} \ . \ \vec{v}_{\ \gamma} \ \right \rbrack^{\ 1/2}
\vspace*{0.1cm} \\
\vec{v}_{\ \alpha} \ = \ \dot{\vec{x}}_{\ \alpha} 
\vspace*{0.1cm} \\
\begin{array}{rlrlll}
m_{\ \alpha} 
& = &
m_{\ \alpha} 
& \left \lbrack 
\ \vec{z}_{\ 1} , \cdots , \vec{z}_{\ N - 1}
\ \right \rbrack
& , &
\mbox{gravitational effective masses}
\vspace*{0.1cm} \\
Q^{\ \alpha}_{\ \beta \gamma} 
& = & 
Q^{\ \alpha}_{\ \beta \gamma} 
& \left \lbrack
\ \vec{z}_{\ 1} , \cdots , \vec{z}_{\ N - 1}
\ \right \rbrack
& , &
\mbox{inertial effective masses}
\end{array}
\end{array}
\end{equation}

\noindent
valid in the c.m. system of the N quarks .

\noindent
The external quark masses $\ M_{\ q} \ $, appropriate multipliers of the scalar
densities $\ \overline{q}  q \ $ composing the mass term in the local (QCD)
Lagrangian

\vspace*{-0.3cm}
\begin{equation}
\label{eq:1-12}
\begin{array}{l}
- \ {\cal{L}}_{\ m} \ = \ \sum_{\ flavors} \ \begin{array}{c}
z_{\ q}
\vspace*{0.1cm} \\ \hline \vspace*{-0.4cm} \\
z_{\ M} 
\end{array}
\hspace*{0.2cm}
M_{\ q} \ \overline{q}^{\ c} \ q^{\ c}
\end{array}
\vspace*{-0.3cm}
\end{equation}

\noindent
appear as constants in the gravitational mass

\vspace*{-0.3cm}
\begin{equation}
\label{eq:1-13}
\begin{array}{l}
m_{\ \alpha} 
\ \left \lbrack
\ M_{\ q} \ , \ \underline{z}
\ \right \rbrack
\ =
\ M_{\ \alpha} \ +
\ m_{\ \alpha} 
\ \left \lbrack
\ M_{\ q} \ = \ 0 \ , \ \underline{z}
\ \right \rbrack
\vspace*{0.1cm} \\
\underline{z} \ = \ \left ( \ \vec{z}_{\ 1} \ , \ \cdots
\ , \ \vec{z}_{\ N = 1} \ \right )
\end{array}
\end{equation}

\noindent
whereas a consistent non-relativistic limit demands

\vspace*{-0.3cm}
\begin{equation}
\label{eq:1-14}
\begin{array}{l}
\begin{array}{rlr}
\left \lbrack
\ Q_{\ \alpha \alpha}^{\ \alpha} 
\ \left (
\ M_{\ q} \ , \ \underline{z}
\ \right )
\ \right \rbrack^{\ 1/2} 
& \stackrel{\longrightarrow}{_{M_{\ q} \rightarrow \infty}} &
M_{\ \alpha} \ + 
\ O \ \left \lbrack
\ m_{\ \beta} \ \left \lbrack
\ M_{\ q} \ = \ 0 \ , \ \underline{z}
\ \right \rbrack
\ \right \rbrack
\vspace*{0.1cm} \\
Q_{\ \beta \gamma \ , \ \beta,\gamma \neq \alpha}^{\ \alpha} 
& \stackrel{\longrightarrow}{_{M_{\ q} \rightarrow \infty}} &
O \ \left \lbrack
\ m_{\ \beta} \ \left \lbrack
\ M_{\ q} \ = \ 0 \ , \ \underline{z}
\ \right \rbrack
\ \right \rbrack
\end{array}
\end{array}
\end{equation}

\noindent
From eqs. \ref{eq:1-13} and \ref{eq:1-14}
we recognize the problem of separation of mass and binding
energy, as relevant, e.g., to the gravitational interaction of the whole
N-quark system, appearing. 

\noindent
In the following m- , Q-  are approximated by the corresponding quantities
for $\ M_{\ q} \ = \ 0 \ $, i.e., in the chiral limit with respect to all the
quark flavors composing the $\ N \ q \ -$ bond .

\noindent
Then in the harmonic long-range limit Q- is determined from m- 
through the relation

\vspace*{-0.3cm}
\begin{equation}
\label{eq:1-15}
\begin{array}{l}
Q_{\ \beta \gamma}^{\ \alpha} \ =
\ \begin{array}{c}
1
\vspace*{0.1cm} \\ \hline \vspace*{-0.4cm} \\
K_{\ N}
\end{array}
\hspace*{0.2cm} \left ( \ m_{\ \alpha} \ \right )^{\ 2} \ \delta_{\ \beta
\gamma}
\hspace*{0.2cm} ; \hspace*{0.2cm}
K_{\ N} \ : \ \mbox{constant}
\end{array}
\end{equation}

\noindent
The kinetic term for the quark -- $\alpha$  -- depends on all the velocities 
$\ \vec{v}_{\ \beta} \ $ and the Lagrangean $\ {\cal{L}}_{\ N} \ $
in eq. \ref{eq:1-11} takes the simplified form

\vspace*{-0.3cm}
\begin{equation}
\label{eq:1-16}
\begin{array}{l}
{\cal{L}}_{N} = - \overline{m} 
\ \left \lbrack
1 - \sum_{\beta} \left ( \ \vec{v}_{\ \beta} \ \right )^{2}
\right \rbrack^{1/2}
\hspace*{0.1cm} ; \hspace*{0.1cm}
\overline{m} = \sum_{\alpha = 1}^{N} m_{\alpha} 
= \overline{m} \left ( \ \vec{x}_{\beta} - \ \vec{X} \ \right )
\vspace*{0.1cm} \\
\vec{X} \ = 
\ \begin{array}{c}
1
\vspace*{0.1cm} \\ \hline \vspace*{-0.4cm} \\
N
\end{array}
\hspace*{0.2cm} \sum_{\ \alpha = 1}^{\ N} \ \vec{x}_{\ \alpha}
\ \rightarrow \ 0 
\end{array}
\end{equation}

\noindent
The meaning of the constant $\ K_{\ N} \ $ is the following :
under the constraint $\ \sum_{\ \alpha} \ \vec{v}_{\ \alpha} \ = \ 0 \ $
the maximum any individual $\ \left ( \ \vec{v}_{\ \beta} \ \right )^{\ 2} \ $ 
can assume for given 
$\ \sum_{\ \gamma} \  \left ( \ \vec{v}_{\ \gamma} \ \right )^{\ 2} \ $
is for the so-called $\ \lambda \ -$ mode, shown in figure 6 below .

\noindent
An inequality for any individual square velocity follows

\vspace*{-0.3cm}
\begin{equation}
\label{eq:1-17}
\begin{array}{l}
v_{\ \alpha} \ = 
\ \left ( \ \left ( \ \vec{v}_{\ \alpha} \ \right )^{\ 2}
\ \right )^{\ 1/2}
\hspace*{0.2cm} \longrightarrow
\vspace*{0.1cm} \\
v_{\ \alpha}^{\ 2} \ \le 
\ \begin{array}{c}
N \ - \ 1
\vspace*{0.1cm} \\ \hline \vspace*{-0.4cm} \\
N
\end{array}
\hspace*{0.2cm} \sum_{\ \gamma} \ v_{\ \gamma}^{\ 2}
\ \le \ \begin{array}{c}
N \ - \ 1
\vspace*{0.1cm} \\ \hline \vspace*{-0.4cm} \\
N
\end{array}
\hspace*{0.2cm} \ K_{\ N} c^{\ 2}
\hspace*{0.3cm} \left ( \ c \ = \ 1 \ \right )
\end{array}
\end{equation}

\vspace*{-0.1cm}
\begin{center}
\hspace*{0.0cm}
\begin{figure}[htb]
 \label{fig6}
\vskip -0.0cm
\hskip -0.0cm
\includegraphics[angle=0,width=9.0cm]{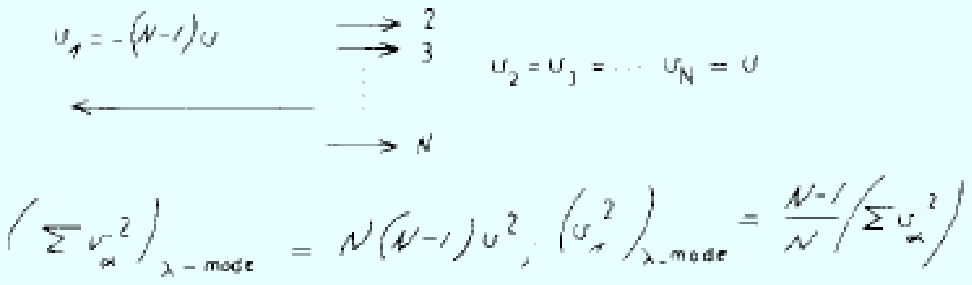}
\vskip -0.0cm
\vspace*{+0.20cm}
{\bf \color{black} \hspace*{0.0cm}
\begin{tabular}{c} Fig. 6 : 
$\ \lambda \ - \ $ mode for N quark bond 
\hspace*{0.2cm} {\color{black} $ \longleftrightarrow$}
\end{tabular}
}
\end{figure}
\end{center}

\vspace*{-0.5cm}
\noindent
${\cal{L}}_{\ N} \ $ in eq. \ref {eq:1-16}
delimits its validity to physical values of $\ v_{\ \alpha}^{\ 2} \ $
for i
\vspace*{0.0cm} \\
$K_{\ N} \ = \ N \ / \ ( \ N \ - 1 \ ) \ $

\vspace*{-0.3cm}
\begin{equation}
\label{eq:1-18}
\begin{array}{l}
v_{\ \alpha}^{\ 2} \ \le \ c^{\ 2}
\hspace*{0.2cm} \longleftrightarrow \hspace*{0.2cm}
K_{\ N} \ = \ \begin{array}{c}
N
\vspace*{0.1cm} \\ \hline \vspace*{-0.4cm} \\
N \ - 1
\end{array}
\end{array}
\end{equation}

\vspace*{-0.3cm}
\noindent
The equation for the conserved energy 
( eq. \ref{eq:1-6} for $\ {\cal{L}}_{\ (2)} \ $ )
for $\ {\cal{L}}_{\ N} \ $ becomes

\vspace*{-0.8cm}
\begin{equation}
\label{eq:1-19}
\begin{array}{l}
\vspace*{0.1cm} \\
{\cal{H}}_{\ N} \ =
\ \vec{v}_{\ \alpha} \ \vec{p}_{\ \alpha} \ - \ {\cal{L}}_{\ N}
\vspace*{0.1cm} \\
\vec{p}_{\ \beta} \ = \ {\cal{L}}_{\ N \ , \ \vec{v}_{\ \beta}}
\ =
\ \begin{array}{c}
\overline{m}
\vspace*{0.1cm} \\ \hline \vspace*{-0.4cm} \\
K_{\ N}
\end{array}
\hspace*{0.2cm} 
\left ( \ \left \lbrack \ 1 \ - \ \omega^{\ 2}
\ \right \rbrack^{\ - 1/2} \ \right )_{\ , \ \vec{v}_{\ \beta}}
\ = 
\ \begin{array}{c}
{\cal{H}}_{\ N}
\vspace*{0.1cm} \\ \hline \vspace*{-0.4cm} \\
K_{\ N}
\end{array}
\hspace*{0.2cm} \vec{v}_{\ \beta}
\vspace*{0.1cm} \\
\omega^{\ 2} \ =
\ \begin{array}{c}
1
\vspace*{0.1cm} \\ \hline \vspace*{-0.4cm} \\
K_{\ N}
\end{array}
\hspace*{0.2cm} \sum_{\ \gamma = 1}^{\ N}
\ v_{\ \gamma}^{\ 2}
\hspace*{0.2cm} ; \hspace*{0.2cm}
K_{\ N} \ = \ \begin{array}{c}
N
\vspace*{0.1cm} \\ \hline \vspace*{-0.4cm} \\
N \ - \ 1
\end{array}
\end{array}
\end{equation}

\noindent
From eq. \ref{eq:1-19} we obtain in canonically conjugate oscillator variables

\vspace*{-0.5cm}
\begin{equation}
\label{eq:1-20}
\begin{array}{l}
\begin{array}{@{\hspace*{0.0cm}}l@{\hspace*{0.0cm}}l@{\hspace*{0.0cm}}
l@{\hspace*{0.0cm}}}
\left ( {\cal{H}}_{N} \right )^{2} 
& = &
\begin{array}[t]{l}
\left \lbrack
\begin{array}{l}
K_{N} \left . \sum_{\alpha = 1}^{N} 
\left ( \ \vec{p}_{\alpha} \ \right )^{2}
\right |_{\ \sum_{\beta = 1}^{N} \ \vec{p}_{\beta} = 0}
+ \overline{m}^{2} \left ( x_{\ \gamma} - X \right )
\end{array}
\right \rbrack
\end{array}
\vspace*{0.4cm} \\
& = &
\begin{array}[t]{l}
\left \lbrack
\begin{array}{l}
K_{N} \left . \sum_{\alpha = 1}^{N}
\ \left ( \ \vec{p}_{\alpha} \ \right )^{2}
\right |_{\sum_{\beta = 1}^{N} \ \vec{p}_{\beta} = 0}
+
\vspace*{0.1cm} \\
{\color{black} \hspace*{0.2cm} + \hspace*{0.1cm} \begin{array}{c}
\Lambda^{2}
\vspace*{0.1cm} \\ \hline \vspace*{-0.4cm} \\
K_{N} 
\end{array}
\left . \sum_{\alpha = 1}^{N} 
\left ( \ \vec{x}_{\alpha} \ \right )^{2}
\right |_{\sum_{\beta = 1}^{N} \ \vec{x}_{\beta} = 0}
}
\end{array}
\right \rbrack
\end{array}
\end{array}
\end{array}
\vspace*{-0.0cm}
\end{equation}

\noindent
$\ {\cal{H}}_{\ N} \ $ is a constant of the motion
by the relations displayed in eq. \ref{eq:1-19} , but it is
$\ \left ( \ {\cal{H}}_{\ N} \ \right )^{\ 2} \ \equiv
\ {\cal{M}}_{\ N}^{\ 2} \ $, which becomes the genuinely canonical dynamic 
operator,
or in the classical framework 'Hamiltonian function' ,  
in the genuinely relativistic situation .

\noindent
The structure of $\ {\cal{M}}_{\ N}^{\ 2} \ $ is derived straightforwardly 
from eqs. \ref{eq:1-19} and \ref{eq:1-20}

\vspace*{-0.3cm}
\begin{equation}
\label{eq:1-21}
\begin{array}{l}
\begin{array}{@{\hspace*{0.0cm}}l@{\hspace*{0.0cm}}l@{\hspace*{0.0cm}}
l@{\hspace*{0.0cm}}}
{\cal{M}}_{N}^{2} 
& = &
\begin{array}[t]{l}
\left \lbrack
\begin{array}{l}
K_{N} \ \left . \sum_{\alpha = 1}^{N}
\left ( \vec{p}_{\alpha} \right )^{2}
\right |_{\sum_{\beta = 1}^{N} \ \vec{p}_{\beta} = 0}
+
\vspace*{0.1cm} \\
{\color{black} \hspace*{0.3cm} + \hspace*{0.0cm} \begin{array}{c}
\Lambda^{2}
\vspace*{0.1cm} \\ \hline \vspace*{-0.4cm} \\
K_{N} 
\end{array}
\left . \sum_{\alpha = 1}^{N} 
\left ( \vec{x}_{\alpha} \right )^{2}
\right |_{\sum_{\beta = 1}^{N} \vec{x}_{\beta} = 0}
}
\end{array}
\right \rbrack
\end{array}
\vspace*{0.4cm} \\
& = &
\begin{array}[t]{l}
\left \lbrack \begin{array}{l}
K_{N} \sum_{\alpha = 1}^{N - 1}
\left ( \ \vec{\pi}_{\alpha} \ \right )^{2}
\hspace*{0.1cm}  + \hspace*{0.1cm}  \begin{array}{c}
\Lambda^{2}
\vspace*{0.1cm} \\ \hline \vspace*{-0.4cm} \\
K_{N}
\end{array}
\hspace*{0.1cm} \sum_{\alpha = 1}^{N - 1}
\left ( \ \vec{z}_{\alpha} \ \right )^{2}
\end{array}
\right \rbrack
\end{array}
\end{array}
\vspace*{0.2cm} \\
\left ( \ \vec{\pi}_{\beta} \ , \ \vec{z}_{\beta} \ \right )
: \mbox{barycentric coordinates defined in eq. \ref{eq:1-1}}
\hspace*{0.1cm} ; \hspace*{0.1cm}
K_{N} = \begin{array}{c}
N
\vspace*{0.1cm} \\ \hline \vspace*{-0.4cm} \\
N - 1
\end{array}
\end{array}
\end{equation}

\noindent
The mass-square spectrum according to $\ {\cal{M}}_{\ N}^{\ 2} \ $
in eq. \ref{eq:1-21} is given by

\vspace*{-0.3cm}
\begin{equation}
\label{eq:1-22}
\begin{array}{l}
\left . {\cal{M}}_{\ N}^{\ 2} \ \right |_{\ spectrum} \ =
\ 2 \ \Lambda \ \sum_{\ \alpha = 1}^{\ 3 N - 3} \ \nu_{\ \alpha}
\ + \ 3 \ \Lambda \ \left ( \ 3 \ N \ - \ 3 \ \right )
\vspace*{0.2cm} \\
\nu_{\ \beta} \ = \ 0 \ , \ 1 \ , \ \cdots
\hspace*{0.2cm} ; \hspace*{0.2cm}
\beta \ = \ 1 \ , \ 2 \ , \ \cdots \ , \ 3 \ N \ - \ 3
\end{array}
\end{equation}

\noindent
Except for the zero point contribution , i. e. for the oscillation level
splittings, it is universal, independent of N, proving the validity
of the universal relation in eq. \ref{eq:1-2} 
$\ \left ( \ \Lambda_{\ N} \ / \ \Lambda \ = \ 1 \ \right ) \ $.

\noindent
Taking the $\ \nu \ = \ \sum_{\ \eta = 1}^{\ 6} \ \nu_{\ \eta} = 2 \ , 
\ P \ = \ + $
nonstrange baryon states , i. e. for $\ N_{\ fl}  \ = \ 2 \ , \ N \ = 3 \ $
and counting in a reduced way, all
corresponding oscillatory modes with positive parity , compatible with overall 
Bose symmetry ,
neglecting color , we obtain the following table of 21 states , not counting
isospin and spin degrees of freedom separately

\vspace*{-0.1cm}
\begin{center}
\hspace*{0.0cm}
\begin{figure}[htb]
 \label{fig7}
\vskip -0.0cm
\hskip -0.0cm
\includegraphics[angle=0,width=8.0cm]{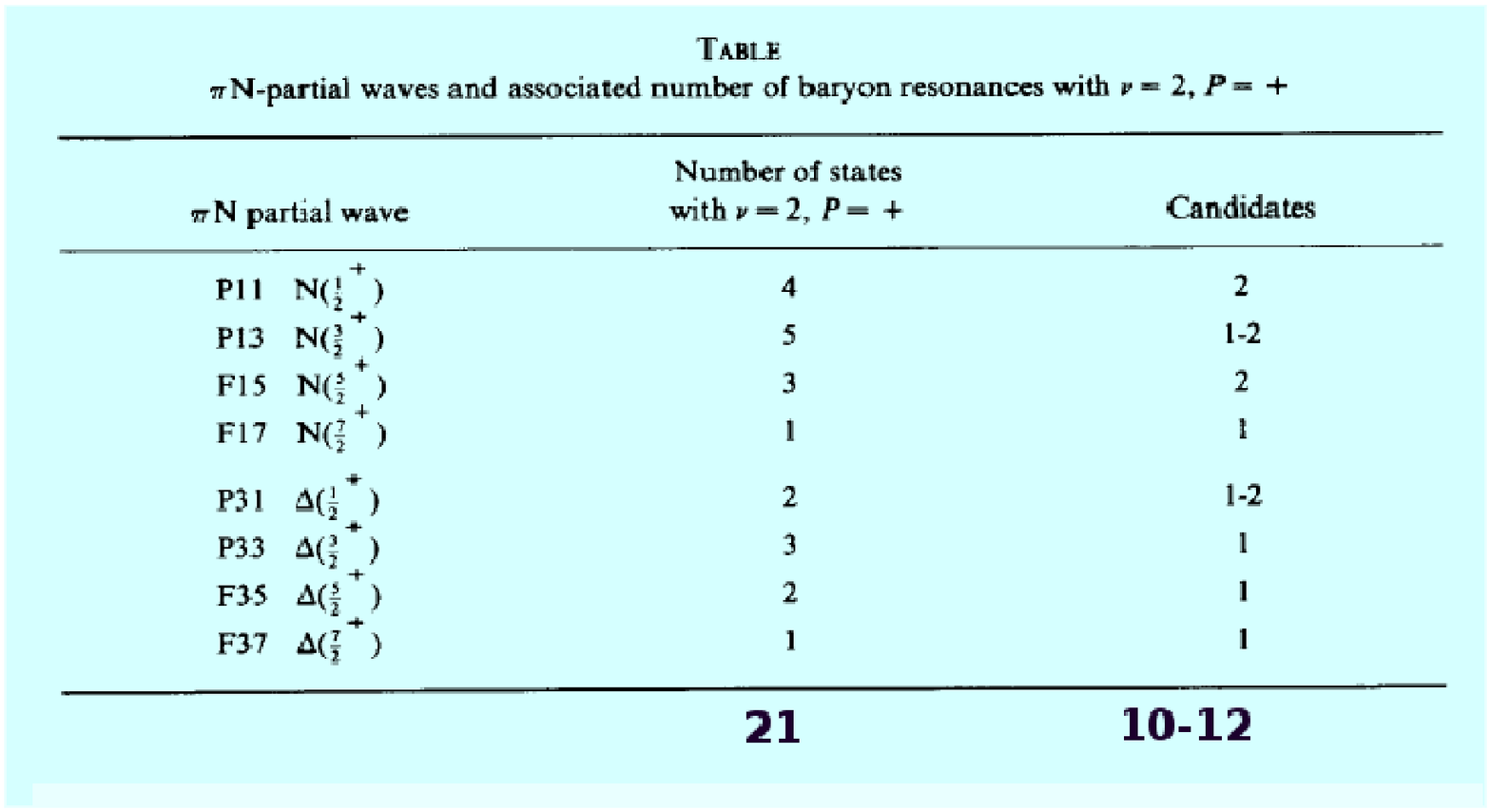}
\vskip -0.0cm
\vspace*{+0.20cm}
{\bf \color{black} \hspace*{0.0cm}
\begin{tabular}{c} Fig. 7 : 
Nonstrange baryons with $\nu \ = \ 2 \ , P = + $
in 1980
\hspace*{0.1cm} {\color{black} $ \longleftrightarrow$}
\end{tabular}
}
\end{figure}
\vspace*{-0.2cm}
\end{center}

\noindent
The candidates were collected in ref. \cite{oscmod1980}
from the PDG tables valid in 1980 . At least then almost 
50 \% of the resonances so characterized were missing , using this way of 
counting .

\begin{center}
\vspace*{-0.3cm}
{\bf \color{black} 5 --  First results from counting 
oscillatory modes in u,d,s flavored baryons 
( \cite{Countoscimodes-fl6-A4a-SKPM} )
} 
\end{center}
\label{'5'}
\vspace*{0.1cm}

\vspace*{-0.2cm}
Perspectives : A. How to count oscillatory modes of quarks in baryons 
\vspace*{-0.1cm} \\
\hspace*{3.4cm}
for 3 quark flavors u , d , s ; $\ t \ \geq \ 0$ . 
\vspace*{0.2cm} 

\noindent
{\color{black} In the 2 figure captions below refs. 
$\ 1 , 2  \ \rightarrow 29 , 30 \ $.}

\vspace*{-0.3cm}
\begin{center}
\hspace*{0.0cm}
\begin{figure}[htb]
\vskip -1.0cm
\hskip 1.7cm
\includegraphics[angle=0,width=8.0cm]{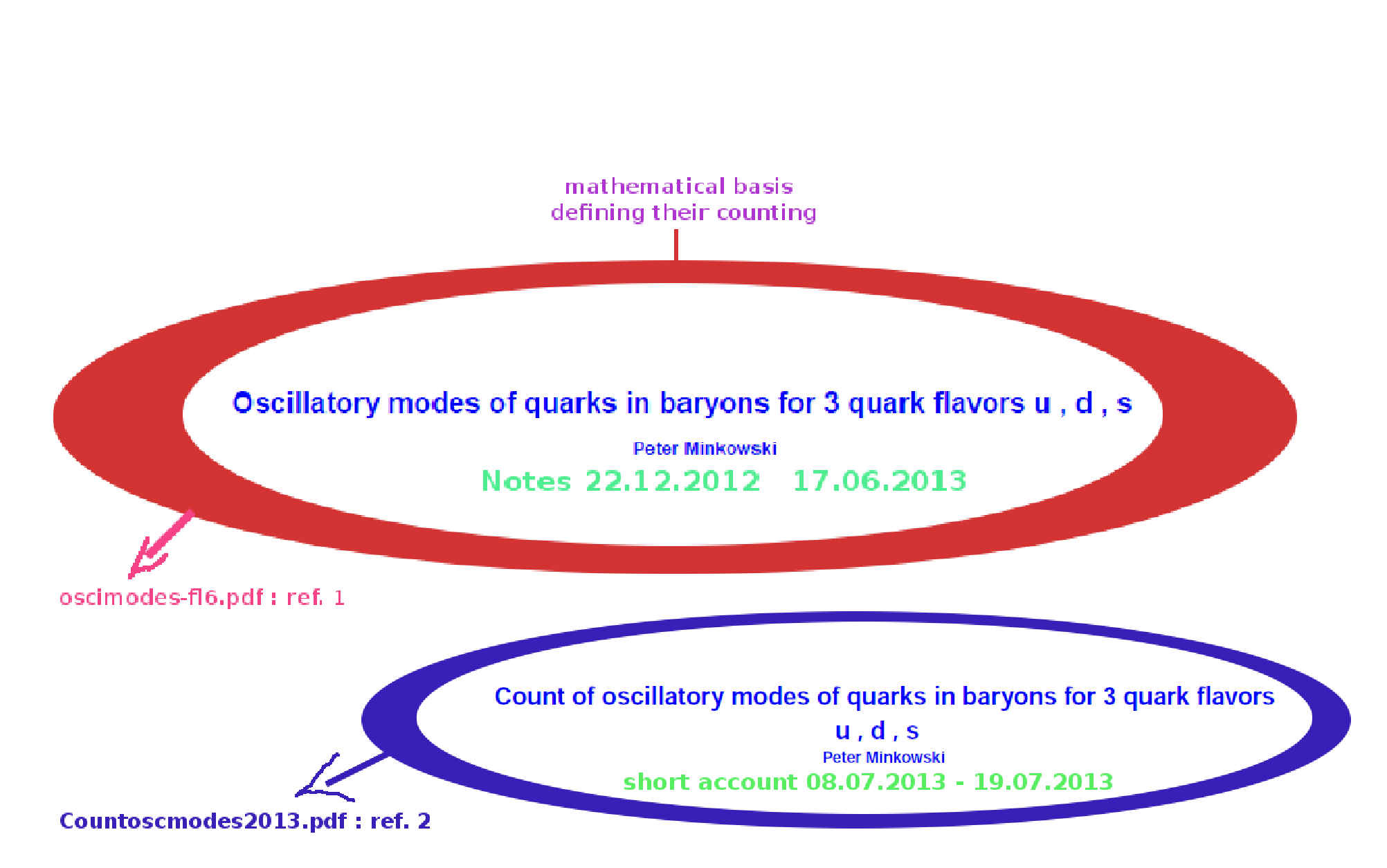}
\vskip -0.2cm
\hskip -2.0cm
\label{fig1}
\vspace*{-0.20cm}
{\bf \color{black} \hspace*{0.0cm}
\begin{tabular}{c} ref. 29 : \cite{oscimodes-fl6-PM} ; 
ref. 30 : \cite{Countoscmodes2013-PM} 
\end{tabular}
}
\end{figure}
\vspace*{-0.15cm}
\end{center}

\vspace*{-0.2cm}
\hspace*{2.3cm} 
{\color{black} B. Tuning to harmonic numbers of oscimodes of baryons}

\vspace*{-0.0cm}
\begin{center}
\hspace*{0.0cm}
\begin{figure}[htb]
\vskip -0.0cm
\hskip 8.7cm
\includegraphics[angle=0,width=1.0cm]{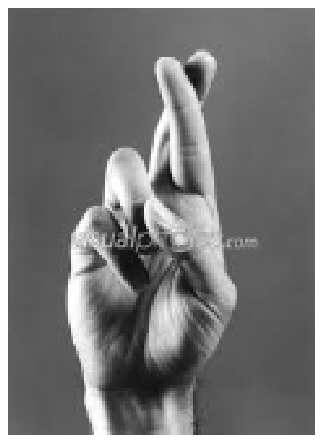}
\label{fig2}
\end{figure}
\vspace*{-0.0cm}
\end{center}

\begin{center}
\vspace*{0.2cm}
{\bf \color{black} 1 - Introduction'
 }
\label{''1''}
\end{center}
\vspace*{0.0cm}

\noindent
The perspectives illustrated in Figs. 1 and 2 are meant to apply to present and
future derivations . In this sense refs. 29 and 30 -- \cite{oscimodes-fl6-PM} 
and \cite{Countoscmodes2013-PM} -- refer to recent results .
The comparison of hadron yields measured at RHIC and LHC with a 
noninteracting hadron resonance gas necessitates the countingt of these
resonances , which is not obvious. This is illustrated in Fig. 1 of ref. 3
\cite{SKPM2010} , reproduced as 
Fig. 3 below . Thus the problem of identifying 'oscillatory modes of light
quark flavors in baryons' , presented in ref. 15 -- \cite{oscmod1980} --
took center-stage .
Having read around the year 1976 a paper on relativistic
oscillator solutions to the Dirac equation , only very recently I could
reconstruct its quotation, which becomes ref. 32 -- \cite{CCritchfield1975}.
\footnote {\color{black} \begin{tabular}[t]{l}
I am indebted to Christoph Greub for reminding me of the 
first article
\vspace*{-0.1cm} \\
by C. L. Critchfield -- \cite{CCritchfield1975} -- on scalar potentials .
\end{tabular}
}

\noindent
Here I focus on the cornerstones , which allow to count these
oscillatory modes , as outlined in
extenso in refs. 29 and 30 , op.cit. , beginning with the classification of
the representations of $\ S_{\ 3} \ $ -- the permutation group of the three
quarks in configuration space -- as they arise through the induced
representation from the associated wave functions in the subsequent sections .
\vspace*{-0.1cm}

\vspace*{-0.1cm}
\begin{center}
\hspace*{0.0cm}
\begin{figure}[htb]
\vskip -0.8cm
\hskip 0.7cm
\includegraphics[angle=0,width=10.0cm]{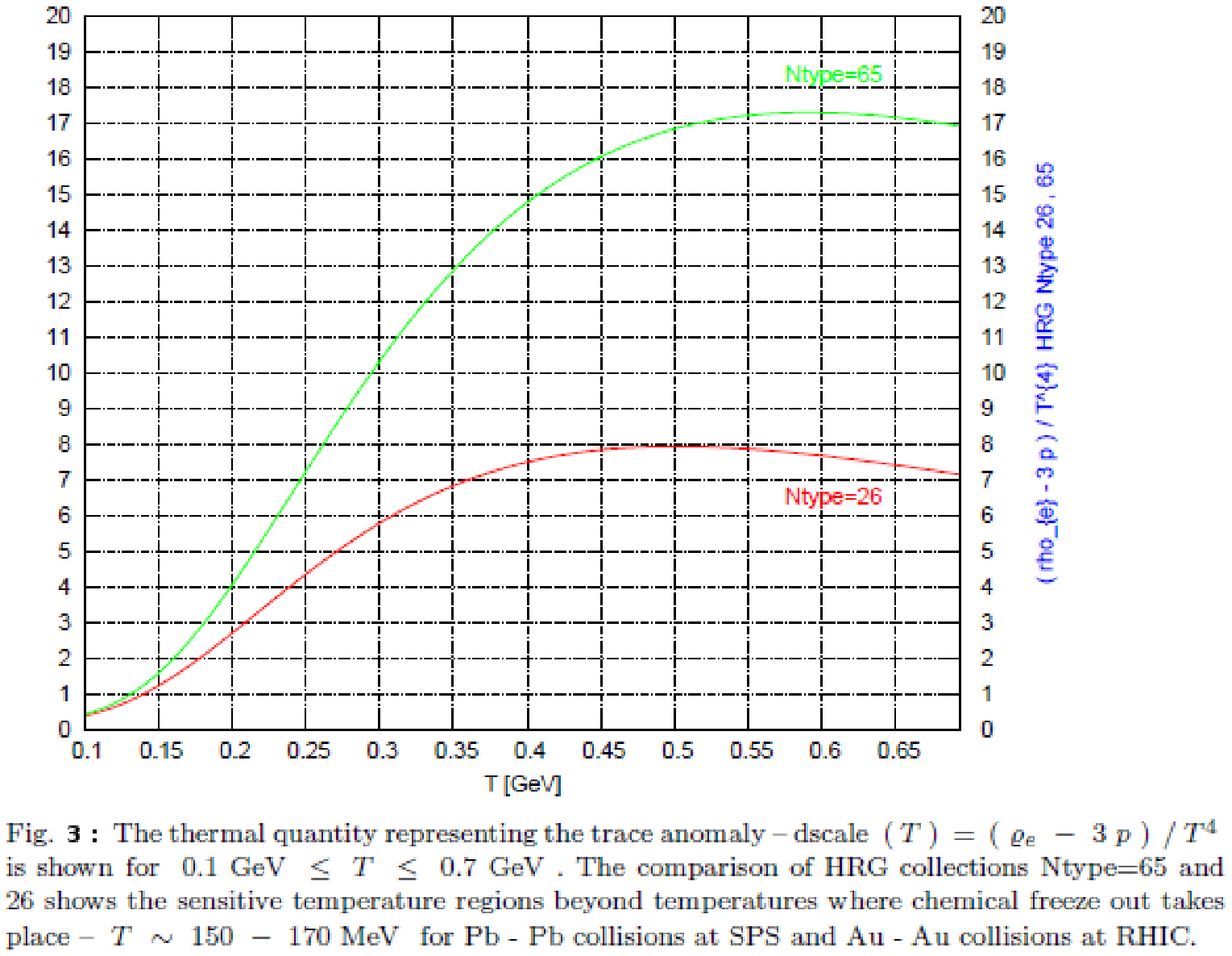}
\label{fig3}
\end{figure}
\vspace*{-0.0cm}
\end{center}

\begin{center}
\vspace*{-0.1cm}
{\bf \color{black} 2 - Factoring out the {\it approximate} symmetry group
in spin-flavor space as well as overall color
 }
\label{''2''}
\end{center}
\vspace*{0.0cm}

\noindent
It is inherent to the path followed to point out the historical formulation of
the 'bootstrap hypothesis' underlying and restricting the full set of 
S-matrix elements pertaining to strongly interacting hadrons , due to G. F.
Chew e.g. in ref. 33 -- \cite{GFChew1962} . A good textbook reference is 
ref. 34 -- \cite{CollinsSquires1968} .

\noindent
In this context all resonances observed or hypothetically to be observed in 
scattering of stable hadrons , irrespective of their width , are considered to 
be hadrons . A central notion within the 'bootstrap'-framework is the density
of hadrons and its limiting behaviour for large  mass-square

\vspace*{-0.2cm}
\begin{equation}
\label{eq:2-1-1}
\begin{array}{l}   
\varrho_{\ n} \ \left ( \ m^{\ 2} \ \right )
\ = \ \ \begin{array}{c} 
\partial \ \mathbb{N} \ \left ( \ m^{\ 2} \ \right ) 
\vspace*{0.2cm} \\ \hline \vspace*{-0.3cm} \\
\partial \ m^{\ 2}
\end{array}
\end{array}
\vspace*{-0.3cm}
\end{equation}
\footnote{\color{black} \begin{tabular}[t]{l}
We use the symbol $\ \mathbb{N} \ $ 
for numbering integers except 
the number of 
\vspace*{-0.1cm} \\
valence quarks forming a baryon ,  denoted $\ N \ $.
\end{tabular}
}

\noindent
In the density with respect to mass square $\ \varrho_{\ n} \ $ defined in eq. 
\ref{eq:2-1-1} the density per phase space of an isolated state of momentum $\
\vec{p} \ $ is not included

\vspace*{-0.4cm}
\begin{equation}
\label{eq:2-1-2}
\begin{array}{l}   
d \ \Phi \ = \ V \ \left ( \ 2 \pi \ \right )^{\ -3} \ d^{\ 3} \ p
\hspace*{0.2cm} ; \hspace*{0.2cm}
V \ : \ \mbox{space volume}
\end{array}
\vspace*{-0.1cm}
\end{equation}

\noindent
The parametrization defined in eq. \ref{eq:2-1-1} is directly applicable to
counting resonances in the particle listings of the PDG
\cite{PDG2012} . To this end a binning in mass square is to be chosen and 
a histogram of resonance counts per bin yields the so approximated density
function $\ \varrho_{\ n} \ $.

\noindent
In 1965 Rolf Hagedorn ( 1919 - 2003 ) wrote an elaborate paper --
\cite{RHagedorn1965} -- centered
around the hypothesis of the limiting behaviour of the quantity 
$\ \varrho_{\ n} \ $ for $\ m \ \rightarrow \ \infty \ $
as a solution to the bootstrap conditions

\vspace*{-0.3cm}
\begin{equation}
\label{eq:2-1-3}
\begin{array}{l}   
\varrho_{\ n} \ \left ( \ m^{\ 2} \ \right ) \ \sim
\ \left ( \begin{array}{c} 
m^{\ 2} 
\vspace*{0.2cm} \\ \hline \vspace*{-0.3cm} \\
m_{\ 0}^{\ 2}
\end{array} \ \right )^{\ a}
\ \exp \ \left ( \ m \ / \ T_{\ 0} \ \right )
\hspace*{0.2cm} \mbox{for} \hspace*{0.2cm} 
m \ \rightarrow \ \infty
\vspace*{0.1cm} \\
m_{\ 0} \ , \ a \ , \ T_{\ 0} \ : \ \mbox{characteristic parameters}
\end{array}
\end{equation}

\noindent
In 1968 Gabriele Veneziano -- \cite{GVeneziano1968} -- arrived at a solution 
to the 'bootstrap' idea starting from the decay amplitude for the process

\vspace*{-0.3cm}
\begin{equation}
\label{eq:2-1-4}
\begin{array}{l}   
\omega \ \rightarrow \ \pi^{\ +} \ \pi^{\ -} \ \pi^{\ 0}
\end{array}
\end{equation}

\noindent
reproducing Regge poles in all two particle channels 
upon suitable extrapolations in their respective momenta , with generalizations
to multiparticle ampltudes called 'dual' .
The structure underlying the totality of dual amplitudes is the quantum
mechanical motion of a one dimensional open (super)string with constant string 
tension 
$\ T \ = \ 1 \ / \ ( \ 2 \pi \ \alpha^{\ '} \ ) \ $,
whose harmonic vibrations generate linear 
$\ \left ( \ J \hspace*{0.2cm}  \mbox{versus} \hspace*{0.2cm} 
\alpha^{\ '} \ m^{\ 2} \ \right ) \ $
bosonic and fermionic
Regge trajectories . A review may be found in ref. 37 , \cite{JSchwarz1982}.  

\noindent
In the domain of numbers the counting of {\it partitions} , i.e. the power of
the set of nonnegative integers 
$\ n_{\ 1} , n_{\ 2} , \ \cdots \ , \ n_{\ \infty} \ \mbox{with}
\ n_{\ k} \ = \ 0,1,\cdots,\infty \ \forall \ k \ $

\vspace*{-0.3cm}
\begin{equation}
\label{eq:2-1-5}
\begin{array}{l}   
\wp \ \left ( \ N \ \right ) \ =
\ \left \lbrace \ \left . n_{\ 1} \ , \ n_{\ 2} \ \cdots \ n_{\ \infty}
\ \right | \ \sum_{\ k = 1}^{\ \infty} \ k \ n_{\ k} \ = \ N \ \right \rbrace
\vspace*{0.1cm} \\
n_{\ 1} \ , \ n_{\ 2} \ \cdots \ n_{\ \infty} \ = \ 0 , 1 , \ \cdots
\end{array}
\end{equation}

\noindent
determines by its asymptotic behaviour for $\ N \ \rightarrow \ \infty \ $
that indeed the free superstring possesses a similar growth as in 
eq. \ref{eq:2-1-3} 
-- modulo multiplicative logarithmic factors in the exponent --
such that a maximal temperature exists in accordance with Hagedorn's 
hypothesis . This is worked out in ref. 38 -- \cite{EAlvarez1986} .
In a specific case of open superstrings on noncommutative space eq.
\ref{eq:2-1-3} is reproduced with $\ a \ = \ - \ \frac{9}{4} \ $ in ref. 39
-- \cite{EWitten2000} .
$\ \wp \ \left ( \ N \ \right ) \ $ in eq. \ref{eq:2-1-5} 
is treated in ref. 40 -- \cite{AbraStegun1970} , pp. 822ff.  

\vspace*{-0.0cm}
\begin{center}
\hspace*{0.0cm}
\begin{figure}[htb]
\vskip -0.6cm
\hskip 2.7cm
\includegraphics[angle=0,width=7.7cm]{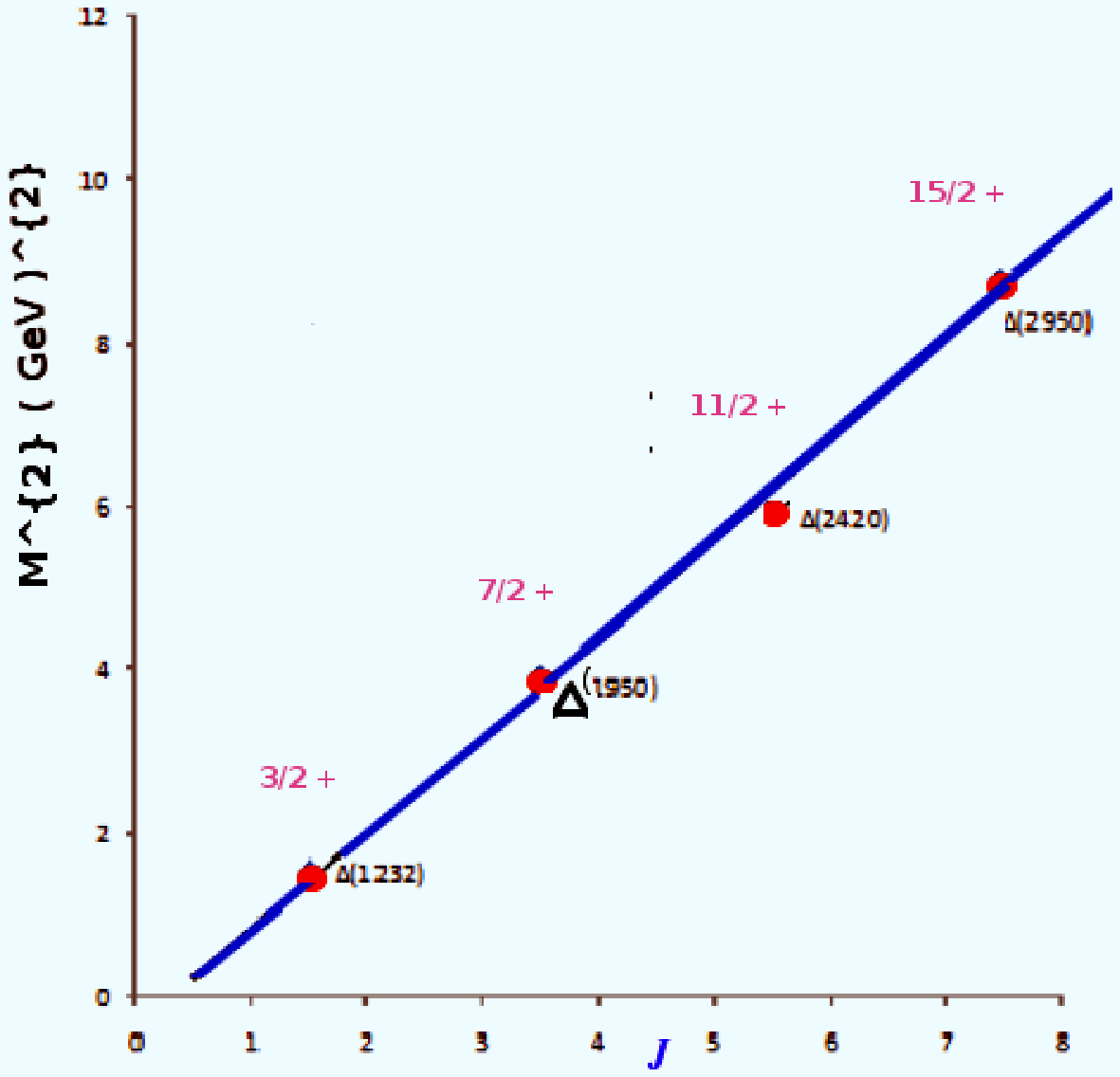}
\vskip -0.0cm
\label{fig4}
\vspace*{+0.20cm}
{\bf \color{black} \hspace*{2.0cm}
\begin{tabular}{c} Fig. 4 : Regge trajectory of $\ \Delta \ $ baryons
\vspace*{0.0cm} \\
comprising $\ 2 \ J^{\ P} \ = 
\ \left ( \ 3 , 7 , 11 , 15 \ \right )^{\ +} \ $
\end{tabular}
}
\vspace*{-0.0cm}
\end{figure}
\vspace*{-0.0cm}
\end{center}

\vspace*{-0.9cm}
\begin{center}
\vspace*{-0.0cm}
{\bf \color{black} 6 - Concluding remarks , outlook
}
\label{'6'}
\end{center}
\vspace*{-0.5cm}

\begin{description}

\item 1) A different view on resonances from QCD sum rules and condensates,
also relating to supersymmetric QCD by Adi Armoni and Mikhail Shifman can be 
found in ref. 41 -- \cite{ShifArmony} .
\vspace*{-0.1cm}

\item 2) The $\ \left ( \ 70 \ \times \ \vec{L} = 1 \ \right )^{\ -} \ $

negative parity 
u , d , s baryon multiplet with $\ \mathbb{N} \ = \ 1 \ $ 
is well described in the current PDG review -- \cite{PDG2012} -- 'Quark Model' 
by C. Amsler, T. De Grand and B. Krusche in ref. 19 -- \cite{AmsGrandKru} .
Extended other baryon- and meson multiplets including heavy quark flavors c , b
are assigned quark and antiquark configurations as well . A pioneering
paper discussing u, d, s flavored $\ \overline{q}^{\ '} \ q \ $ mesons is
due to George Zweig \cite{GZweig1968} .
\vspace*{-0.1cm}

\item 3) The circular pair-mode oscillator basis 

is presented in detail in ref. 29 -- \cite{oscimodes-fl6-PM} . Here 
space
unfortunately does not allow me to cover this topic , instrumental to
establish the counting of oscillatory modes in u. d, s - baryons .
\vspace*{-0.1cm}

\item 4) The $\ \mathbb{N} \ -$ density of baryon states per mass-square

defined in eq. \ref{eq:2-1-1} 

\vspace*{-0.5cm}
\begin{equation}
\label{eq:6-1}
\begin{array}{l}
\varrho_{\ n} \ \left ( \ m^{\ 2} \ \right )
\ = \ \ \begin{array}{c} 
\partial \ \mathbb{N} \ \left ( \ m^{\ 2} \ \right ) 
\vspace*{0.2cm} \\ \hline \vspace*{-0.3cm} \\
\partial \ m^{\ 2}
\end{array}
\end{array} 
\vspace*{-0.2cm}
\end{equation}

behaves for large $\ \mathbb{N} \ $ like 
$\ \mathbb{N}^{\ u} \ ; \ u \ = \ 5 \ $. This is enough to establish 
that string modes and oscillator modes presented are {\it inequivalent} .
\vspace*{-0.2cm}


\vspace*{-0.0cm}
\begin{center}
\hspace*{0.0cm}
\begin{figure}[htb]
\vskip -0.9cm
\hskip 0.0cm
\includegraphics[angle=-90,width=9.5cm]{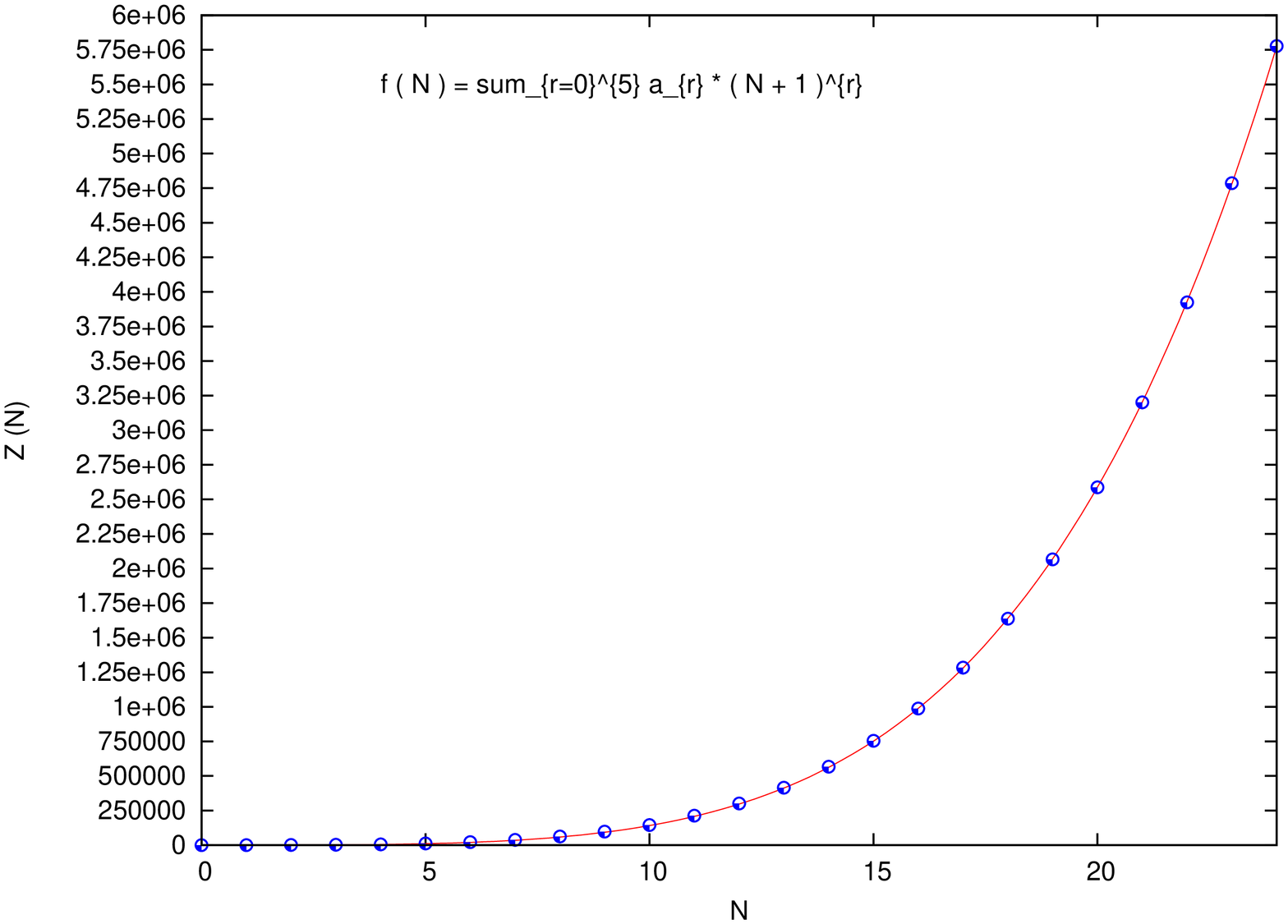}
\label{fig5}
\vspace*{+0.50cm}
\end{figure}
\vspace*{-0.8cm}

Fig. 5 : $\ \# \ ( \ \mathbb{N} \ ) \ $
\end{center}




Counting all u, d, s baryon states in the PDG -- \cite{PDG2012}
with total spin and isospin multiplicities accounted for,
1274 are obtained . This was done omitting a few very doubtful resonances .

The count of states with $\ \mathbb{N} \ \leq \ \mathbb{N}^{\ *} \ $
gives  
\vspace*{0.1cm}

\begin{tabular}[t]{|rr|}
\hline 
$\mathbb{N}^{\ *}$ & \#
\vspace*{0.0cm} \\ \hline \vspace*{-0.5cm} \\
0 & 56
\vspace*{-0.1cm} \\
1 & 266
\vspace*{-0.1cm} \\
2 & 1310
\vspace*{-0.1cm} \\
3 & 4090
\vspace*{0.0cm} \\ \hline
\end{tabular}
\vspace*{0.15cm}

\hspace*{0.0cm}
\begin{tabular}{@{\hspace*{-0.1cm}}l@{\hspace*{0.0cm}}}
with $\mathbb{N}^{\ *} \sim 3 $ beeing a fair estimate of 
resonances 
\vspace*{0.0cm} \\
up to 2.5 - 3 \ GeV .
\end{tabular}
\vspace*{0.1cm}

\item 5) outlook

I hope that at the high energy frontier , despite the odds looking unfavorable ,
exploiting the increased production cross sections of hitherto unobserved 
resonances , the art of resonance spectroscopy -- even of low energy 
resonances --
can witness a new frontier .
\end{description}

\begin{center}
\vspace*{-0.2cm}
{\bf \color{black} Appendix 1: The spin (10) product representations
\\
$\ \left ( \ 16 \ \oplus \ \overline{16} \ \right ) \ \otimes
\ \left ( \ 16 \ \oplus \ \overline{16} \ \right ) \ $
}
\label{'E'}
\end{center}
\vspace*{0.1cm}

\noindent
We follow the spin (10) decomposition discussed in section 2-1 
( eq. 27 repeated below ) 

\vspace*{-0.3cm}
\begin{equation}
\label{eq:2b2E}
\begin{array}{c}
\mbox{spin} \ (10) \ \rightarrow \ \mbox{SU5} \ \times \ \mbox{U1}_{\ J_{\ 5}}
\end{array}
\end{equation}

\noindent
Further let us denote representations of spin (10) as opposed to those
pertaining to SU5 and associated $\ J_{\ 5} \ $
quantum number by

\vspace*{-0.3cm}
\begin{equation}
\label{eq:E1}
\begin{array}{l}
\mbox{spin (10) :} 
\hspace*{0.2cm} \left \lbrack dim \right \rbrack
\hspace*{0.2cm} ; \hspace*{0.2cm}
\mbox{SU5} \ \times \ \mbox{U1}_{\ J_{\ 5}} \ :
\ \left \lbrace dim \right \rbrace_{\ J_{\ 5}}
\end{array}
\end{equation}

\noindent
Thus eq. \ref{eq:2b6} translates to

\vspace*{-0.3cm}
\begin{equation}
\label{eq:E2}
\begin{array}{l}
\begin{array}{lll lll l}
\left \lbrack 16 \right \rbrack
& = & \left \lbrace 1 \right \rbrace_{\ + 5}
& + & \left \lbrace 10 \right \rbrace_{\ + 1}
& + & \left \lbrace \overline{5} \right \rbrace_{\ - 3}
\vspace*{0.2cm} \\
\left \lbrack \overline{16} \right \rbrack
& = & \left \lbrace 1 \right \rbrace_{\ - 5}
& + & \left \lbrace \overline{10} \right \rbrace_{\ - 1}
& + & \left \lbrace 5 \right \rbrace_{\ + 3}
\end{array}
\end{array}
\end{equation}

\noindent
In turn SU5 representations shall be decomposed along the standard model gauge
group $\ \mbox{SU3}_{\ c} \ \otimes \ \mbox{SU2}_{\ L} \ \otimes 
\ \mbox{U1}_{\ {\cal{Y}}} \ $, where $\ {\cal{Y}} \ $ denotes the electroweak
hypercharge $\ \left ( \mbox{with a factor} \frac{1}{2} \ \mbox{included}
\ \right )$

\vspace*{-0.3cm}
\begin{equation}
\label{eq:E3}
\begin{array}{l}
{\cal{Y}} \ = \ Q_{\ e.m.} \ / \ e \ - \ I_{\ 3 \ L}
\end{array}
\end{equation}

\vspace*{-0.3cm}
\begin{equation}
\label{eq:E4}
\begin{array}{l}
\left \lbrace dim \right \rbrace \ \rightarrow
\ \sum
\ \left \rbrack\left (dim\mbox{SU3}_{\ c} \ , \ dim\mbox{SU2}_{\ L}
\right )_{\ {\cal{Y}}}\right \lbrack
\end{array}
\end{equation}

\noindent
The brackets on the right hand side of eq. \ref{eq:E4} are reversed in order
not to confuse spin (10) - and standard model representations.

\noindent
Then the base $\ 16 \ \left ( \ \overline{16} \ \right ) \ $ decompose to

\vspace*{-0.3cm}
\begin{equation}
\label{eq:E5}
\begin{array}{l}
\left \lbrack 16 \right \rbrack \ \rightarrow
\ \left \lbrack \begin{array}{l}
\left \lbrace 1 \right \rbrace_{\ + 5}
\hspace*{0.25cm} \rightarrow \ \left \lbrace \hspace*{0.2cm} \begin{array}{l}
\left \rbrack\left (1 \ , \ 1
\right )_{\ 0}\right \lbrack \hspace*{0.55cm}
\end{array} \right \rbrace_{\ +5}
\vspace*{0.2cm} \\
\left \lbrace 10 \right \rbrace_{\ + 1}
\ \rightarrow \ \left \lbrace \begin{array}{l}
\left \rbrack\left (3 \ , \ 2
\right )_{\ + \hspace*{0.05cm} \frac{1}{6}}\right \lbrack \ +
\vspace*{0.1cm} \\
\left \rbrack\left (\overline{3} \ , \ 1
\right )_{\ - \frac{2}{3}}\right \lbrack \ +
\vspace*{0.1cm} \\
\left \rbrack\left (1 \ , \ 1
\right )_{\ \hspace*{0.1cm} + 1 } \hspace*{0.05cm} \right \lbrack
\end{array} \right \rbrace_{\ +1}
\vspace*{0.2cm} \\
\left \lbrace \overline{5} \right \rbrace_{\ - 3}
\ \rightarrow
\hspace*{0.15cm} \left \lbrace \begin{array}{l}
\left \rbrack\left (\overline{3} \ , \ 1
\right )_{\ + \frac{1}{3}}\right \lbrack \ +
\vspace*{0.1cm} \\
\left \rbrack\left (1 \ , \ 2
\right )_{\ - \frac{1}{2}} \hspace*{0.05cm} \right \lbrack
\end{array} \right \rbrace_{\ -3}
\end{array} \right .
\end{array}
\end{equation}

\noindent
The product representations
$\ \left ( \ 16 \ \oplus \ \overline{16} \ \right ) \ \otimes
\ \left ( \ 16 \ \oplus \ \overline{16} \ \right ) \ $ generate {\it all}
SO (10) antysymmetric tensor ones, of which we encountered
the fivefold antisymmetric in section 2-1 
(eq. \ref{eq:2b11}).

\noindent
To elaborate we specify the n-fold antisymmetric 
tensors obtained from the 10-representation of SO (10)

\vspace*{-0.3cm}
\begin{equation}
\label{eq:E6}
\hspace*{0.2cm} \begin{array}{l}
\left \lbrack t_{0} \right \rbrack
\sim 1
\vspace*{0.2cm} \\
\left \lbrack t_{1} \right \rbrack^{A} 
\sim \ z^{A}
\hspace*{0.1cm} ; \hspace*{0.1cm}
A = 1,2,\cdots,10 
\hspace*{0.1cm} \leftrightarrow \hspace*{0.1cm}
\left \lbrack t_{1} \right \rbrack
= \left \lbrace \overline{5} \right \rbrace_{2} \oplus
\left \lbrace 5 \right \rbrace_{- 2}
\vspace*{0.2cm} \\
\left \lbrack t_{2} 
\right \rbrack^{\left \lbrack  A_{1} \ A_{2} \right \rbrack}
\sim \frac{1}{2} \left ( z_{1}^{A_{1}} z_{2}^{A_{2}}
\ - \ z_{1}^{A_{2}} z_{2}^{A_{1}} \right )
\vspace*{0.0cm} \\
\hspace*{1.0cm}\cdots
\vspace*{-0.2cm} \\
\left \lbrack t_{n}
\right \rbrack^{\ \left \lbrack  A_{1} \ A_{2} \cdots 
A_{n} \right \rbrack}
\sim \frac{1}{n!} \sum \ sgn \left ( \begin{array}{lll}
1 & \cdots & n
\vspace*{0.1cm} \\
\pi_{1} & \cdots & \pi_{n}
\end{array} \right )
z_{1}^{A_{\pi_{ 1}}} z_{2}^{A_{\pi_{ 2}}}
\cdots z_{n}^{A_{\pi_{ n}}}
\vspace*{0.1cm} \\
n \ \leq \ 10
\end{array}
\end{equation}

\noindent
The quantities $\ \left \lbrack \ t_{\ n} \ \right \rbrack \ $ defined in eq.
\ref{eq:E6} form irreducible real representations of SO (10) except
for n = 5 , which is composed of the relatively complex
{\it irreducible} representations $\ 126 \ \mbox{and} \ \overline{126} \ $ 
( eq. \ref{eq:2b11} ) .

\noindent
The tenfold antisymmetric invariant corresponds to 
$\ \left \lbrack \ t_{\ n=10} \ \right \rbrack \ $. The product of 
two full Clifford algebras pertaining to spin (10) contains
all $\left \lbrack t_{n} \right \rbrack \hspace*{0.1cm} ; 
\hspace*{0.1cm} n = 0 \ \cdots \ 10 $ 
representations exactly once .

\vspace*{-0.0cm}
\noindent
Treating the n = 5 tensor as one representation -- it is reducible only over 
$\ \mathbb{C} \ $ -- the dimensions of the 
$\ \left \lbrack \ t_{\ n} \ \right \rbrack \ $ representations follow
Pascal's triangle ( Fig. 3 page C7 ) of binomial coefficients for N = 10,
whereby n even and odd shall be distinguished

\vspace*{-0.3cm}
\begin{equation}
\label{eq:E7}
\begin{array}{l}
\begin{array}{@{\hspace*{0.0cm}}c@{\hspace*{0.0cm}}c@{\hspace*{0.0cm}}c 
@{\hspace*{0.0cm}}c@{\hspace*{0.0cm}}c@{\hspace*{0.0cm}}c @{\hspace*{0.0cm}}
@{\hspace*{0.0cm}}c@{\hspace*{0.0cm}}c@{\hspace*{0.0cm}}c @{\hspace*{0.0cm}}c
@{\hspace*{0.0cm}}c@{\hspace*{0.0cm}}c}
& \left \lbrack t_{0} \right \rbrack 
&
& \left \lbrack t_{2} \right \rbrack
&
& \left \lbrack t_{4} \right \rbrack
&
& \left \lbrack t_{6} \right \rbrack
&
& \left \lbrack t_{8} \right \rbrack
&
& \left \lbrack t_{10} \right \rbrack
\vspace*{0.2cm} \\
& & \left \lbrack t_{1} \right \rbrack
&
& \left \lbrack t_{3} \right \rbrack
&
& \left \lbrack t_{5} \right \rbrack
&
& \left \lbrack t_{7} \right \rbrack
&
& \left \lbrack t_{9} \right \rbrack
&
\vspace*{0.2cm} \\
& 1 & & 45 & & 210 & & 210 & & 45 & & 1
\vspace*{0.2cm} \\
& & 10 & & 120 & & 252 & & 120 & & 10
\end{array}
\end{array}
\end{equation}

\noindent
This corresponds to the following products of $\ 16 \ + \ \overline{16} \ $

\vspace*{-0.3cm}
\begin{equation}
\label{eq:E8}
\begin{array}{l}
\begin{array}{c|c|c}
& \left \lbrack 16 \right \rbrack & \left \lbrack \overline{16} \right \rbrack
\vspace*{0.0cm} \\
\hline \vspace*{-0.4cm} \\
\left \lbrack 16 \right \rbrack  
& s \ : \ \begin{array}{l}
\left \lbrack 10 \right \rbrack \ +
\vspace*{0.1cm} \\
\left \lbrack 126 \right \rbrack
\end{array}
\hspace*{0.1cm} , \hspace*{0.1cm}
a \ : \ \left \lbrack 120 \right \rbrack
& \begin{array}{l} \left \lbrack 1 \right \rbrack \ + 
\ \left \lbrack 45 \right \rbrack \ +
\vspace*{0.1cm} \\
\left \lbrack 210 \ \right \rbrack
\end{array}
\\ \hline \vspace*{-0.4cm} \\
\left \lbrack \overline{16} \right \rbrack
& \begin{array}{l} \left \lbrack 1 \right \rbrack \ +
\ \left \lbrack 45 \right \rbrack \ +
\vspace*{0.1cm} \\
\left \lbrack 210 \ \right \rbrack
\end{array}
& s \ : \ \begin{array}{l}
\left \lbrack 10 \right \rbrack \ +
\vspace*{0.1cm} \\
\left \lbrack \overline{126} \right \rbrack
\end{array}
\hspace*{0.1cm} , \hspace*{0.1cm}
a \ : \ \left \lbrack 120 \right \rbrack
\end{array}
\end{array}
\end{equation}

\noindent
The correspondence of product representations of the 
$\ 16 \ + \ \overline{16} \ = \ 32 \ $ associative Clifford algebra
with the sum of antisymmetric tensor ones follows from the
completeness of all products of $\ \gamma \ $ matrices forming 
the spin (10) algebra i.e. are of dimension

\vspace*{-0.3cm}
\begin{equation}
\label{eq:E9}
\begin{array}{l}
\left ( \ 32 \ \right )^{\ 2} \ = 
\ \left ( \ 2^{\ 5} \ \right )^{\ 2} \ = \ 2^{\ 10} 
\end{array}
\end{equation}

\noindent
We proceed to reduce the 
$\ \left \lbrack 16 \right \rbrack \ \otimes \ \left \lbrack 16 \right \rbrack
\ $ product
with respect to $\ J_{\ 5} \ $ , SU5 and 
$\ \mbox{SU3}_{\ c} \ \times
\ \mbox{SU2}_{\ L} \ \times \mbox{U1}_{\ {\cal{Y}}} \ $. 

\noindent
The individual products are 
$\ \left ( \ _{\ s \ (a)} \ : \mbox{(a)symmetric} \ \right )$ 

\vspace*{-0.5cm}
\begin{equation}
\label{eq:E10}
\begin{array}{l}
\begin{array}{@{\hspace*{0.0cm}}l@{\hspace*{0.1cm}}|@{\hspace*{0.0cm}}c
@{\hspace*{0.0cm}}c @{\hspace*{0.0cm}}c}
& \left \lbrace 1 \right \rbrace_{\ 5} 
& \left \lbrace 10 \right \rbrace_{1}
& \left \lbrace \overline{5} \right \rbrace_{- 3}
\vspace*{-0.3cm} \\ & & & \\ \hline \vspace*{-0.4cm} \\
\left \lbrace 1 \right \rbrace_{5} 
& \left . \left \lbrace 1 \right \rbrace_{ 10}
\right ._{ s}
& \left \lbrace 10 \right \rbrace_{6}
& \left \lbrace \overline{5} \right \rbrace_{2}
\vspace*{-0.3cm} \\ & & & \vspace*{-0.1cm} \\
\left \lbrace 10 \right \rbrace_{1} 
& \left \lbrace 10 \right \rbrace_{6} 
& \left ( \begin{array}{l}
\left \lbrace \overline{5} \right \rbrace_{2} +
\vspace*{0.1cm} \\
\left \lbrace \overline{50} \right \rbrace_{2}
\end{array} \right )_{s} \ \left ( \begin{array}{l}
\left \lbrace \overline{45} \right \rbrace_{2}
\end{array} \right )_{a}
& \left ( \begin{array}{l}
\left \lbrace 5 \right \rbrace_{ - 2}  +
\vspace*{0.1cm} \\
\left \lbrace 45 \right \rbrace_{ - 2}
\end{array} \right )
\vspace*{-0.3cm} \\ & & & \vspace*{-0.1cm} \\
\left \lbrace \overline{5} \right \rbrace_{- 3} 
& \left \lbrace \overline{5} \right \rbrace_{ 2}
& \left ( \begin{array}{l}
\left \lbrace 5 \right \rbrace_{ - 2} +
\vspace*{0.1cm} \\
\left \lbrace 45 \right \rbrace_{ - 2}
\end{array} \right )
& \left ( \begin{array}{l}
\left \lbrace \overline{15} \right \rbrace_{ - 6}
\end{array} \right )_{s} \ \left ( \begin{array}{l}
\left \lbrace \overline{10} \right \rbrace_{ - 6}
\end{array} \right )_{a}
\end{array}
\end{array}
\end{equation}

\noindent
We proceed to decompose the diagonal 
$\ \left \lbrace \mbox{SU5} \right \rbrace_{\ J_{\ 5}} \ $
representations (eq. \ref{eq:E5})

\vspace*{-0.3cm}
\begin{equation}
\label{eq:E11}
\begin{array}{l}
\left ( \ \left \lbrace 10 \right \rbrace_{1} \otimes 
\left \lbrace 10 \right \rbrace_{\ 1} \ \right )_{s}
= \left \lbrace \overline{5} \right \rbrace_{2} +
\left \lbrace \overline{50} \right \rbrace_{2}
\hspace*{1.2cm} {\color{black} \downarrow}
\vspace*{1.1cm} \\
\begin{array}{@{\hspace*{0.0cm}}c@{\hspace*{0.1cm}}|@{\hspace*{0.1cm}}c
@{\hspace*{0.0cm}}c @{\hspace*{0.0cm}}c}
s 
& 
\left \rbrack\left (3 , 2
\right )_{+ \hspace*{0.05cm} \frac{1}{6}}\right \lbrack_{+1} 
& 
\left \rbrack\left (\overline{3} , 1
\right )_{- \frac{2}{3}}\right \lbrack_{+1}
&
\left \rbrack\left (1 , 1
\right )_{\hspace*{0.1cm} + 1 } \hspace*{0.05cm} \right \lbrack_{+1}
\vspace*{-0.3cm} \\ & & & \\ \hline \vspace*{-0.2cm} \\
\left \rbrack\left (3 \ , \ 2
\right )_{+ \hspace*{0.05cm} \frac{1}{6}}\right \lbrack_{+1}
& \left ( \hspace*{-0.1cm} \begin{array}{l}
\left \rbrack \left (6 , 3
\right )_{+ \hspace*{0.05cm} \frac{1}{3}}\right \lbrack_{2} +
\vspace*{0.1cm} \\
\left \rbrack \left (\overline{3} , 1
\right )_{+ \hspace*{0.05cm} \frac{1}{3}}\right \lbrack_{2}
\end{array} \hspace*{-0.2cm} \right )
& \left ( \hspace*{-0.1cm} \begin{array}{l}
\left \rbrack \left (8 , 2
\right )_{- \hspace*{0.05cm} \frac{1}{2}}\right \lbrack_{2} +
\vspace*{0.1cm} \\
\left \rbrack \left (1 , 2
\right )_{- \hspace*{0.05cm} \frac{1}{2}}\right \lbrack_{2}
\end{array} \hspace*{-0.2cm} \right )
& \left \rbrack \left (3 , 2
\right )_{+ \hspace*{0.05cm} \frac{7}{6}}\right \lbrack_{2}
\vspace*{-0.1cm} \\ & & & \vspace*{-0.1cm} \\
\left \rbrack\left (\overline{3} , 1
\right )_{- \frac{2}{3}}\right \lbrack_{+1}
&
& \left \rbrack \left (\overline{6} , 1
\right )_{- \hspace*{0.05cm} \frac{4}{3}}\right \lbrack_{2}
& \left \rbrack \left (\overline{3} , 1
\right )_{+ \hspace*{0.05cm} \frac{1}{3}}\right \lbrack_{2}
\vspace*{-0.1cm} \\ & & & \vspace*{-0.1cm} \\
\left \rbrack\left (1 , 1
\right )_{\hspace*{0.1cm} + 1 } \hspace*{0.05cm} \right \lbrack_{+1}
&
&
& \hspace*{0.0cm} \left \rbrack\left (1 , 1
\right )_{\hspace*{0.1cm} + 2 } \hspace*{0.05cm} \right \lbrack_{2}
\end{array}
\end{array}
\end{equation}

\vspace*{-0.3cm}
\begin{equation}
\label{eq:E12}
\begin{array}{l}
\left ( \left \lbrace \overline{5} \right \rbrace_{- 3} \otimes 
\left \lbrace \overline{5} \right \rbrace_{- 3} \right )_{s}
= \left \lbrace \overline{15} \right \rbrace_{- 6} 
\hspace*{1.2cm} {\color{black} \downarrow}
\vspace*{0.3cm} \\
\begin{array}{c|cc}
s 
& 
\left \rbrack\left (\overline{3} , 1
\right )_{+ \hspace*{0.05cm} \frac{1}{3}}\right \lbrack_{-3} 
& 
\left \rbrack\left (1 , 2
\right )_{- \frac{1}{2}}\right \lbrack_{-3}
\vspace*{-0.3cm} \\ & & \\ \hline \vspace*{-0.2cm} \\
\left \rbrack\left (\overline{3} , 1
\right )_{+ \hspace*{0.05cm} \frac{1}{3}}\right \lbrack_{-3}
& 
\left \rbrack \left (\overline{6} , 1
\right )_{+ \hspace*{0.05cm} \frac{2}{3}}\right \lbrack_{-6}
& 
\left \rbrack \left (\overline{3} , 2
\right )_{- \hspace*{0.05cm} \frac{1}{6}}\right \lbrack_{-6}
\vspace*{-0.1cm} \\ & & \vspace*{-0.1cm} \\
\left \rbrack\left (1 , 2
\right )_{- \frac{1}{2}}\right \lbrack_{-3}
&
& \left \rbrack \left (1 \ , \ 3
\right )_{- \hspace*{0.05cm} 1}\right \lbrack_{-6}
\end{array}
\vspace*{0.2cm} \\
\hspace*{6.0cm} {\color{black} \downarrow}
\vspace*{0.2cm} \\
\hspace*{4.5cm} \mbox{complex e.w. triplet coupling to} 
\vspace*{0.2cm} \\
\hspace*{4.8cm} \frac{1}{2} \ \left ( \ \nu^{\ *}_{\ \dot{F}} 
\ \right )^{\ \alpha} 
\ \left ( \ \nu^{\ *}_{\ \dot{G}} \ \right )_{\ \alpha}
\end{array}
\end{equation}

\noindent
Next we assemble the (anti)symmetric products
$\ \left ( \ \left \lbrack 16 \right \rbrack \ \otimes 
\ \left \lbrack 16 \right \rbrack \ \right )_{\ s} \ = 
\ \left \lbrack 10 \right \rbrack \ \oplus 
\ \left \lbrack 126 \right \rbrack \ $
and
$\ \left ( \ \left \lbrack 16 \right \rbrack \ \otimes
\ \left \lbrack 16 \right \rbrack \ \right )_{\ a} \ =
\ \left \lbrack 120 \right \rbrack \ $
with respect to $\ \mbox{SU5} \ \otimes \ \mbox{U1}_{\ J_{\ 5}} \ $
using eq. \ref{eq:E10}

\vspace*{-0.3cm}
\begin{equation}
\label{eq:E13}
\begin{array}{c}
\left ( \ \left \lbrack 16 \right \rbrack \ \otimes
\ \left \lbrack 16 \right \rbrack \ \right )_{\ s} \ =
\ \left \lbrack 10 \right \rbrack \ \oplus
\ \left \lbrack 126 \right \rbrack
\hspace*{1.0cm} {\color{black} \downarrow}
\vspace*{0.3cm} \\
\ = \ \left \lbrace \begin{array}{c}
\left \lbrack \begin{array}{l} \left \lbrace 5 \right \rbrace_{\ -2} \ +
\vspace*{0.1cm} \\
\left \lbrace \overline{5}_{\ I} \right \rbrace_{\ 2}
\end{array} \right \rbrack 
\vspace*{0.2cm} \\
\oplus \ \left \lbrack \begin{array}{c}
\left \lbrace 1 \right \rbrace_{\ 10}
+ \ \left \lbrace \overline{5}_{\ II} \right \rbrace_{\ 2}
+ \ \left \lbrace 10 \right \rbrace_{\ 6} \ +
\ \left \lbrace \overline{15} \right \rbrace_{\ -6}
\vspace*{0.1cm} \\
+ \ \left \lbrace 45 \right \rbrace_{\ -2} \ +
\ \left \lbrace \overline{50} \right \rbrace_{\ 2}
\end{array} \right \rbrack
\end{array} \right \rbrace
\vspace*{0.3cm} \\
\left ( \ \left \lbrack 16 \right \rbrack \ \otimes
\ \left \lbrack 16 \right \rbrack \ \right )_{\ a} \ =
\ \left \lbrack 120 \right \rbrack 
\hspace*{1.0cm} {\color{black} \downarrow}
\vspace*{0.3cm} \\
\ = \ \left \lbrack \begin{array}{c}
\left \lbrace 5 \right \rbrace_{\ -2} \ +
\ \left \lbrace \overline{5} \right \rbrace_{\ 2}
\vspace*{0.1cm} \\
+ \ \left \lbrace 10 \right \rbrace_{\ 6} \ +
\ \left \lbrace \overline{10} \right \rbrace_{\ -6}
\vspace*{0.1cm} \\
+ \ \left \lbrace 45 \right \rbrace_{\ -2} \ +
\ \left \lbrace \overline{45} \right \rbrace_{\ 2}
\end{array} \right \rbrack
\end{array}
\end{equation}

\noindent
The roman indices $\ _{\ I,II} \ $ in eq. \ref{eq:E13} indicate
that appropriate linear combinations of the {\it two} 
$\ \left \lbrace \overline{5} \right \rbrace_{\ 2} \ $ representations
form parts of $\ \left \lbrack 10 \right \rbrack \ $ and 
$\ \left \lbrack 126 \right \rbrack \ $ respectively .

\noindent
It remains to decompose the $\ \mbox{SU5} \ \otimes \ \mbox{U1}_{\ J_{\ 5}} \ $
representations in eq. \ref{eq:E13} with respect to
$\ \mbox{SU3}_{\ c} \ \times
\ \mbox{SU2}_{\ L} \ \times \mbox{U1}_{\ {\cal{Y}}} \ $.
We do this associating according to the product representations
as they appear in eq. \ref{eq:E13}

\vspace*{-0.3cm}
\begin{equation}
\label{eq:E14}
\begin{array}{c}
\begin{array}{@{\hspace*{0.0cm}}lc@{\hspace*{0.1cm}}|@{\hspace*{0.1cm}}c@{\hspace*{0.0cm}}}
\left \lbrack 10 \right \rbrack \ \left \lbrack 120 \right \rbrack
& \left \lbrace 5 \right \rbrace_{-2}
& \left \rbrack\left (3 , 1
\right )_{- \hspace*{0.05cm} \frac{1}{3}}\right \lbrack_{+3}
+ \left \rbrack\left (1 , \overline{2}
\right )_{\ + \hspace*{0.05cm} \frac{1}{2}}\right \lbrack_{+3}
\vspace*{-0.1cm} \\ & & \vspace*{-0.1cm} \\
\left \lbrack 10 \right \rbrack \left \lbrack 126 \right \rbrack
\left \lbrack 120 \right \rbrack
& \left \lbrace \overline{5} \right \rbrace_{+2}
& \left \rbrack\left (\overline{3} , 1
\right )_{+ \hspace*{0.05cm} \frac{1}{3}}\right \lbrack_{-3}
+ \left \rbrack\left (1 , 2
\right )_{- \hspace*{0.05cm} \frac{1}{2}}\right \lbrack_{+3}
\vspace*{-0.3cm} \\ & & \\ \hline & & \vspace*{-0.3cm} \\
\left \lbrack 126 \right \rbrack
& \left \lbrace 1 \right \rbrace_{+10}
& \left \rbrack\left (1 , 1
\right )_{\hspace*{0.05cm} 0}\right \lbrack_{+10}
\vspace*{-0.1cm} \\ & & \vspace*{-0.1cm} \\
\left \lbrack 126 \right \rbrack \ \left \lbrack 120 \right \rbrack
& \left \lbrace 10 \right \rbrace_{+6}
& \left \rbrack\left (3 , 2
\right )_{- \hspace*{0.05cm} \frac{1}{6}}\right \lbrack_{+6}
+ \left \rbrack\left (\overline{3} , 1
\right )_{- \hspace*{0.05cm} \frac{2}{3}}\right \lbrack_{+6}
+ \left \rbrack\left (1 , 1
\right )_{+ \hspace*{0.05cm} 1}\right \lbrack_{+6}
\vspace*{-0.1cm} \\ & & \vspace*{-0.1cm} \\
\left \lbrack 120 \right \rbrack
& \left \lbrace \overline{10} \right \rbrace_{-6}
& \left \rbrack\left (\overline{3} , \overline{2}
\right )_{+ \hspace*{0.05cm} \frac{1}{6}}\right \lbrack_{-6}
+ \left \rbrack\left (3 , 1
\right )_{+ \hspace*{0.05cm} \frac{2}{3}}\right \lbrack_{-6}
+ \left \rbrack\left (1 , 1
\right )_{- \hspace*{0.05cm} 1}\right \lbrack_{-6}
\end{array}
\end{array}
\end{equation}

\vspace*{-0.6cm}
\begin{equation}
\label{eq:E15}
\begin{array}{c}
\begin{array}{@{\hspace*{0.0cm}}l@{\hspace*{0.1cm}}c@{\hspace*{0.0cm}}|
@{\hspace*{0.1cm}}c@{\hspace*{0.0cm}}}
\left \lbrack 126 \right \rbrack
& \left \lbrace \overline{15} \right \rbrace_{\ -6}
& \left \rbrack \left (\overline{6} , 1
\right )_{+ \hspace*{0.05cm} \frac{2}{3}}\right \lbrack_{-6}
+ \left \rbrack \left (\overline{3} , 2
\right )_{- \hspace*{0.05cm} \frac{1}{6}}\right \lbrack_{-6}
+ \left \rbrack \left (1 , 3
\right )_{- \hspace*{0.05cm} 1}\right \lbrack_{-6}
\vspace*{-0.1cm} \\ & & \vspace*{-0.1cm} \\
\left \lbrack 126 \right \rbrack \ \left \lbrack 120 \right \rbrack
& \left \lbrace 45 \right \rbrace_{-2}
& \begin{array}{l} c.c. \vspace*{-0.3cm} \\
\hspace*{0.5cm} {\color{black} \updownarrow}
\end{array}
\vspace*{-0.3cm} \\ & & \vspace*{-0.1cm} \\
\left \lbrack 120 \right \rbrack
& \left \lbrace \overline{45} \right \rbrace_{+2}
&  \left \lbrack  \begin{array}{c} 
\vspace*{-0.3cm} \\
\left ( \hspace*{-0.1cm} \begin{array}{l}
\left \rbrack \left (6 , 1
\right )_{+ \hspace*{0.05cm} \frac{1}{3}}\right \lbrack_{2} +
\vspace*{0.1cm} \\
\left \rbrack \left (\overline{3} , 3
\right )_{+ \hspace*{0.05cm} \frac{1}{3}}\right \lbrack_{2}
\end{array} \hspace*{-0.2cm} \right )
+ \left ( \hspace*{-0.1cm} \begin{array}{l}
\left \rbrack \left (8 , 2
\right )_{- \hspace*{0.05cm} \frac{1}{2}}\right \lbrack_{2} +
\vspace*{0.1cm} \\
\left \rbrack \left (1 , 2
\right )_{- \hspace*{0.05cm} \frac{1}{2}}\right \lbrack_{2}
\end{array} \hspace*{-0.2cm} \right ) +
\vspace*{0.1cm} \\
\left \rbrack \left (3 , 2
\right )_{+ \hspace*{0.05cm} \frac{7}{6}}\right \lbrack_{2} +
\left \rbrack \left (3 , 1
\right )_{- \hspace*{0.05cm} \frac{4}{3}}\right \lbrack_{2} +
\left \rbrack \left (\overline{3} , 1
\right )_{+ \hspace*{0.05cm} \frac{1}{3}}\right \lbrack_{2}
\vspace*{+0.3cm} \\
\end{array} \right \rbrack 
\vspace*{-0.1cm} \\ & & \vspace*{-0.1cm} \\
\left \lbrack 126 \right \rbrack
& \left \lbrace \overline{50} \right \rbrace_{+2}
& \left \lbrack  \begin{array}{c}
\vspace*{-0.3cm} \\
\left ( \hspace*{-0.2cm} \begin{array}{l}
\left \rbrack \left (6 , 3
\right )_{+ \hspace*{0.05cm} \frac{1}{3}}\right \lbrack_{2} +
\vspace*{0.1cm} \\
\left \rbrack \left (\overline{3} , 1
\right )_{+ \hspace*{0.05cm} \frac{1}{3}}\right \lbrack_{2}
\end{array} \hspace*{-0.2cm} \right )
 +  \left \rbrack \left (8 , 2
\right )_{- \hspace*{0.05cm} \frac{1}{2}}\right \lbrack_{2}  +
 \left \rbrack \left (3 , 2
\right )_{+ \hspace*{0.05cm} \frac{7}{6}}\right \lbrack_{2}
\vspace*{0.2cm} \\
+ \ \left \rbrack \left (\overline{6} , 1
\right )_{- \hspace*{0.05cm} \frac{4}{3}}\right \lbrack_{2} +
\ \left \rbrack\left (1 , 1
\right )_{\hspace*{0.1cm} + 2 } \hspace*{0.05cm} \right \lbrack_{2}
\vspace*{+0.3cm} \\
\end{array} \right \rbrack
\end{array}
\end{array}
\end{equation}

\vspace*{-0.3cm}
\begin{equation}
\label{eq:E16}
\begin{array}{l}
\left ( \left \lbrace 10 \right \rbrace_{\ 1} \otimes 
\left \lbrace 10 \right \rbrace_{\ 1} \ \right )_{s}
= \left \lbrace \overline{45} \right \rbrace_{2}
\hspace*{1.2cm} {\color{black} \downarrow}
\vspace*{1.1cm} \\
\begin{array}{@{\hspace*{0.0cm}}c@{\hspace*{0.0cm}}|@{\hspace*{0.0cm}}c
@{\hspace*{0.0cm}}c @{\hspace*{0.0cm}}c}
a 
& 
\left \rbrack\left (3 , 2
\right )_{+ \hspace*{0.05cm} \frac{1}{6}}\right \lbrack_{+1} 
& 
\left \rbrack\left (\overline{3} , 1
\right )_{- \frac{2}{3}}\right \lbrack_{+1}
&
\left \rbrack\left (1 , 1
\right )_{\hspace*{0.1cm} + 1 } \hspace*{0.05cm} \right \lbrack_{+1}
\vspace*{-0.3cm} \\ & & & \\ \hline \vspace*{-0.2cm} \\
\left \rbrack\left (3 , 2
\right )_{+ \hspace*{0.05cm} \frac{1}{6}}\right \lbrack_{+1}
& \left ( \hspace*{-0.1cm} \begin{array}{l}
\left \rbrack \left (6 , 1
\right )_{+ \hspace*{0.05cm} \frac{1}{3}}\right \lbrack_{2} +
\vspace*{0.1cm} \\
\left \rbrack \left (\overline{3} , 3
\right )_{+ \hspace*{0.05cm} \frac{1}{3}}\right \lbrack_{2}
\end{array} \hspace*{-0.2cm} \right )
& \left ( \hspace*{-0.1cm} \begin{array}{l}
\left \rbrack \left (8 , 2
\right )_{- \hspace*{0.05cm} \frac{1}{2}}\right \lbrack_{2} +
\vspace*{0.1cm} \\
\left \rbrack \left (1 , 2
\right )_{- \hspace*{0.05cm} \frac{1}{2}}\right \lbrack_{2}
\end{array} \hspace*{-0.2cm} \right )
& \left \rbrack \left (3 , 2
\right )_{+ \hspace*{0.05cm} \frac{7}{6}}\right \lbrack_{2}
\vspace*{-0.1cm} \\ & & & \vspace*{-0.1cm} \\
\left \rbrack\left (\overline{3} , 1
\right )_{- \frac{2}{3}}\right \lbrack_{+1}
&
& \left \rbrack \left (3 , 1
\right )_{- \hspace*{0.05cm} \frac{4}{3}}\right \lbrack_{2}
& \left \rbrack \left (\overline{3} , 1
\right )_{+ \hspace*{0.05cm} \frac{1}{3}}\right \lbrack_{2}
\vspace*{-0.1cm} \\ & & & \vspace*{-0.1cm} \\
\left \rbrack\left (1 , 1
\right )_{\hspace*{0.1cm} + 1 } \hspace*{0.05cm} \right \lbrack_{+1}
&
&
& --
\end{array}
\end{array}
\end{equation}

\noindent
I end the collection of representation decompositions with the adjoint 
$\ \left \lbrack 45 \right \rbrack \ $ representation of SO (10)

\vspace*{-0.3cm}
\begin{equation}
\label{eq:E17}
\begin{array}{l}
\left ( \ \left \lbrack 10 \right \rbrack \ \otimes 
\ \left \lbrack 10 \right \rbrack 
\ \right )_{\ a}
\ = \ \left \lbrack 45 \right \rbrack
\hspace*{1.2cm} {\color{black} \downarrow}
\vspace*{0.3cm} \\
\begin{array}{c|cc c}
a
& \left \lbrace 5 \right \rbrace_{\ -2}
& \left \lbrace \overline{5} \right \rbrace_{\ +2}
\vspace*{-0.3cm} \\ & & & \\ \hline \vspace*{-0.2cm} \\
\left \lbrace 5 \right \rbrace_{\ -2}
& \left \lbrace 10 \right \rbrace_{\ -4}
& {\color{black} \left \lbrace \begin{array}{c}
\left \lbrace 1 \right \rbrace_{\ 0} \ \leftrightarrow \ J_{\ 5}
\vspace*{0.1cm} \\
\left \lbrace 24 \right \rbrace_{\ 0} \ \leftrightarrow \ \mbox{adjoint SU5} 
\end{array} \right \rbrace }
\vspace*{-0.1cm} \\ & & & \vspace*{-0.1cm} \\
\left \lbrace \overline{5} \right \rbrace_{\ +2}
& 
& \left \lbrace \overline{10} \right \rbrace_{\ +4}
\end{array}
\end{array}
\end{equation}

\noindent
It should be noted that despite coinciding dimensions the following entities are
most distinct

\vspace*{-0.3cm}
\begin{equation}
\label{eq:E18}
\begin{array}{l}
\left \lbrack 10 \right \rbrack \ \neq \ \left \lbrace 10 \right \rbrace_{\ -4} \ , 
\ \ \left \lbrace 10 \right \rbrace_{\ 6}
\vspace*{0.2cm} \\
\left \lbrack 45 \right \rbrack \ \neq \ \left \lbrace 45 \right \rbrace_{\ -2}
\ ; \ \cdots
\end{array}
\end{equation}

\begin{center}
\vspace*{-0.0cm}
{\bf \color{black} Some conclusions from sections 1-1 and 1-2
}
\label{Sconc}
\end{center}
\vspace*{-0.1cm}

\begin{description}
\item C1) The oscillation phenomena indicate clearly , that a 
{\it genuinely chiral} extension of B - L to a conserved, global symmetry, 
generating a {\it continous} U1 - group of tranformations, is not involved.

\item C2) On the other hand the binary code of a ( minimally) supposed
unifying gauge group SO or spin (10) could, if B - L is {\it not} gauged,
equivalently generate a global symmetry of the vectorlike nature.
The latter however would allow neutrino mass through the 
( electroweak doublet-singlet ) pairing

\vspace*{-0.3cm}
\begin{equation}
\label{eq:b5}
\begin{array}{l}
- \ {\cal{L}}_{\ {\cal{M}}} \ =
\ \mu_{\ F \ G} \ {\cal{N}}^{\ F}_{\ \dot{\gamma}} 
\ \nu^{\ \dot{\gamma} \ G} \ + \ h.c.
\hspace*{0.2cm} ; \hspace*{0.2cm} F,G \ = \ 1,2,3 
\hspace*{0.2cm} \mbox{family}
\end{array}
\end{equation}

\noindent
without symmetry restrictions on the mass matrix $\ \mu_{\ F \ G} \ $
in eq. \ref{eq:b5}.
\vspace*{-0.0cm}

\item C3) Then however the question arises, why the mass matrix
$\ \mu \ $, involving the scalar doublet(s) within the electroweak gauge group,
also generating masses of charged spin $\frac{1}{2}$ fermions, gives rise to
very small physical neutrino masses.
Thus we follow the {\it hypothesis} that SO (10) {\it is} gauged and that 
it is the {\it large} mass scale of the gauge boson associated with B - L 
in particular, which distinguishes neutrino flavors \cite{SO10}
\cite{MGMPMHF}, \cite{PMHFosc} .

\end{description}

\begin{center}
\vspace*{1.2cm}
{\bf \color{black} 2-1+ The Majorana logic \cite{FukuYana} 
and mass from mixing -- \\
setting within the 'tilt to the left'
or 'seesaw' of type I ( $\cdots$ )
characterized by $\ {\cal{N}}_{\ F} \ $
}
\label{'2-1+'}
\end{center}
\vspace*{-0.3cm}

\noindent
Within the subgroup decompositions of SO (10) the 'tilt to the left'
does not appear obvious 

\vspace*{-0.7cm}
\begin{equation}
\label{eq:2b1}
\begin{array}{c}
\begin{array}{@{\hspace*{0.0cm}}c@{\hspace*{0.0cm}}c@{\hspace*{0.0cm}}c
@{\hspace*{0.0cm}}}
& \mbox{spin} \ (10) &
\vspace*{-0.2cm} \\
\swarrow & & \searrow
\vspace*{0.0cm} \\
\mbox{spin} \ (6) \ \equiv \ \mbox{SU4} & \times & \mbox{spin} \ (4) 
\ \equiv \ \mbox{SU2}_{\ L} \ \times \ \mbox{SU2}_{\ R} \\
\begin{array}{l}
\mbox{lepton number}
\vspace*{0.1cm} \\
\mbox{as 4th color \cite{PaSa}}
\end{array}
& & \\
\downarrow & & \downarrow
\vspace*{0.0cm} \\
{\color{black} \mbox{SU3}_{\ c}} \ \times \ \mbox{U1}_{\ B \ - \ L} 
& \times & \mbox{SU2}_{\ L} \ \times \ \mbox{U1}_{\ I_{\ 3 \ R}} \\
\searrow & 
& \swarrow 
\vspace*{-0.2cm} \\
& \hspace*{0.0cm} {\color{black} \mbox{SU3}_{c} \times 
\ \mbox{U1}_{\ Q_{\ e.m.}}}
&
\end{array}
\vspace*{0.1cm} \\
Q_{\ e.m.} \ / \ e \ = \ I_{\ 3 \ L} \ + \ I_{\ 3 \ R} \ + \ \frac{1}{2} 
\ ( \ B \ - \ L \ )
\end{array}
\vspace*{-0.3cm}
\end{equation}

\noindent
In eq. \ref{eq:2b1} the conserved charge-like gauges are marked especially.

\noindent
The large scale breaking of {\it gauged} B - L or 'tilt to the left'
was not assumed essential in refs. \cite{SO10} - \cite{PMHFosc} and
brings about a definite 'mass from mixing' scenario 
\cite{PMmfm} , \cite{CHPM} , \cite{mfmothers} 
to which we turn below.

\begin{center}
\vspace*{-0.2cm}
{\bf \color{black} The Majorana logic characterized by 
$\ {\cal{N}}_{\ F} \ $+
}
\label{Maj+}
\end{center}
\vspace*{-0.3cm}

\noindent
Here we consider the alternative subgroup decomposition

\vspace*{-0.3cm}
\begin{equation}
\label{eq:2b2}
\begin{array}{c}
\mbox{spin} \ (10) \ \rightarrow \ \mbox{SU5} \ \times \ \mbox{U1}_{\ J_{\ 5}}
\end{array}
\vspace*{-0.2cm}
\end{equation}

\noindent
Among the 3 generators of spin (10) commuting with $\mbox{SU3}_{\ c}$ ,
$ I_{\ 3 \ L} , I_{\ 3 \ R} $ and $B - L $ and forming part of the
Cartan subalgebra of spin (10)
there is one combination, denoted $\ J_{\ 5} \ $ in eq. \ref{eq:2b2},
commuting with its subgroup SU5 .

\noindent
The 16 representation in the left-chiral basis 
displays the charges pertinent to $J_{5}$ normalized to integer
values {\it modulo an overall sign}, as in the discussion of genuinely chiral U1-charges in
eq. \ref{eq:b4} -- but referring to N = 16
\vspace*{0.0cm}

{\color{black} 
\noindent
While the Majorana logic indeed opens a 'path' to trace the
origin of the 'tilt to the left' , the origin of three families is
unexplained at this stage.}

\vspace*{-0.7cm}
\begin{center}
\hspace*{0.0cm}
\begin{figure}[htb]
\vskip -0.3cm
\hskip -0.0cm
\includegraphics[angle=0,width=6.0cm]{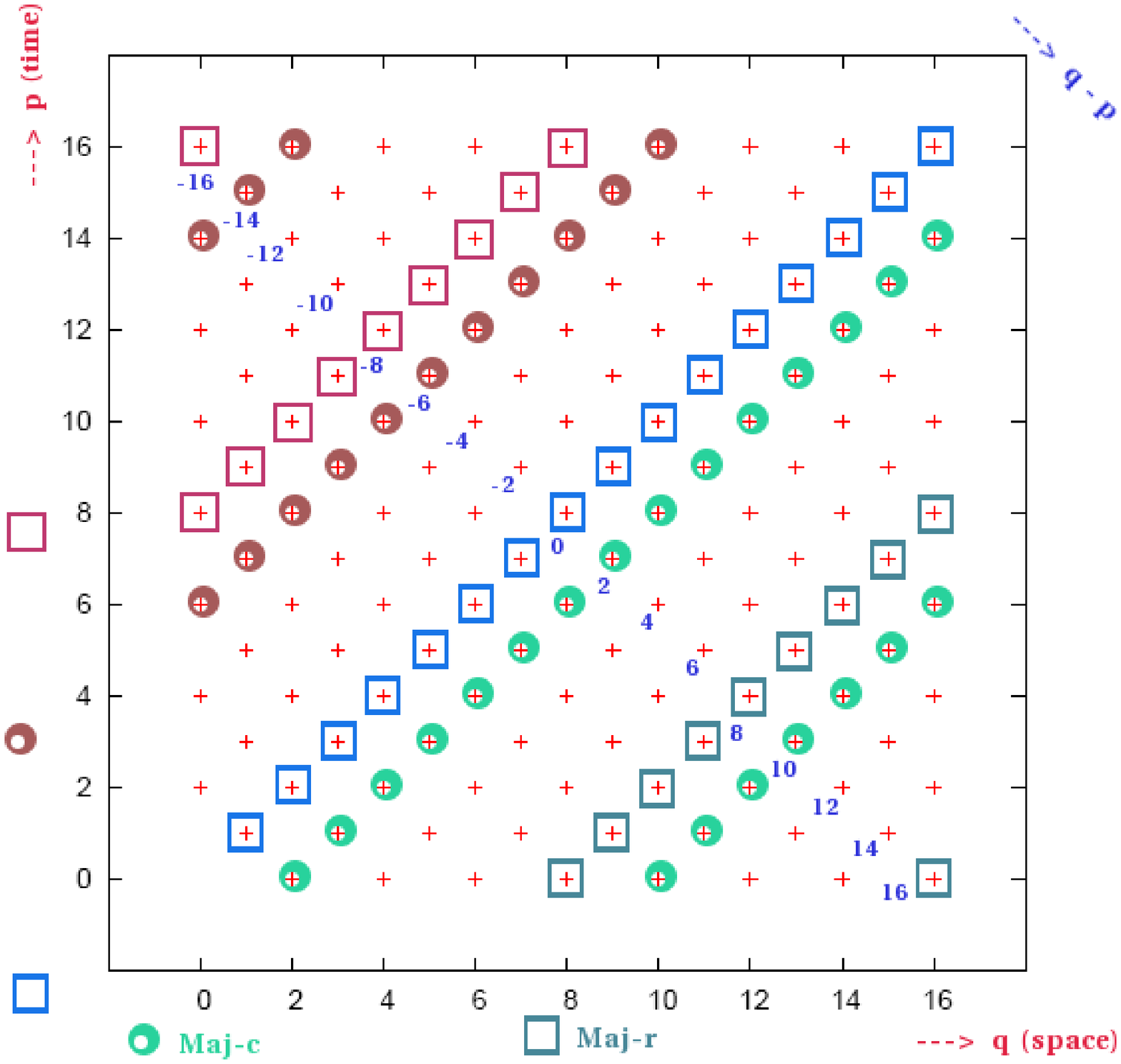}
\vskip -0.0cm
\vspace*{+0.20cm}
{\bf \color{black} \hspace*{0.0cm}
\begin{tabular}{c} Fig B1 :
The complex and real Majorana representations  
 \vspace*{-0.1cm} \\
$\ \mbox{MajCR} \ ( \ p \ , \ q \ ) \ $ \hspace*{0.2cm} 
{\color{black} $\ \longleftrightarrow$}
\end{tabular}
}
\end{figure}
\vspace*{-0.2cm}
\end{center}

\noindent
The associative Clifford algebras 
$\left \lbrace \ \Gamma_{\ p \ , \ q} \ ; \ \mathbb{C} \ \right \rbrace
\ \supset \ \left \lbrace \ \Gamma_{\ \widetilde{p} \ , \ \widetilde{q}} 
\ ; \ \mathbb{R} \ \right \rbrace \ $ are constructed in sections 
4-1a $\ \rightarrow \ $  4-1c, 4-2 and Appendices A, B forming
complementary material to the present outline . 

\noindent
p , q denote
time like ( p ) and spacelike ( q ) dimensions of space-time .

\noindent
Fig. B1 shows the repartition of real ( Maj-r ) and complex ( Maj-c )
character of irreducible 
{\it associative} , real ( Majorana ) Clifford algebras with
their characteristic mod 8 property relative to q - p \cite{coquereaux}.
These representations form the roots of the 'Majorana logic' discussed below .

\vspace*{-0.3cm}
\begin{equation}
\label{eq:2b3}
\begin{array}{l}
\left ( \ f \ \right )^{\ \dot{\gamma}} \ =
\ \left ( 
\begin{array}{llll l llll}
\color{red} u^{\ 1} & \color{green} u^{\ 2} & \color{cyan} u^{\ 3}
& \color{blue} \nu \hspace*{0.2cm} & \color{blue} | & \color{blue} {\cal{N}}
& \color{cyan} \widehat{u}^{\ 3} & \color{green} \widehat{u}^{\ 2}
& \color{red} \widehat{u}^{\ 1}
\vspace*{0.2cm} \\
\color{red} d^{\ 1} & \color{green} d^{\ 2} & \color{cyan} d^{\ 3}
& \color{blue} e^{\ -} \hspace*{0.2cm} & \color{blue} | & \color{blue} e^{\ +}
& \color{cyan} \widehat{d}^{\ 3} & \color{green} \widehat{d}^{\ 2}
& \color{red} \widehat{d}^{\ 1} \
\end{array} \color{blue} \right ) \vspace*{-0.5cm} \hspace*{-0.3cm}
\begin{array}{l}
\vspace*{-1.5cm} \\
^{\ \color{blue} \dot{\gamma} \ \rightarrow \ L}
\end{array}
\vspace*{0.7cm} \\
J_{\ 5} \ \rightarrow
\hspace*{0.5cm} \left (
\begin{array}{llll l llll}
\color{red} \hspace*{0.2cm} 1 & \color{green} \hspace*{0.2cm} 1 
& \color{cyan} \hspace*{0.2cm} 1
& \color{blue} - 3 \hspace*{0.2cm} & \color{blue} | 
& \color{blue} \hspace*{0.2cm} 5 
& \color{cyan} \hspace*{0.2cm} 1 & \color{green} \hspace*{0.2cm} 1
& \color{red} \hspace*{0.2cm} 1
\vspace*{0.2cm} \\
\color{red} \hspace*{0.2cm} 1 & \color{green} \hspace*{0.2cm} 1 
& \color{cyan} \hspace*{0.2cm} 1
& \color{blue} - 3 \hspace*{0.2cm} & \color{blue} | 
& \color{blue} \hspace*{0.2cm} 1 
& \color{cyan} - 3 & \color{green} - 3 \hspace*{0.1cm}
& \color{red} - 3 \hspace*{0.1cm}
\end{array} \color{blue} \right )
\end{array}
\end{equation}

\noindent
The assignment of $\ J_{\ 5} \ - $ charges in eq. \ref{eq:2b3} follows from
the fermionic oscillator representation of the spin (2n) associated
$\ \Gamma \ $ algebra through n such oscillators and the associated embedding
$\ \mbox{spin (10)} \ \supset \ \mbox{SU5} \ $ \cite{GeoGla} for n \ = \ 5 here 
\cite{PMvarenna}

\vspace*{-0.3cm}
\begin{equation}
\label{eq:2b4}
\begin{array}{l}
\left \lbrace a_{\ s} , a_{\ t}^{\dagger} \right \rbrace = 
\delta_{\ s t} 
\hspace*{0.1cm} ; \hspace*{0.1cm} 
s,t = 1,2 \cdots,n
\hspace*{0.1cm} ; \hspace*{0.1cm}
\left \lbrace a_{\ s} , a_{\ t} \right \rbrace = 0
= \left \lbrace a_{\ s}^{\dagger} , a_{\ t}^{\dagger} 
\right \rbrace
\hspace*{0.1cm} {\color{black} \rightarrow}
\vspace*{0.2cm} \\
J_{\ n} = \sum_{s = 1}^{\ n} \ \left ( \begin{array}{r}
a_{\ s}^{\ \dagger} \ a_{\ s}
\vspace*{0.1cm} \\
- \ a_{\ s} \ a_{\ s}^{\ \dagger}
\end{array} \right )
= 2 \widehat{n} - n \ \P_{\ 2^{n} \ \times \ 2^{n}}
\hspace*{0.1cm} ; \hspace*{0.1cm}
\widehat{n} = \sum_{s = 1}^{\ n} \ a_{\ s}^{\dagger} \ a_{\ s}
\end{array}
\end{equation}

\noindent
The eigenvalues (X) {\it and} multiplicities  (\#) of $\ J_{\ n} \ $

\vspace*{-0.3cm}
\begin{equation}
\label{eq:2b5}
\begin{array}{l}
\begin{array}{c|cc ccc c}
(X) & n & n-2 & n-4 & \cdots & - n + 2 & - n
\vspace*{0.0cm} \\ \hline \vspace*{-0.3cm} \\
(\#) & \left ( \begin{array}{c}
n \\
0
\end{array} \right ) &
\left ( \begin{array}{c}
n \\
1
\end{array} \right ) 
& \left ( \begin{array}{c}
n \\
2
\end{array} \right )
& \cdots
& \left ( \begin{array}{c}
n \\
n - 1
\end{array} \right )
& \left ( \begin{array}{c}
n \\
n 
\end{array} \right )
\end{array}
\end{array}
\end{equation}

\noindent
The orthogonal series for n even $\ \leftrightarrow \ $ real 
( spin (8) , spin(12) 
$\ \cdots \ $ ) has another decompostion within the associated $\ \Gamma \ $
algebra , than the one with 

\newpage

\vspace*{-0.3cm}
\noindent
n odd $\ \leftrightarrow \ $ complex
( spin (10) , spin (14) 
$\ \cdots \ $ ) . We give here the explicit numbers according to
eq. \ref{eq:2b5} for n = 5 , i.e. spin (10)

\vspace*{-0.3cm}
\begin{equation}
\label{eq:2b6}
\begin{array}{l}
\begin{array}{@{\hspace*{0.0cm}}c|cc ccc c@{\hspace*{0.0cm}}}
(X) & 5 & {\color{black} 3} & 1 & {\color{black} - 1} & - 3  & 
{\color{black} - 5}
\vspace*{0.0cm} \\ \hline \vspace*{-0.3cm} \\
(\#) & \left ( \begin{array}{c}
5 \\
0
\end{array} \right ) &
{\color{black} \left ( \begin{array}{c}
5 \\
1
\end{array} \right )} 
& \left ( \begin{array}{c}
5 \\
2
\end{array} \right )
& {\color{black}  \left ( \begin{array}{c}
5 \\
3 
\end{array} \right )}
& \left ( \begin{array}{c}
5 \\
4
\end{array} \right )
& {\color{black} \left ( \begin{array}{c}
5 \\
5 
\end{array} \right )}
\vspace*{0.0cm} \\ 
& & & & & &
\vspace*{-0.3cm} \\
\mbox{SU5} & \left \lbrace 1 \right \rbrace
& {\color{black} \left \lbrace 5 \right \rbrace}
& \left \lbrace 10 \right \rbrace
& {\color{black} \left \lbrace \overline{10} \right \rbrace}
& \left \lbrace \overline{5} \right \rbrace
& {\color{black} \left \lbrace \overline{1} \right \rbrace}
\end{array}
\end{array}
\end{equation}

\noindent
The subset of states in blue in eq. \ref{eq:2b6} 
$\ (X) \ = \ \left \lbrace \ 5 \ , \ 1 \ , \ - 3 \ \right \rbrace \ $ 
forms the 16 representation
of spin 10, while those in red 
$\ (X) \ = \ {\color{black} \left \lbrace \ 3 \ , \ - 1 \ , \ - 5 
\ \right \rbrace} 
\ $ the complex conjugate {\color{black} $\ \overline{16} \ $} .

\noindent
This opens the 'path' of linking the 'tilt to the left' with a substructure 
based on the primary in strength breakdown of the local gauged chargelike 
symmetry associated with 

\vspace*{-0.3cm}
\begin{equation}
\label{eq:2b7}
\begin{array}{l}
J_{\ 5} \ = \ - 4 \ I_{\ 3 \ R} \ + \ 3 \ \left ( \ B \ - \ L \ \right )
\end{array}
\end{equation}

\noindent
$J_{\ 5} \ $ as defined through integer eigenvalues
$\ (X) \ $ given in eqs. \ref{eq:2b3} and \ref{eq:2b6} is normalized
differently from the other Cartan subalgebra charges 
$\ I_{\ 3 \ L} \ , \ I_{\ 3 \ R} \ , \ B \ - L \ $

\vspace*{-0.3cm}
\begin{equation}
\label{eq:2b8}
\begin{array}{c}
\left | \ Q_{\ C} \ \right |^{\ 2} \ = 
\ \sum_{\ \left \lbrace 16 \right \rbrace} 
\ \left ( \ Q_{\ C} \ ( \ f \ ) \ \right )^{\ 2}
\hspace*{0.2cm} , \hspace*{0.2cm}
\left | \ I_{\ 3 \ L} \ \right |^{\ 2} \ = \ 2
\hspace*{0.2cm} , \hspace*{0.2cm}
\left | \ I_{\ 3 \ R} \ \right |^{\ 2} \ = \ 2
\vspace*{0.2cm} \\
\left | \ B \ - \ L \ \right |^{\ 2} \ = \ \frac{16}{ 3}
\hspace*{0.2cm} , \hspace*{0.2cm}
{\color{black} \left | \ J_{\ 5} \ \right |^{\ 2} \ = \ 80}
\end{array}
\end{equation}

\noindent
The consequence as far as neutrino-mass and mixing is concerned follows
from identifying the $\ {\color{black} J_{\ 5}} \ $ direction with a 
major axis of primary spontaneous gauge-symmetry breaking , bringing
about the \\ 'tilt to the left' from eq. \ref{eq:b5}

\vspace*{-0.3cm}
\begin{equation}
\label{eq:2b9}
\begin{array}{c}
{\cal{H}}_{\ {\cal{M}}} \ =
\ \mu_{\ F \ G} \ {\cal{N}}^{\ F}_{\ \dot{\gamma}} 
\ \nu^{\ \dot{\gamma} \ G} \ + \ h.c. \ {\color{black} + \ {\cal{H}}_{\ M}}
\vspace*{0.2cm} \\
{\cal{H}}_{\ M} \ = \ \frac{1}{2} \ M_{\ F \ G} 
\ {\cal{N}}^{\ F}_{\ \dot{\gamma}}
\ {\cal{N}}^{\ \dot{\gamma} \ G} \ + \ h.c.
\hspace*{0.2cm} ; \hspace*{0.2cm} F,G \ = \ 1,2,3
\vspace*{0.2cm} \\
M_{\ F \ G} \ = \ M_{\ G \ F} 
\hspace*{0.2cm} : \hspace*{0.2cm}
\mbox{complex arbitrary otherwise}
\hspace*{0.2cm} ; \hspace*{0.2cm}
{\color{black} \left | \ M \ \right | \ \gg \ } \ \left | \ \mu \ \right |
\end{array}
\end{equation}

\noindent
It is the primary breakdown along the direction of $\ {\color{black} 
J_{\ 5}} \ $
which contrary to all 'mirror complexes' brings on the level of 
(pseudo-) scalar fields to the foreground the complex bosonic 126 
and $\overline{126}$ representations of SO10 

\vspace*{-0.6cm}
\begin{equation}
\label{eq:2b10}
\begin{array}{l}
\hspace*{2.5cm} {\cal{H}}_{\ M} \ \longleftarrow 
\vspace*{0.2cm} \\
\left ( \Phi^{\ \overline{126} \ F \ G} \right )^{ \overline{\xi}}
\ \left ( f_{\ a \ 16 \ F} \right )_{\dot{\gamma}} 
\ \left ( f_{\ b \ 16 \ G} \right )^{\dot{\gamma}}
\ C \ \left ( \begin{array}{c|cc}
\begin{array}{c} 126
\\
\xi
\end{array} 
& \begin{array}{c} 16
\\
a
\end{array}
& \begin{array}{c} 16
\\
b
\end{array}
\end{array} \right ) + h.c.
\vspace*{0.2cm} \\
\left ( \ \Phi^{\ \overline{126} \ F \ G} \ \right )^{ \overline{\xi}}
\ : \ \mbox{(p-) scalar fields in the} \ \overline{126}
\ \mbox{representation of SO (10)}
\end{array}
\vspace*{-0.2cm}
\end{equation}

\noindent
In eq. \ref{eq:2b10} $\ C \ \left ( \ \begin{array}{c|cc}
\begin{array}{c} 126
\\
\xi
\end{array}
& \begin{array}{c} 16
\\
a
\end{array}
& \begin{array}{c} 16
\\
b
\end{array}
\end{array} \right ) \ $ denotes the coupling coefficients, projecting 
the symmetric product of two 16-representations of spin (10) to the 126 representation of SO (10) .

\noindent
The 126 {\it complex} representation of SO (10) is singled out by the value of 
$\ {\color{black} J_{\ 5}} \ $
of $ \ 10 \ = \ 2 \ \times 5 \ {\cal{N}} \ {\cal{N}} \ $. 

\noindent
The relatively complex conjugate representations $\ 126 \ \oplus \ \overline{126} \ $
are contained in the {\it real , reducible} fivefold antisymmetric tensor representation of SO (10)
decomposing into the irreducible pair upon the duality conditions

\vspace*{-0.3cm}
\begin{equation}
\label{eq:2b11}
\begin{array}{l}
t^{\ \left \lbrack \ A_{\ 1} \ A_{\ 2} \ \cdots \ A_{\ 5} \ \right \rbrack }
\hspace*{0.2cm} ; \hspace*{0.2cm} 
A_{\ 1 \ \cdots \ 5} \ = \ 1,2,\cdots,10
\vspace*{0.2cm} \\
t^{\ \left \lbrack \ A_{\ \pi_{\ 1}} \ A_{\ \pi_{\ 2}} \ \cdots \ A_{\ \pi_{\ 5}} \ \right \rbrack }
 = sgn \ \left ( \begin{array}{cccc}
1 & 2 & \cdots & 5
\vspace*{0.1cm} \\
\pi_{\ 1} & \pi_{\ 2} & \cdots & \pi_{\ 5}
\end{array} \right ) \ t^{\ \left \lbrack \ A_{\ 1} \ A_{\ 2} \ \cdots \ A_{\ 5} \ \right \rbrack }
\vspace*{0.3cm} \\
\ \frac{1}{5!} \ \varepsilon_{\ A_{\ 1} \ \cdots \ A_{\ 5} \ B_{\ 1} \ \cdots \ B_{\ 5}}
\ t^{\ \left \lbrack \ B_{\ 1} \ B_{\ 2} \ \cdots \ B_{\ 5} \ \right \rbrack }_{\ \pm}
\ = \ \left ( \ \pm \ i \ \right )
\ t^{\ \left \lbrack \ A_{\ 1} \ A_{\ 2} \ \cdots \ A_{\ 5} \ \right \rbrack }_{\ \pm}
\vspace*{0.3cm} \\
\varepsilon_{\ A_{\ 1} \ \cdots \ A_{\ 5} \ A_{\ 6} \ \cdots \ A_{\ 10}}
\ = 
\vspace*{0.2cm} \\
\hspace*{0.5cm} = \  sgn \ \left ( \begin{array}{cccc}
1 & 2 & \cdots & 10
\vspace*{0.1cm} \\
\pi_{\ 1} & \pi_{\ 2} & \cdots & \pi_{\ 10}
\end{array} \right )
\ \varepsilon_{\ A_{\ \pi_{\ 1}} \ \cdots \ A_{\ \pi_{\ 5}} \ A_{\ \pi_{\ 6}} \ \cdots \ A_{\ \pi_{\ 10}}}
\vspace*{0.2cm} \\
\varepsilon_{\ 1 \ 2 \ \cdots \  10} \ = \ 1
\end{array}
\end{equation}

\noindent
Within the complex spin $\ \left ( 2 \nu \ = \ 4 \tau + 2 \right ) \ , 
\ \tau \ = \ 2,3, \ \cdots \ $ series -- $\ \tau \ = \ 2 \ \leftrightarrow \ \mbox{spin (10)}\ $ --
the relatively complex conjugate spinorial pair of representations with 
dimension 
$\ 4^{\ \tau} \ \leftarrow \ 16 (64, \  \cdots) \ $ and the complex selfdual-antiselfdual pair of representations with dimension 
$\ \frac{1}{2} \ \left ( \begin{array}{c} 
4 \ \tau \ + \ 2 \\
2 \ \tau \ + \ 1
\end{array} \right ) \ \leftarrow \ 126 \ ( 11 . 12 . 13 = 1716, \ \cdots) \ $ 
are intrinsically related for $\ \tau \ = \ 2,3,4,\cdots \ $.

\begin{center}
\vspace*{-0.0cm}
{\bf \color{black} Some conclusions and questions from section 2-1 .
}
\end{center}
\vspace*{-0.1cm}

\begin{description}
\item Q1) Is it enough to consider the primary breakdown and its characteristic, the 'tilt to the left'
concerning 3 families, as due {\it essentially} to spin (10) , which is the {\it lowest} simple spin
group along the complex {\it orthogonal} chain ? 

It has been argued interestingly by Feza Gursey and collaborators
\cite{Gursey}, that it is the chain of exceptional groups which encode intrinsically the number 3,
which in turn underlies the 3  as the number of (left-chiral) families as well as the
strong interaction gauge group $\ \mbox{SU3}_{\ c} \ $.

\item A1) I think the answer is to the affirmative, since all higher 
gauge groups , including the
exceptional chain and especially E8 , but also spin (14) , (18) do {\it not} 
explain
the \#3 of families , rather generate together with even the apparently correct 3 families -- for E8 --
also mirror families -- 3 for E8 , and powers of 2 for the orthogonal chain 
with $\ \tau \ \geq \ 3 \ $.


The tentative conclusion remains, that the structure of families has to be 
explained
outside \\
spin (10) and {\it also} outside larger unifying gauge groups 
containing spin (10) ,
whereas the origin of neutrino mass is layed out by the lowest member of 
the complex orthogonal chain $\ \rightarrow \ $ spin (10) .

\item C4) The two apparently different phenomena of a) 'tilt to the left'
and b) baryon number violation are intrinsically associated with
the {\it unusual} sequence of (pseudo)scalar fields
generating primary breakdown . We use the notation ( eq. \ref{eq:2b2} )

\vspace*{-0.3cm}
\begin{equation}
\label{eq:2b12}
\begin{array}{c}
\begin{array}{c}
\mbox{spin} \ (10) \rightarrow \mbox{SU5}  \times \mbox{U1}_{J_{\ 5}}
\rightarrow \mbox{SU3}_{\ c} \times \mbox{SU2}_{\ L} 
 \times \mbox{U1}_{{\cal{Y}}} = \mbox{G}_{ s.m.}
\end{array}
\vspace*{0.2cm} \\
\begin{array}{@{\hspace*{0.0cm}}l@{\hspace*{0.0cm}}l@{\hspace*{0.0cm}}l 
@{\hspace*{0.0cm}}l@{\hspace*{0.0cm}}l@{\hspace*{0.0cm}}l @{\hspace*{0.0cm}}l}
\left \lbrack 16 \right \rbrack
& = & \left \lbrace 1 \right \rbrace_{\ + 5}
& + & \left \lbrace 10 \right \rbrace_{\ + 1}
& + & \left \lbrace \overline{5} \right \rbrace_{\ - 3}
\vspace*{0.2cm} \\
\left \lbrack \overline{16} \right \rbrack
& = & \left \lbrace 1 \right \rbrace_{\ - 5}
& + & \left \lbrace \overline{10} \right \rbrace_{\ - 1}
& + & \left \lbrace 5 \right \rbrace_{\ + 3}
\vspace*{0.4cm} \\
\left \lbrace \ \overline{5} \ \right \rbrace_{\ -3}
& = & \left \rbrack \ \left ( \ \overline{3} \ , \ 1 
\ \right )_{\ +\frac{1}{3}} \ \right \lbrack_{\ -3}
& + & \left \rbrack \ \left ( \ 1 \ , \ 2 
\ \right )_{\ -\frac{1}{2}} \ \right \lbrack_{\ -3}
& &
\end{array}
\end{array}
\end{equation}


\vspace*{-0.3cm}
\begin{equation}
\label{eq:2b13}
\begin{array}{l}
\begin{array}{@{\hspace*{0.0cm}}c@{\hspace*{0.0cm}}c@{\hspace*{0.0cm}}c
@{\hspace*{0.0cm}}c@{\hspace*{0.0cm}}}
\begin{array}{c}
\mbox{(p)scalar}r 
\vspace*{-0.0cm} \\
\mbox{SO (10)} 
\vspace*{0.0cm} \\
\mbox{reprsnt.}
\end{array}
& \begin{array}{c}
\mbox{active}
\vspace*{-0.2cm} \\
\mbox{components}
\end{array}
& \begin{array}{c}
\mbox{induced}
\vspace*{-0.2cm} \\
{\color{black} \mbox{(a)symmetries}}
\end{array}
& \begin{array}{c}
\mbox{preserved}
\vspace*{-0.2cm} \\
\mbox{gauge}
\vspace*{-0.2cm} \\
\mbox{group}
\end{array}
\vspace*{-0.2cm} \\
\vspace*{-0.5cm} \\ & & & \\ \hline \vspace*{-0.3cm} \\
\left .
\begin{array}{c}
\left \lbrack 126 \right \rbrack_{\hspace*{0.0cm} \mathbb{C}}
\vspace*{-0.0cm} \\
\left \lbrack \overline{126} \right \rbrack_{\overline{\mathbb{C}}}
\end{array} \right \rbrace \ \rightarrow
&
\begin{array}{l}
\left \lbrace 1 \right \rbrace_{+10}
\\
\left \lbrace \overline{1} \right \rbrace_{-10}
\end{array}
& \begin{array}{c}
P : \mbox{'tilt to the left'} 
\\ 
{\cal{N}} {\cal{N}} - \mbox{mass}, B - L
\\
CP \hspace*{0.1cm} \downarrow
\end{array}
& \mbox{SU5}
\vspace*{-0.1cm} \\ & & & \vspace*{-0.1cm} \\
\left .
\begin{array}{c}
\left \lbrack 45 \right \rbrack_{ \mathbb{R}}
\end{array} \right \rbrace \ \nearrow \hspace*{-0.0cm} \rightarrow
& \begin{array}{l}
\left \lbrace 24 \right \rbrace_{ 0} \hspace*{0.1cm} \downarrow
\\
\left \rbrack \left ( \ 1 , 1 \ \right )_{0} \ \right \lbrack_{0}
\end{array}
& \begin{array}{c}
B , L \ , \hspace*{0.1cm} \updownarrow \hspace*{0.1cm} CP
\end{array}
& \mbox{G}_{\ s.m.}
\vspace*{-0.3cm} \\ & & & \\ \hline \vspace*{-0.3cm} \\
\left .
\begin{array}{c}
\left \lbrack 10 \right \rbrack_{\mathbb{R}}
\end{array} \right \rbrace \ \nearrow \hspace*{-0.0cm} \rightarrow
& \begin{array}{l}
\left \rbrack \ \left ( \ 1 , 2 \ \right )_{-\frac{1}{2}} 
\ \right \lbrack_{-3} 
\\
\left \rbrack \ \left ( \ 1 , \overline{2} \ \right )_{+\frac{1}{2}} 
\ \right \lbrack_{+3}
\end{array}
& \begin{array}{c}
{\cal{N}} {\cal{\nu}} \ \mbox{mass} \hspace*{0.1cm} \uparrow
\\
\overline{q} q \ \mbox{mass} \hspace*{0.1cm} CKM 
\\
CP \hspace*{0.1cm} \uparrow
\end{array}
& \begin{array}{c}
\mbox{SU3}_{\ c} \times \\
\mbox{U1}_{\ e.m.}
\end{array}
\end{array}
\end{array}
\end{equation}

\end{description}

\newpage

\vspace*{-0.1cm}
\noindent
Pascals triangle $\ \left ( \begin{array}{l} 
n \\ k
\end{array} \right ) \ $ for $\ n = 1,2, \cdots ,16 \ $

\vspace*{-0.2cm}
\begin{center}
\hspace*{0.0cm}
\begin{figure}[htb]
\vskip -0.6cm
\hskip 0.0cm
\includegraphics[angle=0,width=7.0cm]{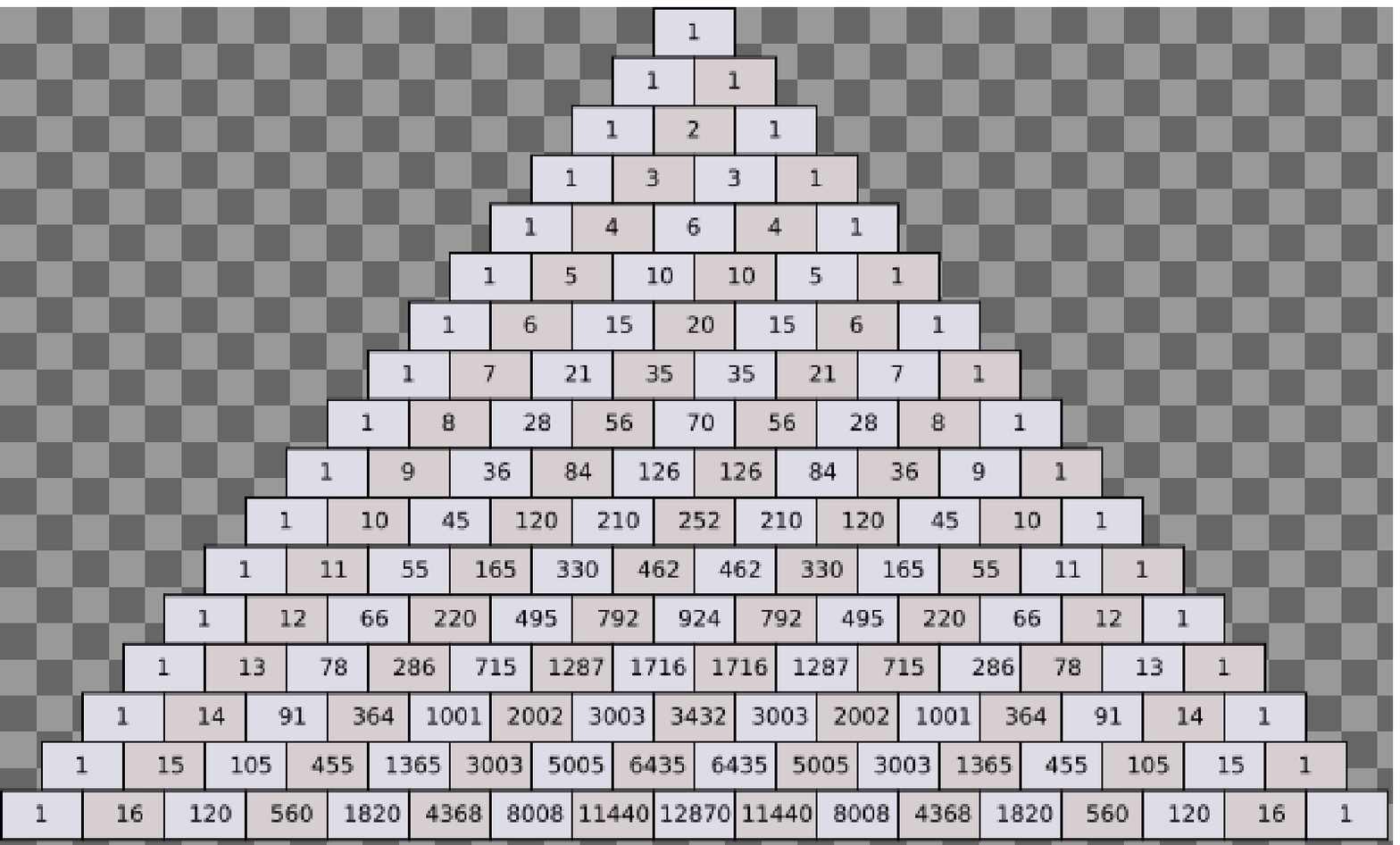}
\vskip -0.2cm
\vspace*{+0.20cm}
{\bf \color{black} \hspace*{0.0cm}
\begin{tabular}{c} Fig 3 :
Pascal's triangle  
\end{tabular}
}
\end{figure}
\end{center}

\begin{center}
\vspace*{-0.7cm}
{\bf \color{black} 1-1a There does not exist a symmetry -- within
the standard model including gravity 
and containing only chiral spin $\frac{1}{2} \ $ 16 families of SO (10) -- 
which could enforce the vanishing of neutrino mass(es) .
 }
\label{'1-1a'}
\end{center}
\vspace*{-0.4cm}

\noindent
The divergence of the current associated to the global charge B - L
for three standard model families of 15 base fields -- in the left chiral basis
removing -- to infinite mass -- the 16-th components 
$\ \left ( \ {\cal{N}} \ \right ) $
pertaining to one full 16-representation of SO (10) 
[ \ spin (10) ] 

\vspace*{-0.3cm}
\begin{equation}
\label{eq:a-1}
\begin{array}{c}
\left ( 
\begin{array}{llll l llll}
\color{red} u^{\ 1} & \color{green} u^{\ 2} & \color{cyan} u^{\ 3}
& \color{blue} \nu \hspace*{0.2cm} & \color{blue} | & \color{blue} {\cal{N}}
& \color{cyan} \widehat{u}^{\ 3} & \color{green} \widehat{u}^{\ 2}
& \color{red} \widehat{u}^{\ 1}
\vspace*{0.2cm} \\
\color{red} d^{\ 1} & \color{green} d^{\ 2} & \color{cyan} d^{\ 3}
& \color{blue} e^{\ -} \hspace*{0.2cm} & \color{blue} | & \color{blue} e^{\ +}
& \color{cyan} \widehat{d}^{\ 3} & \color{green} \widehat{d}^{\ 2}
& \color{red} \widehat{d}^{\ 1} \
\end{array} \color{blue} \right ) \vspace*{-0.5cm} \hspace*{-0.3cm}
\begin{array}{l}
\vspace*{-1.5cm} \\
^{\ \color{blue} \dot{\gamma} \ \rightarrow \ L}
\end{array}
\vspace*{0.5cm} \\
\color{blue}
= \ \left ( \ f \ \right )^{\ \dot{\gamma}}
\end{array}
\vspace*{-0.4cm}
\end{equation}

\noindent
and admitting a gravitational background field is in this minimal neutrino
flavor embedding anomalous , i.e. the global symmetry is broken by
winding gravitational fields \cite{PMang} .

\vspace*{-0.3cm}
\begin{equation}
\label{eq:a-2}
\begin{array}{l}
\left . j_{\ \varrho} \ ( \ B \ - \ L \ ) \right |_{\ 3 \times 15} \ =
\vspace*{0.2cm} \\
\ \sum_{\ fmlies} \ \left \lbrack \begin{array}{c}
\left ( \ u^{\ *} \ \right )^{\ \alpha \ \dot{c}} 
\ \left ( \ \sigma_{\ \mu} \ \right )_{\ \alpha \ \dot{\gamma}}
\ \left ( \ u \ \right )^{\ \dot{\gamma} \ c}
\ {\color{black} -} 
\vspace*{0.1cm} \\
\hspace*{1.0cm} 
- \ \left ( \ \widehat{u}^{\ *} \ \right )^{\ \alpha \ c}
\ \left ( \ \sigma_{\ \mu} \ \right )_{\ \alpha \ \dot{\gamma}}
\ \left ( \ \widehat{u} \ \right )^{\ \dot{\gamma} \ \dot{c}}
\vspace*{0.2cm} \\
+ \ \left ( \ d^{\ *} \ \right )^{\ \alpha \ \dot{c}}
\ \left ( \ \sigma_{\ \mu} \ \right )_{\ \alpha \ \dot{\gamma}}
\ \left ( \ d \ \right )^{\ \dot{\gamma} \ c}
\ {\color{black} -} 
\vspace*{0.1cm} \\
\hspace*{1.0cm} 
- \ \left ( \ \widehat{d}^{\ *} \ \right )^{\ \alpha \ c}
\ \left ( \ \sigma_{\ \mu} \ \right )_{\ \alpha \ \dot{\gamma}}
\ \left ( \ \widehat{d} \ \right )^{\ \dot{\gamma} \ \dot{c}}
\vspace*{0.2cm} \\
- \ \left ( \ e^{\ -} \ \right )^{\ * \ \alpha}
\ \left ( \ \sigma_{\ \mu} \ \right )_{\ \alpha \ \dot{\gamma}}
\ \left ( \ e^{\ -} \ \right )^{\ \dot{\gamma}}
\ {\color{black} +} 
\vspace*{0.1cm} \\
\hspace*{1.0cm} 
+ \ \left ( \ e^{\ +} \ \right )^{\ * \ \alpha}
\ \left ( \ \sigma_{\ \mu} \ \right )_{\ \alpha \ \dot{\gamma}}
\ \left ( \ e^{\ +} \ \right )^{\ \dot{\gamma}}
\vspace*{0.2cm} \\
- \ \left ( \ \nu \ \right )^{\ * \ \alpha}
\ \left ( \ \sigma_{\ \mu} \ \right )_{\ \alpha \ \dot{\gamma}}
\ \left ( \ \nu \ \right )^{\ \dot{\gamma}}
\end{array} \right \rbrack
 {\color{black} e_{\ \varrho}^{\ \mu}}
\vspace*{0.3cm} \\
g_{\ \varrho \ \tau} \ = \  e_{\ \varrho}^{\ \mu} \ \eta_{\ \mu \ \nu}
\  e_{\ \tau}^{\ \nu} \ \mbox{: metric}
\hspace*{0.2cm} ; \hspace*{0.2cm} 
e_{\ \varrho}^{\ \mu} \ \mbox{: vierbein}
\hspace*{0.2cm} ; 
\vspace*{0.1cm} \\
^{\ *} \ \mbox{: hermitian operator conjugation}
\hspace*{0.1cm} ; \hspace*{0.1cm}
\left ( \ u^{\ *} \ \right )^{\ \alpha \ \dot{c}} \ \equiv 
\ \left ( u^{\ \dot{\alpha} \ c} \ \right )^{\ *}
\vspace*{0.1cm} \\
\eta_{\ \mu \nu} \ = \ diag \ (\ 1 , -1, -1, -1 \ )
\ \mbox{: tangent space metric}
\vspace*{0.2cm} \\
^{\ c} \ \left ( ^{\ \dot{c}} \ \right ) \ \mbox{: color and anticolor}
\ ; \ c \ = \ 1,2,3
\end{array}
\end{equation}

\noindent
The contribution of charged fermion (pairs) 
$\ q \ , \widehat{q} \ ; \ e^{\ \mp} \ $ 
can be combined to vector currents -- Dirac doubling --
$\ \overline{q} \ \gamma_{\ \mu} \ q \ ; 
\ \overline{e} \ \gamma_{\ \mu} \ e \ $ with 
$\ q \ \rightarrow u,d,c,s,t,b \ ; \ e \ \rightarrow \ e^{\ -},\mu^{\ -},
\tau^{\ -} \ $ .

\noindent
The anomalous Ward
identy for the B - L current ( - density ) defined in eq. \ref{eq:a-2}
takes the form

\vspace*{-0.2cm}
\begin{equation}
\label{eq:a-3}
\begin{array}{l}
d^{\ 4} \ x \ \sqrt{ | \ g \ | \ } \ D^{\ \varrho} 
\ \left . j_{\ \varrho} \ ( \ B \ - \ L \ ) \right |_{\ 3 \times 15}
\ = 3 \ \widehat{A}_{\ 1} \ ( \ X \ )
\vspace*{0.2cm} \\
\widehat{A}_{\ 1} \ ( X ) \ = \ - \frac{1}{24} \ tr X^{\ 2}
\hspace*{0.1cm} ; \hspace*{0.1cm}
\left ( \ X \ \right )^{\ a}_{\hspace*{0.4cm} b} \ =
\ \frac{1}{2 \ \pi} \ \frac{1}{2} \ d \ x^{\ \varrho} \ \wedge \ d \ x^{\ \tau}
\ \left ( \ R^{\ a}_{\hspace*{0.4cm} b} \ \right )_{\ \varrho \ \tau}
\vspace*{0.2cm} \\
\left ( \ R^{\ a}_{\hspace*{0.4cm} b} \ \right )_{\ \varrho \ \tau} \ :
\ \left \lbrace \begin{array}{l}
\mbox{Riemann curvature tensor}
\vspace*{-0.1cm} \\
\mbox{\bf mixed components :} \ ^{\ a} \ _{\ b} \ \rightarrow 
\mbox{\bf tangent space}
\vspace*{-0.1cm} \\
\hspace*{0.4cm} _{\ \mu \ \nu} \ \rightarrow \ \mbox{\bf covariant space}
\end{array} \right .
\vspace*{0.2cm} \\
\color{black}
D^{\ \varrho} \ \left . j_{\ \varrho} \ ( \ B \ - \ L \ ) 
\ \right |_{\ 3 \times (16)} \ = \ 0
\end{array}
\end{equation}

\noindent
Before discussing the extension 

\begin{displaymath}
\left . j_{\ \varrho} \ ( \ B \ - \ L \ )
\ \right |_{\ 3 \times (15)} \ {\color{black} \rightarrow
\ \left . j_{\ \varrho} \ ( \ B \ - \ L \ )
\ \right |_{\ 3 \times (16)}} 
\end{displaymath}

\noindent
which renders the latter current conserved,
lets define the quantities appearing in eq. \ref{eq:a-3} :

\vspace*{-0.2cm}
\begin{equation}
\label{eq:a-4}
\begin{array}{l}
\left ( \ R^{\ {\color{red} a}}_{\hspace*{0.4cm} {\color{black} b}} 
\ \right )_{\ \varrho \ \tau} \ = 
\ e^{\ {\color{red} a}}_{\ \mu} \ e_{\ {\color{black} b} \ \nu}
\left ( \ R^{\ \mu}_{\hspace*{0.4cm} \nu} \ \right )_{\ \varrho \ \tau}
\hspace*{0.2cm} ; \hspace*{0.2cm} e_{\ {\color{black} b} \ \nu} \ =
\ \eta_{\ {\color{black} b b'}} \ e^{\ {\color{black} b'}}_{\ \nu}
\vspace*{0.2cm} \\
\left ( \ R^{\ \mu}_{\hspace*{0.4cm} \nu} \ \right )_{\ \varrho \ \tau}
\ = \ \left ( \ \partial_{\ \varrho} \ \Gamma_{\ \tau} 
\ - \ \partial_{\ \tau} \ \Gamma_{\ \varrho}
\ + \ \Gamma_{\ \varrho} \ \Gamma_{\ \tau} \ -
\ \Gamma_{\ \tau} \ \Gamma_{\ \varrho} \ \right )^{\ \mu}_{\hspace*{0.4cm} \nu}
\vspace*{0.2cm} \\
\left ( \ \Gamma^{\ \mu}_{\hspace*{0.4cm} \nu} \ \right )_{\ \tau}
\ \mbox{: matrix valued 
$\ \left ( \ GL \ ( \ 4 \ , \ \mathbb{R} \ ) \ \right ) \ $
connection ; {\color{black} minimal here}}
\end{array}
\end{equation}

\noindent
For clarity eq. \ref{eq:a-3} is repeated below

\vspace*{-0.2cm}
\begin{equation}
\label{eq:a3c}
\begin{array}{l}
d^{\ 4} \ x \ \sqrt{ | \ g \ | \ } \ D^{\ \varrho} 
\ \left . j_{\ \varrho} \ ( \ B \ - \ L \ ) \right |_{\ 3 \times 15}
\ = 3 \ \widehat{A}_{\ 1} \ ( \ X \ )
\vspace*{0.2cm} \\
\widehat{A}_{\ 1} \ ( X ) \ = \ - \frac{1}{24} \ tr X^{\ 2}
\hspace*{0.1cm} ; \hspace*{0.1cm}
\left ( X \right )^{\ a}_{\hspace*{0.4cm} b} =
\frac{1}{2 \pi} \ \frac{1}{2} \ d \ x^{\ \varrho} \ \wedge \ d \ x^{\ \tau}
\ \left ( \ R^{\ a}_{\hspace*{0.4cm} b} \ \right )_{\ \varrho \ \tau}
\vspace*{0.2cm} \\
\left ( \ R^{\ a}_{\hspace*{0.4cm} b} \ \right )_{\ \varrho \ \tau} \ :
\ \left \lbrace \begin{array}{l}
\mbox{Riemann curvature tensor}
\vspace*{-0.1cm} \\
\mbox{\bf mixed components :} \ ^{\ a} \ _{\ b} \ \rightarrow 
\mbox{\bf tangent space}
\vspace*{-0.1cm} \\
\hspace*{0.4cm} _{\ \mu \ \nu} \ \rightarrow \ \mbox{\bf covariant space}
\end{array} \right .
\vspace*{0.2cm} \\
\color{black}
D^{\ \varrho} \ \left . j_{\ \varrho} \ ( \ B \ - \ L \ ) 
\ \right |_{\ 3 \times (16)} \ = \ 0
\end{array}
\end{equation}

\noindent
In eq. \ref{eq:a-3} \hspace*{0.1cm}
$\widehat{A} \ ( \ X \ \rightarrow \ \lambda \ ) \ = \ \frac{1}{2} \ \lambda \
/ \ \sinh \ ( \ \frac{1}{2} \ \lambda \ )$
denotes the Atiyah - Hirzebruch character or $\widehat{A} \ -$ genus
\cite{Hirzebr} with its integral over a compact,
euclidean signatured closed manifold $M_{\ 4}$ , capable of carrying on
SO4 - spin structure , becomes the index of the associated
{\it elliptic} Dirac equation

\vspace*{-0.3cm}
\begin{equation}
\label{eq:a-5}
\begin{array}{l}
{\displaystyle{\int}} \ \widehat{A} \ ( \ X_{\ E} \ ) \ = 
\ n_{\ R} \ - \ n_{\ L} \ = \ \mbox{integer}
\end{array}
\end{equation}

\noindent
In eq. \ref{eq:a-5} $n_{\ R,L}$ denote the 
numbers of right - and left - chiral solutions of the Dirac equation 
on $M_{\ 4}$ . The index $_{\ E} \ \rightarrow \ X_{\ E}$ shall indicate the
euclidean transposed curvature 2 - form , and is {\it adapted}
here to physical curved and uncurved space time . 

\noindent
For the latter case the first 
relation in eq. \ref{eq:a-3} \color{black} yields the integrated form
-- in the limit of infinitely heavy \\
${\cal{N}}_{\ F}$ ( eq. \ref{eq:a-1} ) --

\vspace*{-0.3cm}
\begin{equation}
\label{eq:a-6}
\begin{array}{l}
\Delta_{\ R-L} \ n_{\ \nu} \ = 
\ {\displaystyle{\int}} \ d^{\ 4} \ x \ \sqrt{ | \ g \ | \ } \ D^{\ \mu} \ j_{\
\mu}^{\ B \ - \ L \ (15)} \ = \ \color{red} 3 \ \color{blue} 
\ \Delta \ n \ ( \ \widehat{A} \ )
\vspace*{0.2cm} \\
\color{black}
3 \ = \ \mbox{\bf number of families = odd}
\hspace*{0.4cm} ; \hspace*{0.4cm} m_{\ \nu_{\ F}} \ \rightarrow \ 0
\end{array}
\end{equation}

\noindent
In eq. \ref{eq:a-6} \hspace*{0.05cm} 
$\Delta_{\ R-L} \ n_{\ \nu}$ \hspace*{0.05cm} denotes
the difference of right - chiral $\left ( \ \widehat{\nu} \ \right )$
\footnote{\color{black} \hspace*{0.1cm} 
$\ \widehat{\nu}_{\ \alpha} \ \equiv \ \varepsilon_{\ \alpha \ \beta}
\ \left ( \ \nu^{\ *} \ \right )^{\ \gamma} \ ; \ \varepsilon \ = \ i \
\sigma_{\ 2} \ ; \ \left ( \ \mbox{2nd Pauli matrix} \ \right ) \ $ 
stands for the left-chiral
neutrino fields transformed to the right-chiral basis .}
and left - chiral $\left ( \ \nu \ \right )$ flavors 
between times \hspace*{0.05cm} $t \ \rightarrow \ \pm \ \infty$ .

\noindent
Here a subtlety arises {\it precisely} because the number of families on the
level of $G_{\ SM}$ is odd , and the light neutrino flavors are not 'Dirac -
doubled' , which according to eq. \ref{eq:a-6} 
could potentially lead to a change in fermion number being odd , which
violates the rotation by \hspace*{0.05cm} $2 \ \pi$ \hspace*{0.05cm} symmetry ,
equivalent to \hspace*{0.05cm} $\widehat{\Theta}^{\ 2}$ 
$\ \left ( \ CPT^{\ 2} \ \right ) \ $ , {\it unless} \footnote{
\color{black} \hspace*{0.1cm} The obviously nontrivial relation between 
the compact Euclidean -  and noncompact asymptotic and locality restricted 
form of the index theorem involves not clearly formulated 
{\it boundary conditions} .
}

\vspace*{-0.3cm}
\begin{equation}
\label{eq:a-7}
\begin{array}{l}
\Delta \ n \ ( \ \widehat{A} \ ) \ \color{black} = \ \mbox{even}
\hspace*{0.5cm} ( \ \surd \hspace*{0.2cm} \mbox{for} \ dim \ = \ 4 \ \mbox{mod} \ 8 \ )
\end{array}
\vspace*{-0.0cm}
\end{equation}

\noindent
We now turn to the SO (10) inspired cancellation of the gravity induced
anomaly, giving rise to the completion of neutrino flavors to
3 families of 16-plets , sometimes called 'right-handed' neutrino flavors,
denoted $\ {\cal{N}} \ $ in the left-chiral basis in eq. \ref{eq:a-1}
\cite{PMalbufeira}

\vspace*{-0.1cm}
\begin{equation}
\label{eq:a1c}
\begin{array}{c}
\left ( 
\begin{array}{llll l llll}
\color{red} u^{\ 1} & \color{green} u^{\ 2} & \color{cyan} u^{\ 3}
& \color{blue} \nu \hspace*{0.2cm} & \color{blue} | & \color{blue} {\cal{N}}
& \color{cyan} \widehat{u}^{\ 3} & \color{green} \widehat{u}^{\ 2}
& \color{red} \widehat{u}^{\ 1}
\vspace*{0.2cm} \\
\color{red} d^{\ 1} & \color{green} d^{\ 2} & \color{cyan} d^{\ 3}
& \color{blue} e^{\ -} \hspace*{0.2cm} & \color{blue} | & \color{blue} e^{\ +}
& \color{cyan} \widehat{d}^{\ 3} & \color{green} \widehat{d}^{\ 2}
& \color{red} \widehat{d}^{\ 1} \
\end{array} \color{blue} \right ) \vspace*{-0.5cm} \hspace*{-0.3cm}
\begin{array}{l}
\vspace*{-1.5cm} \\
^{\ \color{blue} \dot{\gamma} \ \rightarrow \ L}
\end{array}
\vspace*{0.5cm} \\
\color{blue}
= \ \left ( \ f \ \right )^{\ \dot{\gamma}}
\end{array}
\end{equation}

\vspace*{-0.3cm}
\begin{equation}
\label{eq:a-8}
\begin{array}{l}
\left . j_{\ \varrho} \ ( \ B \ - \ L \ ) \right |_{\ 3 \times 15} 
\ {\color{black} \rightarrow
\ \left . j_{\ \varrho} \ ( \ B \ - \ L \ ) \right |_{\ 3 \times 16}}
\end{array}
\end{equation}

\vspace*{-0.2cm}
\begin{equation}
\label{eq:a3cc}
\begin{array}{l}
d^{\ 4} \ x \ \sqrt{ | \ g \ | \ } \ D^{\ \varrho} 
\ \left . j_{\ \varrho} \ ( \ B \ - \ L \ ) \right |_{\ 3 \times 15}
\ = 3 \ \widehat{A}_{\ 1} \ ( \ X \ )
\vspace*{0.2cm} \\
\widehat{A}_{\ 1} \ ( X ) \ = \ - \ \frac{1}{24} \ tr X^{\ 2}
\hspace*{0.1cm} ; \hspace*{0.1cm}
\left ( X \right )^{\ a}_{\hspace*{0.4cm} b} \ =
\ \frac{1}{2 \ \pi} \ \frac{1}{2} \ d \ x^{\ \varrho} \ \wedge \ d \ x^{\ \tau}
\ \left ( \ R^{\ a}_{\hspace*{0.4cm} b} \ \right )_{\ \varrho \ \tau}
\vspace*{0.2cm} \\
\left ( \ R^{\ a}_{\hspace*{0.4cm} b} \ \right )_{\ \varrho \ \tau} \ :
\ \left \lbrace \begin{array}{l}
\mbox{Riemann curvature tensor}
\vspace*{-0.1cm} \\
\mbox{\bf mixed components :} \ ^{\ a} \ _{\ b} \ \rightarrow 
\mbox{\bf tangent space}
\vspace*{-0.1cm} \\
\hspace*{2.4cm} _{\ \mu \ \nu} \ \rightarrow \mbox{\bf covariant space}
\end{array} \right .
\vspace*{0.2cm} \\
\color{black}
D^{\ \varrho} \ \left . j_{\ \varrho} \ ( \ B \ - \ L \ ) 
\ \right |_{\ 3 \times (16)} \ = \ 0
\hspace*{2.0cm} {\color{black} \rightarrow}
\end{array}
\end{equation}

\vspace*{-0.3cm}
\begin{equation}
\label{eq:a-9}
\begin{array}{l}
\left . j_{\ \varrho} \ ( \ B \ - \ L \ ) \right |_{\ 3 \times 15} 
\ \rightarrow \ {\color{black} 
\left . j_{\ \varrho} \ ( \ B \ - \ L \ ) \right |_{\ 3 \times 16} } \ =
\vspace*{0.2cm} \\
\ \sum_{\ fmlies} \ \left \lbrack \begin{array}{c}
\left ( \ u^{\ *} \ \right )^{\ \alpha \ \dot{c}} 
\ \left ( \ \sigma_{\ \mu} \ \right )_{\ \alpha \ \dot{\gamma}}
\ \left ( \ u \ \right )^{\ \dot{\gamma} \ c}
\ {\color{black} -} 
\vspace*{0.1cm} \\
\hspace*{1.0cm} 
- \ \left ( \ \widehat{u}^{\ *} \ \right )^{\ \alpha \ c}
\ \left ( \ \sigma_{\ \mu} \ \right )_{\ \alpha \ \dot{\gamma}}
\ \left ( \ \widehat{u} \ \right )^{\ \dot{\gamma} \ \dot{c}}
\vspace*{0.2cm} \\
+ \ \left ( \ d^{\ *} \ \right )^{\ \alpha \ \dot{c}}
\ \left ( \ \sigma_{\ \mu} \ \right )_{\ \alpha \ \dot{\gamma}}
\ \left ( \ d \ \right )^{\ \dot{\gamma} \ c}
\ {\color{black} -} 
\vspace*{0.1cm} \\
\hspace*{1.0cm} 
- \ \left ( \ \widehat{d}^{\ *} \ \right )^{\ \alpha \ c}
\ \left ( \ \sigma_{\ \mu} \ \right )_{\ \alpha \ \dot{\gamma}}
\ \left ( \ \widehat{d} \ \right )^{\ \dot{\gamma} \ \dot{c}}
\vspace*{0.2cm} \\
- \ \left ( \ e^{\ -} \ \right )^{\ * \ \alpha}
\ \left ( \ \sigma_{\ \mu} \ \right )_{\ \alpha \ \dot{\gamma}}
\ \left ( \ e^{\ -} \ \right )^{\ \dot{\gamma}}
\ {\color{black} +} 
\vspace*{0.1cm} \\
\hspace*{1.0cm} 
+ \ \left ( \ e^{\ +} \ \right )^{\ * \ \alpha}
\ \left ( \ \sigma_{\ \mu} \ \right )_{\ \alpha \ \dot{\gamma}}
\ \left ( \ e^{\ +} \ \right )^{\ \dot{\gamma}}
\vspace*{0.2cm} \\
- \ \left ( \ \nu \ \right )^{\ * \ \alpha}
\ \left ( \ \sigma_{\ \mu} \ \right )_{\ \alpha \ \dot{\gamma}}
\ \left ( \ \nu \ \right )^{\ \dot{\gamma}}
\hspace*{0.1cm} {\color{black} +} 
\vspace*{0.1cm} \\
\hspace*{1.0cm} 
{\color{black} \underbrace
{\color{black} \left ( \ {\cal{N}} \ \right )^{\ * \ \alpha}
+ \ \left ( \ \sigma_{\ \mu} \ \right )_{\ \alpha \ \dot{\gamma}}
\ \left ( \ {\cal{N}} \ \right )^{\ \dot{\gamma}}}
}
\end{array} \right \rbrack
 {\color{black} e_{\ \varrho}^{\ \mu}}
\vspace*{0.3cm} \\
g_{\ \varrho \ \tau} \ = \  e_{\ \varrho}^{\ \mu} \ \eta_{\ \mu \ \nu}
\  e_{\ \tau}^{\ \nu} \ \mbox{: metric}
\hspace*{0.2cm} ; \hspace*{0.2cm} 
e_{\ \varrho}^{\ \mu} \ \mbox{: vierbein}
\hspace*{0.2cm} ; 
\vspace*{0.1cm} \\
^{\ *} \ \mbox{: hermitian operator conjugation}
\hspace*{0.1cm} ; \hspace*{0.1cm}
\left ( \ u^{\ *} \ \right )^{\ \alpha \ \dot{c}} \ \equiv 
\ \left ( u^{\ \dot{\alpha} \ c} \ \right )^{\ *}
\hspace*{0.3cm} ; 
\vspace*{0.1cm} \\
\eta_{\ \mu \nu} \ = \ diag \ (\ 1 , -1, -1, -1 \ )
\ \mbox{: tangent space metric}
\vspace*{0.2cm} \\
^{\ c} \ \left ( ^{\ \dot{c}} \ \right ) \ \mbox{: color and anticolor}
\ ; \ c \ = \ 1,2,3
\vspace*{0.3cm} \\
\color{black}
D^{\ \varrho} \ \left . j_{\ \varrho} \ ( \ B \ - \ L \ )
\ \right |_{\ 3 \times (16)} \ = \ 0
\end{array}
\end{equation}

\noindent
Let me illustrate the triple doubling in the elimination
of the anomaly in the covariant divergence of 
$\ \left . j_{\ \varrho} \ ( \ B \ - \ L \ ) \right |_{\ 3 \times 15} \ $ in
eq. \ref{eq:a-2} as seen through the left-chiral basis ,
repeating only the $\ \nu \ , \ {\cal{N}} \ $ components of
the B - L current in eq. \ref{eq:a-9}

\vspace*{-0.3cm}
\begin{equation}
\label{eq:a-10}
\begin{array}{c}
{\color{black}
\left . j_{\ \varrho} \ ( \ B \ - \ L \ ) \right |_{\ 3 \times 16} } \ =
\vspace*{0.2cm} \\
\ \sum_{\ fmlies} \ \left \lbrack \begin{array}{c}
\cdots 
\vspace*{0.2cm} \\
\begin{array}{l}
- \ \left ( \ \nu \ \right )^{\ * \ \alpha}
\ \left ( \ \sigma_{\ \mu} \ \right )_{\ \alpha \ \dot{\gamma}}
\ \left ( \ \nu \ \right )^{\ \dot{\gamma}} \ +
\vspace*{0.1cm} \\
\hspace*{0.3cm}
\underbrace
{\color{black} \left ( \ {\cal{N}} \ \right )^{\ * \ \alpha}
\ \left ( \ \sigma_{\ \mu} \ \right )_{\ \alpha \ \dot{\gamma}}
\ \left ( \ {\cal{N}} \ \right )^{\ \dot{\gamma}}}
\end{array}
\end{array} \right \rbrack
\vspace*{0.4cm} \\
\begin{array}{|l|ll|}
\hline
\vspace*{-0.47cm} \\ & & \vspace*{-0.2cm} \\
 & \nu^{\ \dot{\gamma}}_{\ F} & {\cal{N}}^{\ \dot{\gamma}}_{\ F}
\vspace*{-0.3cm} \\ & & \vspace*{0.0cm} \\ \hline
\vspace*{-0.5cm} \\ & & \vspace*{-0.2cm} \\
B \ - \ L & -1 & +1 
\vspace*{-0.2cm} \\ & & \\
\hline
\end{array}
\hspace*{0.2cm} ; \hspace*{0.2cm} 
F \ = \ 1,2,3 \ \mbox{family}
\end{array}
\end{equation}


\begin{center}
\vspace*{-0.1cm}
{\bf \color{black}
1-2a There does not exist a symmetry -- within
the standard model including gravity 
and containing only chiral 16 families of SO (10) -- 
enforcing the vanishing of neutrino mass(es), 
yet chiral extensions can accomplish this 
}
\label{'1-2a'}
\end{center}
\vspace*{0.1cm}

\noindent
Here I briefly describe one such extension. It consists
of replacing in each family the SO (10) induced $\ {\cal{N}}_{\ F} \ $ flavors
by four alternative ( sterile ) 
$\ {\cal{X}}_{\ J \ = \ 2,3,4,5 \ ; \ F} \ $ flavors, singlets under the 
electroweak gauge group with genuinely chiral B - L charges, 
changing the structure in eq. \ref{eq:a-10} to

\vspace*{-0.5cm} 
\begin{equation}
\label{eq:b1}
\begin{array}{c}
{\color{black}
\left . j_{\ \varrho} \ ( \ B \ - \ L \ ) \right |_{\ 3 \times 19} } \ =
\vspace*{0.0cm} \\
\ \sum_{\ F} \ \left \lbrack \begin{array}{c}
\cdots 
\vspace*{0.0cm} \\
\begin{array}{l}
- \ \left ( \ \nu \ \right )^{\ * \ \alpha}
\ \left ( \ \sigma_{\ \mu} \ \right )_{\ \alpha \ \dot{\gamma}}
\ \left ( \ \nu \ \right )^{\ \dot{\gamma}}
\hspace*{-0.0cm} {\color{black} +} 
\vspace*{0.1cm} \\
{\color{black} \hspace*{1.0cm} +} \ {\color{black} \underbrace  
\sum_{\ J=2}^{\ 5} \ (\chi)_{\ J} 
\ \left ( \ {\cal{X}}_{\ J} \ \right )^{\ * \ \alpha}
\ \left ( \ \sigma_{\ \mu} \ \right )_{\ \alpha \ \dot{\gamma}}
\ \left ( \ {\cal{X}}_{\ J} \ \right )^{\ \dot{\gamma}}}
\end{array}
\end{array} \right \rbrack
\vspace*{0.5cm} \\
\begin{array}{|l|cll ll|}
\hline
\vspace*{-0.47cm} \\ & & & & & \vspace*{-0.2cm} \\
& \nu^{\ \dot{\gamma}}_{\ F} \ = \ {\cal{X}}^{\ \dot{\gamma}}_{\ 1 , F}
& {\cal{X}}^{\ \dot{\gamma}}_{\ 2 , F}
& {\cal{X}}^{\ \dot{\gamma}}_{\ 3 , F}
& {\cal{X}}^{\ \dot{\gamma}}_{\ 4 , F}
& {\cal{X}}^{\ \dot{\gamma}}_{\ 5 , F}
\vspace*{-0.3cm} \\ & & & & & \vspace*{0.0cm} \\ \hline
\vspace*{-0.5cm} \\ & & & & & \vspace*{-0.2cm} \\
\begin{array}{c}B \ - \ L 
\\ = \ (\chi)_{\ J} 
\end{array} & -1 & -5 & -9 & \ 7 & 8 
\vspace*{-0.2cm} \\ & & & & & \\
\hline
\end{array}
\hspace*{0.2cm} ; 
\vspace*{0.2cm} \\ 
\begin{array}{c}
F \ = \ 1,2,3 \ \mbox{family} \\
J \ = \ 1,2\cdots,5
\end{array}
\end{array}
\vspace*{-0.0cm}
\end{equation}

\noindent
The genuinely chiral couplings 
$(\chi)_{\ J=1,\cdots,5} \ = 
\ \left \lbrack \ -1 \ , \ -5 \ , \ -9 \ ; \ 7 \ , \ 8 
\ \right \rbrack \ $ 
for neutrino flavors as shown in eq. \ref{eq:b1} with 5 chiral base flavors
merit some comments :

\begin{description}
\item 1) a sequence of charges $(\chi)_{\ J} \ , \ J \ = \ 1,\cdots,N$
with respect to the left-chiral basis -- to be specific --
shall be called {\it genuinely} chiral , if none of the charges vanishes and 
no pairs of opposite charge 
$\ \left \lbrack \ \pm \ (\chi) \ \right \rbrack \ $ 
are admitted.

\item 2) the absence of an anomaly of the associated chiral current , 
of the form
given for neutrino flavors in eqs. \ref{eq:a-2} , \ref{eq:a-8} and \ref{eq:b1}
including also gravitational fields leads {\it in 4 dimensions} 
to the two conditions

\vspace*{-0.3cm}
\begin{equation}
\label{eq:b2}
\begin{array}{l}
\sum_{\ J}^{\ N} \ (\chi)_{\ J} \ = \ 0 
\hspace*{0.2cm} , \hspace*{0.2cm}
\sum_{\ J}^{\ N} \ \left \lbrack \ (\chi)_{\ J} \ \right \rbrack^{\ 3} \ = \ 0
\end{array}
\end{equation}

\item 3)
there does not exist a genuinely chiral set 
$\ \left \lbrace \ (\chi)_{\ J} \ , \ J \ = \ 1,\cdots,N \ \right \rbrace \ $
for $\ N \ < \ 5 \ $ .

For N = 3,4 it is equivalent to show that the two equations

\vspace*{-0.3cm}
\begin{equation}
\label{eq:b3}
\begin{array}{l}
A \ + \ B \ = \ C \ + \ D 
\hspace*{0.2cm} , \hspace*{0.2cm}
A^{\ 3} \ + \ B^{\ 3} \ = 
\ C^{\ 3} \ + \ D^{\ 3} 
\hfill \rightarrow
\vspace*{0.2cm} \\
A \ = \ x \ - a \ , \ B \ = \ x \ + \ a \ , \ C \ = \ x \ - \ b 
\ , \ D \ = \ x \ + \ b
\hfill \rightarrow
\vspace*{0.2cm} \\
x \ a^{\ 2} \ = \ x \ b^{\ 2} 
\hspace*{0.2cm} \rightarrow \hspace*{0.2cm}
\left \lbrace \ \begin{array}{l}
x \ = \ 0 
\hspace*{0.2cm} \mbox{or} \hspace*{0.2cm}
x \ \neq \ 0 \ ; \ b \ = \ \pm \ a
\end{array} \right .
\end{array}
\end{equation}
 
have no solution, satisfying the conditions for genuine chirality .
\end{description}

\begin{description}
\item 4) There are infinitely many solutions for $\ N \ \geq \ 5 \ $,
with chiral charges relatively irrational as well as rational .
For integer values and N = 5 with the norm
$\ \left | \ (\chi) \ \right | \ = \ \sum \ \left | \ (\chi)_{\ J} \ \right |
\ $ the solution with smallest norm is unique up to an overall change 
of sign \footnote{\color{black} \hspace*{0.1cm} 
It is due to Paul Frampton , on a beautiful morning in 1993 , along the coastal
range above the mediterranean sea near Cassis, France .
}

\vspace*{-0.3cm}
\begin{equation}
\label{eq:b4}
\begin{array}{l}
(\chi)_{\ J} \ = \ \left \lbrack \ -1 \ , \ -5 \ , \ -9 \ ; \ 7 \ , \ 8 
\ \right \rbrack
\end{array}
\end{equation}

\end{description}
\newpage

{\color{black}


\noindent
References for Appendix 1 and Supplementary material

}

\end{document}